\documentclass[aps,superscriptaddress,showpacs,showkeys,nofootinbib,floatfix]{revtex4}
\usepackage{graphicx,epsfig,wrapfig,bbm}
\usepackage{bm}
\usepackage{amssymb,amsmath,slashed,slashbox,multirow,amsfonts,latexsym}
\usepackage[normalem]{ulem} 
\usepackage{dsfont}
\usepackage{pdfpages}

\newcommand{\be}{\begin{equation}}
\newcommand{\ee}{\end{equation}}
\newcommand{\ba}{\begin{eqnarray}}
\newcommand{\ea}{\end{eqnarray}}

\newcommand{\ud}{\mathrm{d}}

\newcommand{\uTr}{\mathrm{Tr}}

\newcommand{\sgn}{\text{sgn}}

\newlength\savedwidth

\newcommand\whline{\noalign{\global\savedwidth\arrayrulewidth
\global\arrayrulewidth 1pt}%
\hline
\noalign{\global\arrayrulewidth\savedwidth}}
\newcommand{\uvec}[1]{\boldsymbol{#1}}

\begin{document}
%
%
\newcommand*{\Ecolepolytechnique}{Centre de Physique Th\'eorique, \'Ecole polytechnique, CNRS, Universit\'e Paris-Saclay, F-91128 Palaiseau, France}\affiliation{\Ecolepolytechnique}
\newcommand*{\Pavia}{Dipartimento di Fisica, 
  Universit\`a degli Studi di Pavia, Pavia, Italy}\affiliation{\Pavia}
\newcommand*{\INFN}{Istituto Nazionale di Fisica Nucleare, 
  Sezione di Pavia, Pavia, Italy}\affiliation{\INFN}

\title{Multipole decomposition of the nucleon transverse phase space}
\author{C.~Lorc\'e}\affiliation{\Ecolepolytechnique}
\author{B.~Pasquini}\affiliation{\Pavia}\affiliation{\INFN}
\date{\today}
\vspace{0.5in}
\begin{abstract}{
We present a complete study of the leading-twist quark Wigner distributions in the nucleon, discussing both the $\mathsf T$-even and $\mathsf T$-odd sector, along with all the possible configurations of the quark and nucleon polarizations. We identify the basic multipole structures associated with each distribution in the transverse phase space, providing a transparent interpretation of the spin-spin and spin-orbit correlations of quarks and nucleon encoded in these functions.  Projecting the multipole parametrization of the Wigner functions onto the transverse-position and  the transverse-momentum spaces, we find a natural link with the corresponding multipole parametrizations for the generalized parton distributions and transverse-momentum dependent parton distributions, respectively. Finally,  we show results for all the distributions in the transverse phase space, introducing a representation that allows one to visualize simultaneously the multipole structures in both  the transverse-position and transverse-momentum spaces.} 
\end{abstract}
\maketitle

\section{Introduction}

The concept of phase-space distributions borrowed from Classical Mechanics has been transposed to Quantum Mechanics~\cite{Wigner:1932eb}, where it finds numerous applications~\cite{Balazs:1983hk,Hillery:1983ms, Lee:1995}. Phase-space distributions have also been defined in the context of Relativistic Field Theory~\cite{Carruthers:1982fa,Hakim:1976bn,DeGroot:1980dk} and more specifically in Quantum ChromoDynamics~\cite{Elze:1986qd,Ochs:1998qj,Heinz:1983nx,Heinz:1984yq,Elze:1986hq}. The six-dimensional version of these phase-space distributions has been discussed for the first time in connection with Generalized Parton Distributions (GPDs) in Refs.~\cite{Ji:2003ak,Belitsky:2003nz}. However, in this case the physical interpretation is plagued by relativistic corrections. This issue has been solved in the light-front formalism by integrating over the longitudinal spatial dimension~\cite{Soper:1976jc,Burkardt:2000za,Burkardt:2002hr,Burkardt:2005hp}, leading to five-dimensional phase-space distributions~\cite{Lorce:2011kd} which are related \emph{via} a proper Fourier transform to Generalized Transverse-Momentum dependent Distributions (GTMDs)~\cite{Meissner:2009ww,Lorce:2011dv,Lorce:2013pza}.

The GTMDs recently received increasing attention due to the fact that they can be considered as the \emph{mother distributions} of GPDs and Transverse-Momentum dependent Distributions (TMDs)~\cite{Meissner:2009ww,Lorce:2011dv,Lorce:2013pza}. Moreover, it turned out that they are naturally related to the parton orbital angular momentum (OAM)~\cite{Lorce:2011kd,Hatta:2011ku,Lorce:2011ni,Liu:2015xha}. Except possibly at low-$x$~\cite{Martin:1999wb,Khoze:2000cy,Martin:2001ms,Albrow:2008pn,Martin:2009ku}, no experimental process directly sensitive to GTMDs has been identified so far. Nevertheless, these distributions can be studied using phenomenological or perturbative models~\cite{Meissner:2009ww,Lorce:2011kd,Kanazawa:2014nha,Mukherjee:2014nya,Liu:2014vwa,Liu:2015eqa,Miller:2014vla,Mukherjee:2015aja,Burkardt:2015qoa}, and can also in principle be computed on a lattice~\cite{Ji:2013dva}.

In total, there are at leading twist 32 quark phase-space distributions among which half originate from naive $\mathsf T$-odd GTMDs. In a former work~\cite{Lorce:2011kd}, we studied the four naive $\mathsf T$-even distributions associated with longitudinal polarization. Here, we present for the first time a complete study of all the 32 distributions.

Even though the number of independent functions is fixed by hermiticity and space-time symmetries, the parametrization of the correlator is not unique. In some sense, choosing a particular parametrization amounts to choosing a particular basis for decomposing the correlator. One can change the basis, but not the number of independent basis elements. The choice of a particular decomposition is arbitrary and is often motivated by the simplicity of the mathematical expressions. However, simple mathematical expressions often turn out to have rather obscure physical interpretation. 

In this work, we choose  natural combinations of GTMDs corresponding to distributions for all the possible configurations of the target and quark polarizations, and perform a multipole decomposition of each of these distributions  in the transverse phase space.   
This multipole analysis allows us to identify in a clear way all the possible spin-spin and spin-orbit correlations of quarks and nucleon in  phase space, and has a direct connection with the spin densities in impact-parameter space described by GPDs and the transverse-momentum densities described by TMDs.
\newline

The plan of the manuscript is as follows.
 In Sec.~\ref{section:1} we review the definition of the Wigner distributions obtained
by Fourier transform of the GTMDs to the impact-parameter
space, and we summarize the transformation properties of these functions under time-reversal, parity and hermitian conjugation.
 In Sec.~\ref{sec:multipole}, we outline the general method for the decomposition of the Wigner functions in basic multipoles in the transverse phase space, and we identify all the possible correlations between target polarization, quark polarization and quark OAM encoded in these phase-space distributions. 
In Sec.~\ref{section:3} we introduce a new representation of the transverse phase space, which allows one to visualize the multipole structures simultaneously in both the transverse-momentum and transverse-position spaces.
 In Sec.~\ref{section:4} we present and discuss the results of both the  $\mathsf T$-even and $\mathsf T$-odd distributions, for all the possible quark and target polarizations.
  Although the calculation is performed within  a specific relativistic light-front constituent quark model~\cite{Lorce:2011dv}, we can draw 
 general and model-independent conclusions about the physical information encoded in these functions. Finally, we summarize our results in Sec.~\ref{section:5}.

\section{Polarized relativistic phase-space distributions}
\label{section:1}
We introduce two lightlike four-vector $n_\pm$ satisfying $n_+\cdot n_-=1$. Any four-vector $a^\mu$ can then be decomposed as
\be
a^\mu=a^+n^\mu_++a^- n^\mu_-+a^\mu_T,
\ee
where $a^\pm=a\cdot n_\mp$ and $a^\mu_T=-\delta^{\mu\nu}_Ta_\nu$ with
\be\label{inner}
\delta^{\mu\nu}_T\equiv n^\mu_+n^\nu_-+n^\mu_-n^\nu_+-g^{\mu\nu}.
\ee
Writing the light-front components of $a^\mu$ as $[a^+,a^-,\uvec a_T]$, we have $a^2_T=-\uvec a^2_T$. The transverse skewed product is then given by
\be
 \epsilon^{\mu\nu}_T\equiv\epsilon^{\mu\nu\alpha\beta}n_{-\alpha}n_{+\beta}
\ee 
with $\epsilon_{0123}=1$ so that $\epsilon^{12}_T=-\epsilon^{21}_T=1$. Denoting by $\vec P=\tfrac{1}{2}(\vec p\,'+\vec p)$ the average hadron three-momentum and working in a frame where $\uvec P_T=\uvec 0_T$, any spatial three-vector $\vec a$ can similarly be decomposed as
\be\label{vecdec}
\vec a=a_L\,\hat P+\vec a_T,
\ee
where $a_L=\vec a\cdot\hat P$ with $\hat P=\vec P/|\vec P|$, and $a^i_T=\delta^{ij}_Ta^j$. For later convenience, we shall also denote the longitudinal component of the skewed product as $(\vec a\times\vec b)\cdot\hat P=\epsilon^{ij}_Ta^i_Tb^j_T=(\uvec a_T\times\uvec b_T)_L$.

The quark GTMD correlator is defined as~\cite{Meissner:2009ww,Lorce:2013pza}
\begin{equation}\label{GTMDcorr-def}
W^{ab}_{\Lambda'\Lambda}\equiv\int\ud k^-\int\frac{\ud^4z}{(2\pi)^4}\,e^{ik\cdot z}\,\langle P+\tfrac{\Delta}{2},\Lambda'|\overline\psi_b(-\tfrac{z}{2})\,\mathcal W\,\psi_a(\tfrac{z}{2})|P-\tfrac{\Delta}{2},\Lambda\rangle,
\end{equation}
where $\mathcal W$ is an appropriate Wilson line ensuring color gauge invariance, $k$ is the quark average four-momentum conjugate to the quark field separation $z$, and $|p,\Lambda\rangle$ is the spin-$1/2$ target state with four-momentum $p$ and light-front helicity $\Lambda$. The correlator $W^{ab}_{\Lambda'\Lambda}$ can be thought of as a $2\times 2$ matrix in target polarization space and as a $4\times 4$ matrix in Dirac space. At leading twist, one can interpret
\begin{equation}\label{GTMDcorr}
W_{\vec S\vec S^q}=\tfrac{1}{8}\sum_{\Lambda',\Lambda}(\mathds 1+\vec S\cdot\vec \sigma)_{\Lambda\Lambda'}\,\uTr[W_{\Lambda'\Lambda}\Gamma_{\vec S^q}]
\end{equation}
with $\Gamma_{\vec S^q}=\gamma^++S^q_L\,\gamma^+\gamma_5+S^{qj}_T\,i\sigma^{j+}_T\gamma_5$,
as the GTMD correlator describing the distribution of quarks with polarization $\vec S^q$ inside a target with polarization $\vec S$~\cite{Lorce:2011zta}.

The corresponding phase-space distribution is obtained by performing an appropriate Fourier transform~\cite{Lorce:2011kd}
\begin{equation}
\rho_{\vec S\vec S^q}(x,\uvec k_T,\uvec b_T;\hat P,\eta)=\int\frac{\ud^2\Delta_T}{(2\pi)^2}\,e^{-i\uvec\Delta_T\cdot\uvec b_T}\,W_{\vec S\vec S^q}(P,k,\Delta;n_-)\big|_{\xi=0},
\end{equation}
where $x=k^+/P^+$ and $\uvec k_T$ are, respectively, the longitudinal fraction and transverse component of the quark average momentum, $\uvec b_T$ is the quark average impact parameter conjugate to the transverse-momentum transfer $\uvec\Delta_ T$, $\xi=-\Delta^+/2P^+$ is the fraction of longitudinal momentum transfer, and $\eta=\sgn(n^0_-)$. This phase-space distribution can be interpreted semi-classically as giving the quasi-probability of finding a quark with polarization $\vec S^q$, transverse position $\vec b_T$ and light-front momentum $(xP^+,\vec k_T)$ inside a spin-$1/2$ target with polarization $\vec S$~\cite{Lorce:2011kd}. The hermiticity property of the GTMD correlator~\eqref{GTMDcorr} ensures that these phase-space distributions are always real-valued~\cite{Lorce:2011ni}, see Table~\ref{Transfprop}, which is 
consistent with their quasi-probabilistic interpretation. The behavior of the variables $x$, $\uvec k_T$, $\uvec b_T$, $\hat P$, $\eta$, $\vec S$, and $\vec S^q$ under parity and time-reversal\footnote{We work here with the \emph{passive} form of parity and time-reversal transformations so that the two lightlike four-vectors $n_\pm$ also undergo the transformations. In light-front 
quantization, one often choose instead the \emph{active} form so that these four-vectors remain invariant, with the annoying consequence that the components $a^\pm$ are then transformed into 
each other. This can be cured by performing an additional $\pi$-rotation about \emph{e.g.} the $x$-axis, \emph{i.e.} by defining light-front parity and time-reversal as 
$\mathsf P_\text{LF}=R_x(\pi)\mathsf P$ and $\mathsf T_\text{LF}=R_x(\pi)\mathsf T$, see~\cite{Soper:1972xc,Carlson:2003je,Brodsky:2006ez,Lorce:2013pza}.} can also be read from Table~\ref{Transfprop} by looking at the arguments of the functions.

There are 16 independent polarization configurations~\cite{Lorce:2011kd,Lorce:2013pza} which correspond to particular linear combinations of the 16 independent complex-valued GTMDs~\cite{Meissner:2009ww,Lorce:2013pza}.
By construction, the real and imaginary parts of the GTMDs have opposite behavior under naive time-reversal transformation~\cite{Meissner:2009ww,Lorce:2013pza}, which is defined as usual time-reversal but without interchange of initial and final states. Similarly, we can separate each phase-space distribution into naive $\mathsf T$-even and $\mathsf T$-odd contributions
\begin{equation}
\rho_{\vec S\vec S^q}=\rho^e_{\vec S\vec S^q}+\rho^o_{\vec S\vec S^q},
\end{equation}
where 
\begin{equation}
\rho^{e,o}_{\vec S\vec S^q}(x,\uvec k_T,\uvec b_T;\hat P,\eta)=\pm\rho^{e,o}_{\vec S\vec S^q}(x,\uvec k_T,\uvec b_T;\hat P,-\eta)=\pm\rho^{e,o}_{-\vec S-\vec S^q}(x,-\uvec k_T,\uvec b_T;-\hat P,\eta). 
\end{equation}
In some sense, the naive $\mathsf T$-even contributions describe the \emph{intrinsic} distribution of quarks inside the target, whereas the naive $\mathsf T$-odd contributions describe how initial- and final-state interactions modify this distribution. 

So, based on hermiticity and space-time symmetries, we find in total 32 (leading-twist) phase-space distributions. We stress that this counting is completely model-independent, though it may appear that some linear combinations of these distributions vanish in particular models or theories for deeper symmetry reasons.

\begin{table}[t!]
\begin{center}
\caption{\footnotesize{Transformation properties of the polarized GTMD correlator and phase-space distribution. For a generic four-vector $a^\mu$ with light-front components $[a^+,a^-,\uvec a_T]$, the light-front components of $\bar a^\mu$ are $[a^+,a^-,-\uvec a_T]$.}}\label{Transfprop}
\begin{tabular}{@{\quad\!}c@{\quad}|@{\quad}c@{\quad}|@{\quad}c@{\quad\!}}\whline
&$W_{\vec S\vec S^q}(P,k,\Delta;n_-)$&$\rho_{\vec S\vec S^q}(x,\uvec k_T,\uvec b_T;\hat P,\eta)$\\
\hline
Hermiticity&$W^*_{\vec S\vec S^q}(P,k,-\Delta;n_-)$&$\rho_{\vec S\vec S^q}(x,\uvec k_T,\uvec b_T;\hat P,\eta)$\\
Parity&$W_{\vec S\vec S^q}(\bar P,\bar k,\bar \Delta;\bar n_-)$&$\rho_{\vec S\vec S^q}(x,-\uvec k_T,-\uvec b_T;-\hat P,\eta)$\\
Time-reversal&$W^*_{-\vec S-\vec S^q}(\bar P,\bar k,\bar\Delta;-\bar n_-)$&$\rho_{-\vec S-\vec S^q}(x,-\uvec k_T,\uvec b_T;-\hat P,-\eta)$\\
\whline
\end{tabular}
\end{center}
\end{table}

\section{Multipole decomposition}
\label{sec:multipole}

\begin{figure}[t!]
	\centering
		\includegraphics[width=.8\textwidth]{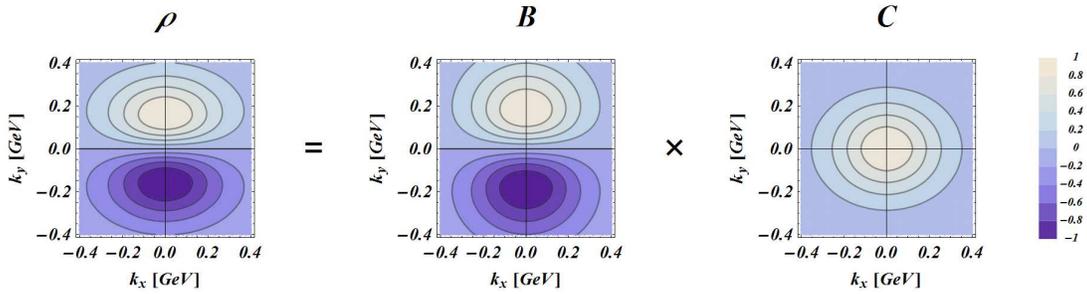}
\caption{\footnotesize{Simple illustration of the decomposition~\eqref{Decomposition} at fixed $x$ and $\uvec b_T$. The phase-space distribution $\rho$ can be written as a product of a basic multipole $B$ (here a dipole in $\uvec k_T$-space) with an oval-shaped coefficient function $C$.}}
		\label{Illustration}
\end{figure}

The relativistic phase-space distribution is linear in $\vec S$ and $\vec S^q$ 
\begin{equation}
\begin{aligned}
\rho_{\vec S\vec S^q}&=\rho_{UU}+S_L\,\rho_{LU}+S^q_L\,\rho_{UL}+S_LS^q_L\,\rho_{LL}\\
&\phantom{=}+S^i_T\,(\rho_{T^iU}+S^q_L\,\rho_{T^iL})+S^{qi}_T\,(\rho_{UT^i}+S_L\,\rho_{LT^i})+S^i_TS^{qj}_T\,\rho_{T^iT^j},
\end{aligned}
\end{equation}
and can further be decomposed into two-dimensional multipoles in both $\uvec k_T$ and $\uvec b_T$ spaces. While there is  no limit in the multipole order\footnote{Indeed, the multiplication by $(\uvec k_T\cdot\uvec b_T)^2$ increases the transverse multipole order in $\uvec k_T$ and $\uvec b_T$ spaces, but does not change the transformation properties under parity and time-reversal.}, parity and time-reversal impose certain constraints on the allowed  multipoles. It is therefore more sensible to decompose the phase-space distributions $\rho_X$ with $X=UU, LU,\cdots$ as follows
\begin{align}
\rho_{X}(x,\uvec k_T,\uvec b_T;\hat P,\eta)&=\sum_{m_k,m_b} \rho^{(m_k,m_b)}_{X}(x,\uvec k_T,\uvec b_T;\hat P,\eta),\\
\rho^{(m_k,m_b)}_{X}(x,\uvec k_T,\uvec b_T;\hat P,\eta)&=B^{(m_k,m_b)}_{X}(\hat k_T,\hat b_T;\hat P,\eta)\,C^{(m_k,m_b)}_{X}[x,\uvec k^2_T,(\uvec k_T\cdot\uvec b_T)^2,\uvec b^2_T],\label{Decomposition}
\end{align}
where $B^{(m_k,m_b)}_{X}$ represent the basic (or simplest) multipoles allowed by parity and time-reversal symmetries. These basic multipoles are multiplied by the coefficient functions $C^{(m_k,m_b)}_{X}$ which depend on $\mathsf P$ and $\mathsf T$-invariant variables only. The couple of integers $(m_k,m_b)$ gives the basic multipole order in both $\uvec k_T$ and $\uvec b_T$ spaces. An illustration of the decomposition of a phase-space density into basic multipole and coefficient function is given in Fig.~\ref{Illustration}.

The basic multipoles can be expressed in terms of transverse multipoles in $\uvec k_T$-space
\begin{equation}
\begin{aligned}
m_k&=0\qquad\qquad&M_k&=1,\\
m_k&=1&D^i_k&=\hat k^i_T,\\
m_k&=2&Q^{ij}_k&=\hat k^i_T\hat k^j_T-\tfrac{1}{2}\,\delta^{ij}_T,\\
m_k&=3&O^{ijl}_k&=\hat k^i_T\hat k^j_T\hat k^l_T-\tfrac{1}{4}\left(\delta^{ij}_T\hat k^l_T+\delta^{jl}_T\hat k^i_T+\delta^{li}_T\hat k^j_T\right),\\
&\,\,\,\vdots&&\,\,\,\vdots
\end{aligned}
\end{equation}
and the corresponding ones in $\uvec b_T$-space. For example, for the spin-independent contribution $\rho_{UU}$, the simplest basic multipole one can build is obviously in terms of the transverse monopoles $B^{(0,0)}_{UU}=M_kM_b=1$. The only possibility\footnote{Indeed, the other combination $\epsilon^{ij}_T D^i_kD^j_b=(\hat k_T\times\hat b_T)_L$ is $\mathsf P$-odd.} involving the transverse dipoles is $B^{(1,1)}_{UU}=\eta\,D^i_kD^i_b=\eta\,(\hat k_T\cdot\hat b_T)$, where the $\eta$ factor ensures time-reversal invariance. Higher transverse multipoles do not lead to new basic multipoles since they always reduce to $B^{(0,0)}_{UU}$ and $B^{(1,1)}_{UU}$ multiplied by some function of $\uvec k^2_T$, $(\uvec k_T\cdot\uvec b_T)^2$ and $\uvec b^2_T$. This analysis is consistent with the fact that there exists only one spin-independent complex-valued GTMD denoted as $F_{11}$~\cite{Meissner:2009ww}, leading to two different real-valued phase-space distributions. Note also that, from the explicit expressions for the basic multipoles $B^{(m_k,m_b)}_{UU}$, we find that $\rho^e_{UU}=\rho^{(0,0)}_{UU}$ and $\rho^o_{UU}=\rho^{(1,1)}_{UU}$. All the basic multipoles associated with the other contributions $\rho_X$ are obtained following the same strategy.

Note that only the multipoles with $m_b=0$ survive integration over $\uvec b_T$ and reduce to TMD amplitudes. Similarly, only the multipoles with $m_k=0$ survive integration over $\uvec k_T$ and  reduce to  impact-parameter distributions. Since GPDs do not depend on $\eta$, only the naive $\mathsf T$-even multipoles correspond to the Fourier transforms of GPD amplitudes~\cite{Diehl:2005jf,Lorce:2011dv}. Interestingly, the naive $\mathsf T$-odd ones represent \emph{new} contributions, just like new contributions were obtained in the general parametrization of the light-front energy-momentum tensor~\cite{Lorce:2015lna}.
\newline

\begin{table}[t!]
\begin{center}
\caption{\footnotesize{Correlations between target polarization ($S_L,\uvec S_T$), quark polarization ($S^q_L,\uvec S^q_T$) and quark OAM ($\ell^q_L,\uvec{\ell}^q_T$) encoded in the various phase-space distributions $\rho_X$. We then see that \emph{e.g.} $\rho_{UL}$ encodes the spin-orbit correlation $\langle S^q_L\ell^q_L\rangle$, and $\rho_{T_xT_y}$ encodes the double spin-orbit correlation $\langle S_x\ell^q_xS^q_y\ell^q_y\rangle$.}}\label{angcorr}
\begin{tabular}{@{\quad\!}c@{\quad}|@{\quad}c@{\quad}c@{\quad}c@{\quad}c@{\quad\!}}\whline
$\rho_X$&$U$&$L$&$T_x$&$T_y$\\
\hline
$U$&$\langle 1\rangle$&$\langle S^q_L\ell^q_L\rangle$&$\langle S^q_x\ell^q_x\rangle$&$\langle S^q_y\ell^q_y\rangle$\\
$L$&$\langle S_L\ell^q_L\rangle$&$\langle S_LS^q_L\rangle$&$\langle S_L\ell^q_LS^q_x\ell^q_x\rangle$&$\langle S_L\ell^q_LS^q_y\ell^q_y\rangle$\\
$T_x$&$\langle S_x\ell^q_x\rangle$&$\langle S_x\ell^q_xS^q_L\ell^q_L\rangle$&$\langle S_xS^q_x\rangle$& $\langle S_x\ell^q_xS^q_y\ell^q_y\rangle$\\
$T_y$&$\langle S_y\ell^q_y\rangle$&$\langle S_y\ell^q_yS^q_L\ell^q_L\rangle$& $\langle S_y\ell^q_yS^q_x\ell^q_x\rangle$&$\langle S_yS^q_y\rangle$\\
\whline
\end{tabular}
\end{center}
\end{table}

Remarkably, it turns out that all the contributions $\rho_X$ can be understood as encoding all the possible correlations between target and quark angular momenta, see Table~\ref{angcorr}. We stress in particular that $\vec \ell^q$ refers to the  \emph{canonical} quark OAM, since it is defined in terms of the canonical quark momentum $\vec k$~\cite{Lorce:2012ce}. As will be discussed in more detail in Sec.~\ref{section:4}, the relation with all the possible angular correlations becomes more transparent once one sees the 5-dimensional relativistic phase-space distributions as 6-dimensional phase-space distributions integrated over the quark average longitudinal position
\begin{equation}
\rho_X(x,\uvec k_T,\uvec b_T;\hat P,\eta)=\int\ud b_L\,\rho_X(\vec k,\vec b;\hat P,\eta).
\end{equation}
Noting that $b_L$ is even under parity and odd under time-reversal, one can perform a similar multipole expansion for the 6-dimensional distributions. Naturally, the integral over $b_L$ of the 6-dimensional multipoles can be expressed in terms of the 5-dimensional ones 
\begin{equation}
\int\ud b_L\,\mathcal B(\vec k,\vec b;\hat P,\eta)\,\mathcal C[x,\vec k^2,(\vec k\cdot\vec b)^2,\vec b^2,b_L(\vec k\cdot\vec b)]=\sum_i B^i(\hat k_T,\hat b_T;\hat P,\eta)\,C^i[x,\uvec k^2_T,(\uvec k_T\cdot\uvec b_T)^2,\uvec b^2_T].
\end{equation}
For convenience, this correspondence will simply be denoted in Sec.~\ref{section:4} as
\begin{equation}
\int\ud b_L\,\mathcal B\sim\sum_i B^i.
\end{equation}
 We shall also implicitly use the fact that 
\begin{align}
b_L\,\mathcal C[x,\vec k^2,(\vec k\cdot\vec b)^2,\vec b^2,b_L(\vec k\cdot\vec b)]&=(\vec k\cdot\vec b)\,\frac{b_L(\vec k\cdot\vec b)}{(\vec k\cdot\vec b)^2}\,\mathcal C[x,\vec k^2,(\vec k\cdot\vec b)^2,\vec b^2,b_L(\vec k\cdot\vec b)]\nonumber\\
&\equiv(\vec k\cdot\vec b)\,\mathcal C'[x,\vec k^2,(\vec k\cdot\vec b)^2,\vec b^2,b_L(\vec k\cdot\vec b)],
\end{align}
so that we can write \emph{e.g.} $\int\ud b_L\,b_L\sim(\hat k_T\cdot\hat b_T)$.  As we will explicitly show in the following, working at the level of phase-space distributions gives us much more insight about the physics encoded in the various GPDs and TMDs.

\section{Representation of transverse phase space}
\label{section:3}

\begin{figure}[t!]
	\centering
		\includegraphics[width=.4\textwidth]{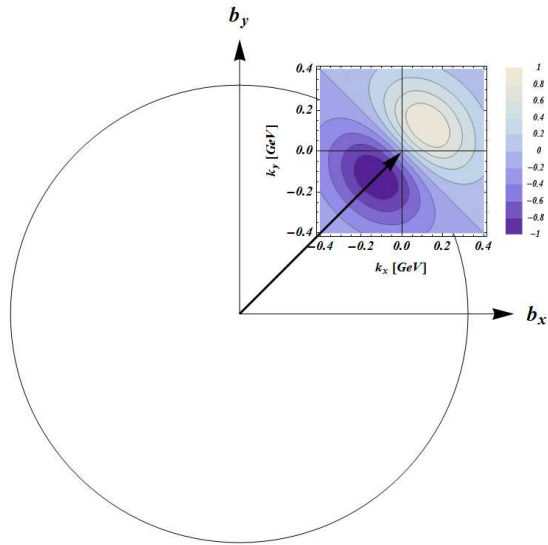}
\caption{\footnotesize{Representation of the transverse phase space. The circle represents the points in impact-parameter space at a fixed distance $|\uvec b_T|$ from the center of the target. To each point on this circle is associated a distribution in transverse-momentum space. See text for more details.}}
		\label{Tphasespace}
\end{figure}

The relativistic phase-space distributions are functions of five continuous variables. It is therefore particularly difficult to represent them on a two-dimensional space. Since we are mainly interested in the transverse direction, we reduce the number of variables by
\begin{enumerate}
\item integrating these phase-space distributions over $x$;
\item discretizing the polar coordinates of $\uvec b_T$.
\end{enumerate}
For further convenience, we also set $\eta=+1$ and choose $\hat P=\vec e_z=(0,0,1)$ so that $\hat b_T=(\cos\phi_b,\sin\phi_b,0)$ and $\hat k_T=(\cos\phi_k,\sin\phi_k,0)$. The resulting transverse phase-space distributions are then represented as sets of distributions in $\uvec k_T$-space 
\begin{equation}
\rho_X(\uvec k_T|\,\uvec b_T)=\int\ud x\,\rho_X(x,\uvec k_T,\uvec b_T;\hat P=\vec e_z,\eta=+1)\big|_{\uvec b_T\text{ fixed}}
\end{equation}
with the origin of axes lying on circles of radius $|\uvec b_T|$ at polar angle $\phi_b$ in impact-parameter space, see Fig.~\ref{Tphasespace}. In this way, one can see how the transverse momentum is distributed at some point in the impact-parameter space. In the language of differential geometry, the $\uvec b_T$-space plays the role of a base space and the $\uvec k_T$-space plays the role of the corresponding tangent space. All we do is just drawing the tangent spaces associated with a couple of points  in the base space and situated at a fixed distance from the center. Naturally, one can also represent the same transverse phase-space distributions in terms of $\uvec b_T$-distributions \begin{equation}
\rho_X(\uvec b_T|\,\uvec k_T)=\int\ud x\,\rho_X(x,\uvec k_T,\uvec b_T;\hat P=\vec e_z,\eta=+1)\big|_{\uvec k_T\text{ fixed}}
\end{equation}
with the origin of axes lying on circles of radius $|\uvec k_T|$ at polar positions $\phi_k$ in transverse-momentum space. In this case, one sees how some specific transverse momentum is distributed in impact-parameter space. In the following, we shall only consider the discrete $\uvec b_T$ representation $\rho_X(\uvec k_T|\,\uvec b_T)$. 

The above representation of transverse phase space has the advantage to make the multipole structure in both $\uvec k_T$ and $\uvec b_T$ spaces particularly clear. For example, the basic multipole $B^{(m_k,m_b)}_X$ simply displays a $m_k$-pole in transverse-momentum space at any transverse position $\uvec b_T$. The orientation of this $m_k$-pole is determined by $m_b$ and $\phi_b=\arg\hat b_T$. More precisely, by going once around the circle, the $m_k$-pole will undergo $\tfrac{m_b}{m_k}$ complete rotations. The case $m_k=0$ does not cause any problem since a monopole is invariant under rotations. 

\section{Discussion}
\label{section:4}

Since the focus of this paper is on the multipole decomposition of the transverse phase space, we choose for all the figures in the following  to represent only eight points in impact-parameter space lying on a circle with radius $|\uvec b_T|=0.4$ fm. Also, for a better legibility, the $\uvec k_T$-distributions are normalized to the absolute maximal value over the whole circle in impact-parameter space
\begin{equation}
\max_{|\uvec b_T|=0.4\text{ fm}}|\rho_X(\uvec k_T|\uvec b_T)|=1.
\end{equation}

The results presented in the following are obtained using the light-front constituent quark model (LFCQM)~\cite{Lorce:2011dv} for up quarks, by computing directly the Fourier transform of the helicity amplitudes associated with the GTMD correlator. Light and dark regions represent, respectively, positive and negative domains of the transverse phase-space distributions. Since our purpose at this point is simply to illustrate the multipole structure, we computed only the naive $\mathsf T$-even contributions in this model. The fact that the calculated distributions perfectly match the expected multipole decomposition presented in Sec.~\ref{sec:multipole} proves the consistency of the approach\footnote{Note that alternative definitions of the GTMDs including a soft factor contribution~\cite{paper} modifies only the $\bold{k}_{\perp}^2$-dependence, and so does not alter the following multipole analysis.}. The naive $\mathsf T$-odd contributions have been obtained by extracting the coefficient functions from the naive $\mathsf T$-even part and multiplying them by the appropriate basic multipoles. We stress that the global sign of these naive $\mathsf T$-odd contributions has been chosen arbitrarily. Only a proper calculation including initial- and/or final-state interactions can determine these global signs.

\subsection{Unpolarized target}

\subsubsection{Unpolarized quark}
\label{subsect:UU}

\begin{figure}[t!]
\centerline{\includegraphics[width=7cm]{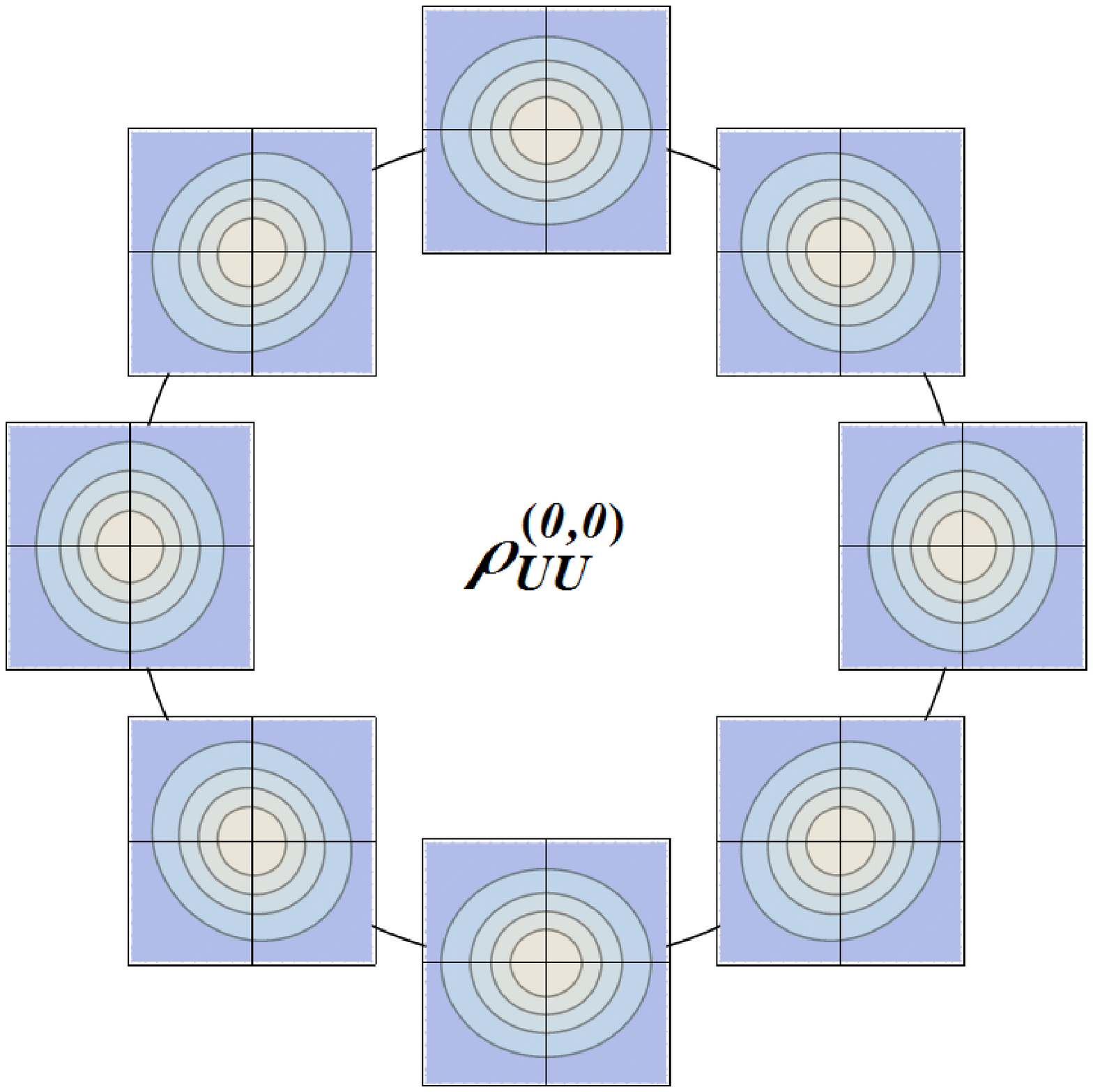}\hspace{1.5cm}\includegraphics[width=7cm]{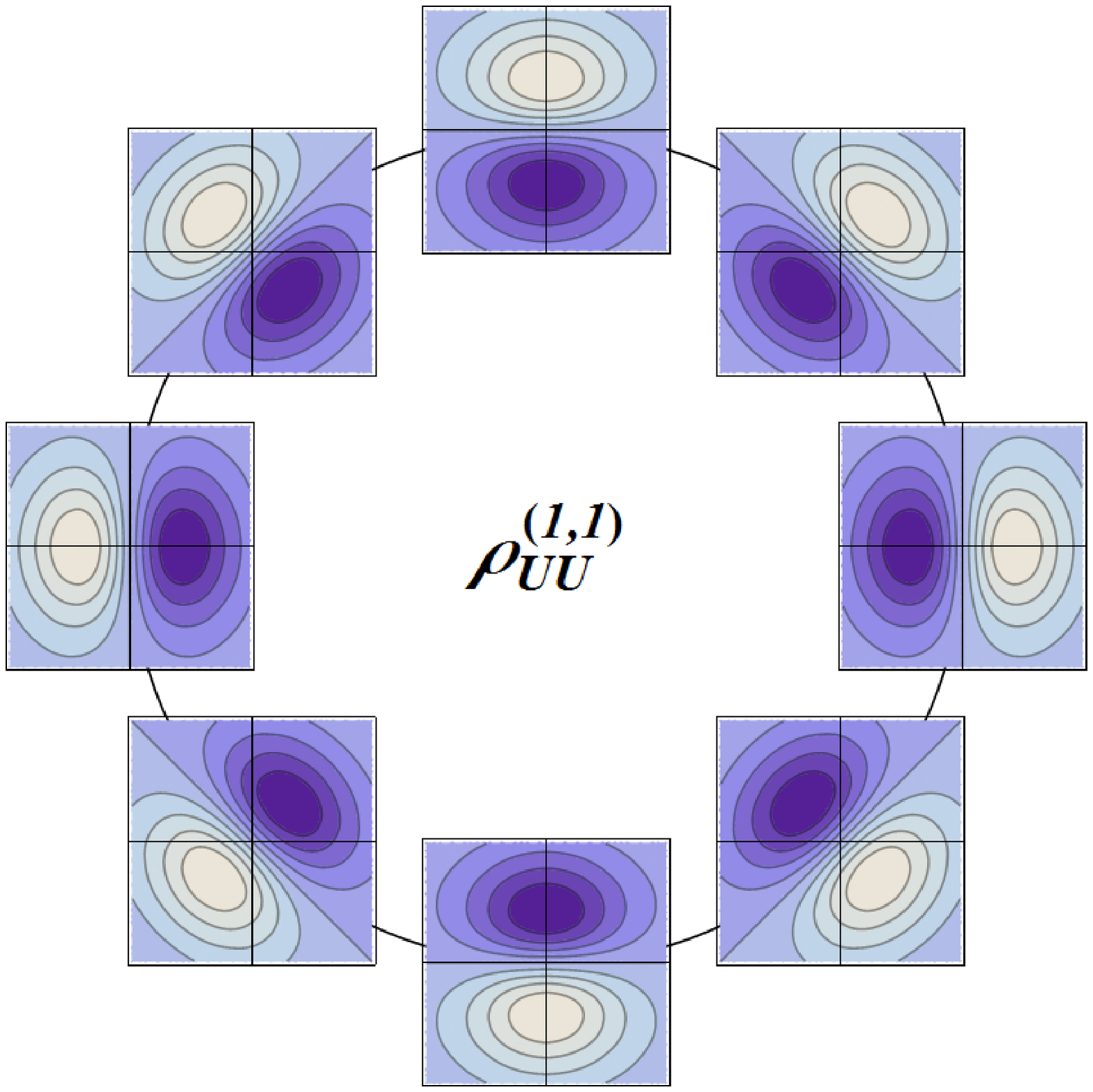}}
\vspace*{8pt}
\caption{Naive $\mathsf T$-even (left) and $\mathsf T$-odd (right) contributions to the transverse phase-space distribution $\rho_{UU}$. See text for more details. \label{fig1}}
\end{figure}

The simplest contribution is $\rho_{UU}$. It describes the distribution of unpolarized quarks inside an unpolarized target. As already discussed at the end of Sec.~\ref{sec:multipole}, there exist only two spin-independent phase-space distributions
\begin{equation}
\rho^e_{UU}=\rho^{(0,0)}_{UU},\qquad\rho^o_{UU}=\rho^{(1,1)}_{UU},
\end{equation}
which are represented in Fig.~\ref{fig1}. 
The corresponding basic multipoles are
\begin{align}
B^{(0,0)}_{UU}(\hat k_T,\hat b_T;\hat P,\eta)&=M_kM_b=1,\\
B^{(1,1)}_{UU}(\hat k_T,\hat b_T;\hat P,\eta)&=\eta\,D^i_kD^i_b=\eta\,(\hat k_T\cdot\hat b_T).
\end{align}
Only $\rho^{(0,0)}_{UU}$ survives integration over $\uvec k_T$ or $\uvec b_T$ and is then naturally related to both the unpolarized GPD $H$ and the unpolarized TMD $f_1$~\cite{Diehl:2005jf,Lorce:2011dv}. Contrary to its $\uvec k_T$- and $\uvec b_T$-integrated versions, $\rho^{(0,0)}_{UU}$ is not circularly symmetric. The reason is that $\rho^{(0,0)}_{UU}$ also contains information about the \emph{correlation} between $\uvec k_T$ and $\uvec b_T$, which is lost under integration over one of the transverse variables~\cite{Lorce:2011kd}. As one can see from Fig.~\ref{fig1}, the $\uvec k_T$-distribution is elongated in the direction orthogonal to the transverse position. This means that a polar flow ($\uvec k_T\perp\uvec b_T$) is preferred over a radial flow ($\uvec k_T\parallel\uvec b_T$), which is expected because the quarks are bound in the target. In other words, the preferred flow of quarks is along  circles around the center of the target. The quark motion is of course not limited to the transverse plane. So, for fixed quark momentum $\vec k$, $\rho^{(0,0)}_{UU}$ should better be thought of as the projection of a 3-dimensional distribution in position space onto the transverse plane
\begin{equation}
\int\ud b_L\,1\sim 1.
\end{equation}
Note however that the \emph{net} OAM is zero in this case, because there is no preferred direction in $\rho_{UU}$. Quarks tend to follow circular motion equally in both clockwise and anti-clockwise directions.
\newline

Since it integrates to zero in both $\uvec k_T$ and $\uvec b_T$-spaces, $\rho^{(1,1)}_{UU}$ represents a completely new piece of information which is  not accessible \emph{via} GPDs or TMDs at leading twist. The dipole in $\uvec k_T$-space signals the presence of a net flow in the transverse radial direction $(\hat k_T\cdot\hat b_T)$, which can be seen as the projection of a 3-dimensional radial flow $(\hat k\cdot\hat b)$ onto the transverse position space
\begin{equation}
\int\ud b_L\,(\vec k\cdot\vec b)\sim (\hat k_T\cdot\hat b_T).
\end{equation}
For a stable target, this must obviously be zero. A non-vanishing net radial flow therefore originates purely from initial- and final-state interactions, in agreement with the naive $\mathsf T$-odd nature of $\rho^{(1,1)}_{UU}$. The coefficient function $C^{(1,1)}_{UU}$ then represents in some sense the strength of the spin-independent part of the force felt by the quark due to initial- and final-state interactions.

\subsubsection{Longitudinally polarized quark}
\label{subsect:UL}

\begin{figure}[t!]
\centerline{\includegraphics[width=7cm]{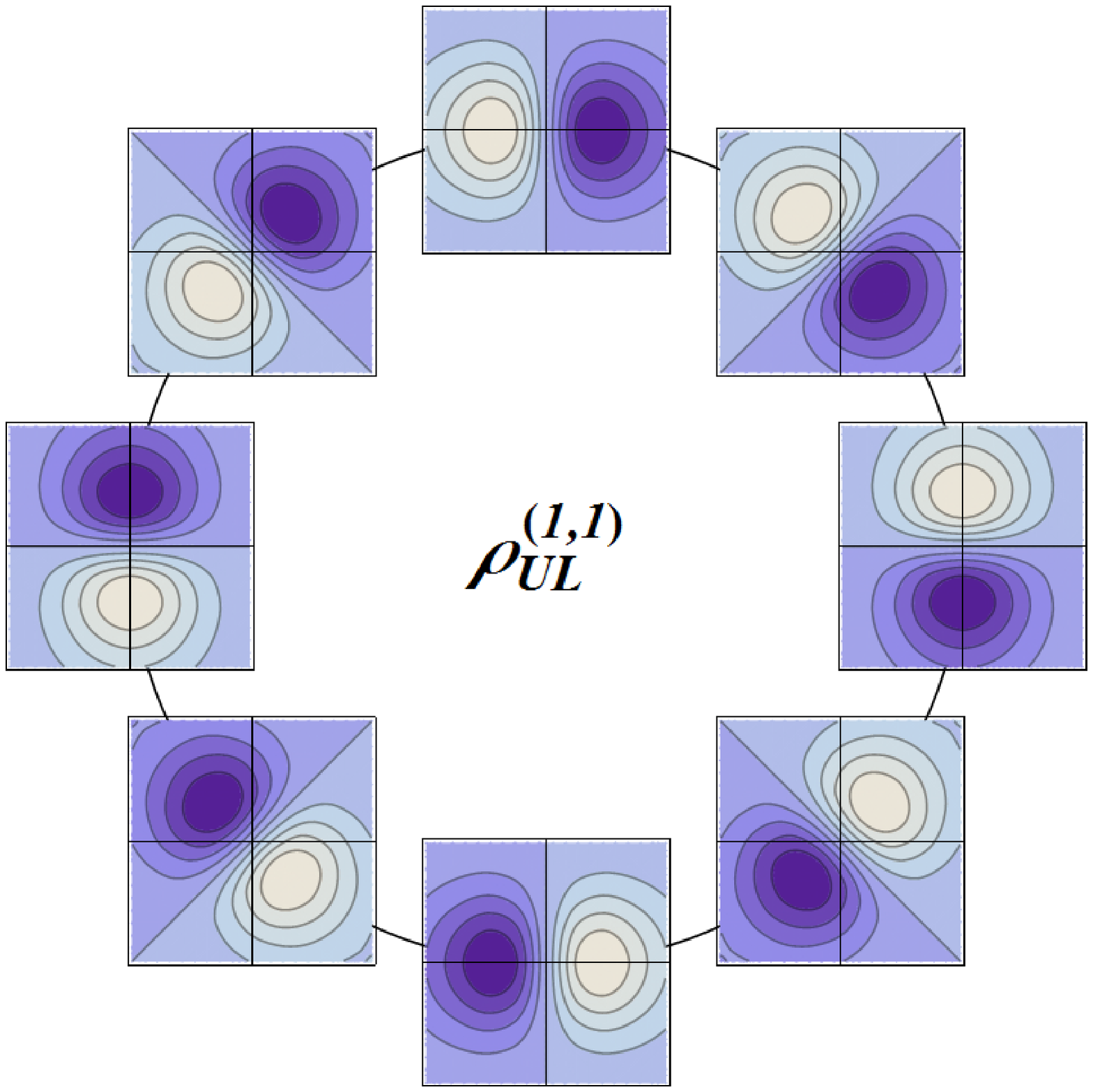}\hspace{1.5cm}\includegraphics[width=7cm]{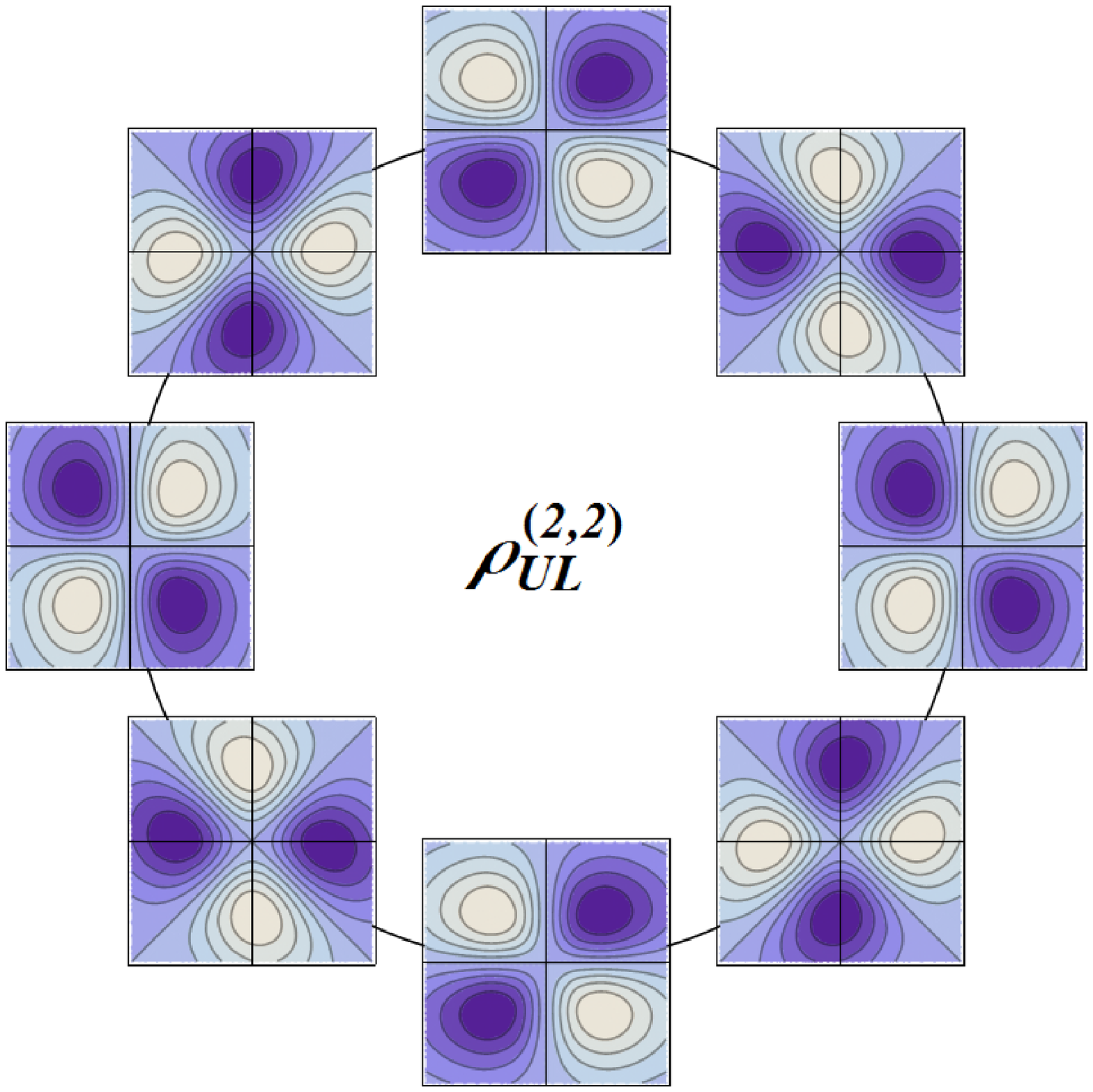}}
\vspace*{8pt}
\caption{Naive $\mathsf T$-even (left) and $\mathsf T$-odd (right) contributions to the transverse phase-space distribution $\rho_{UL}$. See text for more details. \label{fig2}}
\end{figure}

The contribution $\rho_{UL}$ describes how the distribution of quarks inside an unpolarized target is affected by the quark longitudinal polarization. We find only two phase-space distributions
\begin{equation}
\rho^e_{UL}=\rho^{(1,1)}_{UL},\qquad\rho^o_{UL}=\rho^{(2,2)}_{UL},
\end{equation}
which are represented in Fig.~\ref{fig2}. 
The corresponding basic multipoles are
\begin{align}
S^q_L\,B^{(1,1)}_{UL}(\hat k_T,\hat b_T;\hat P,\eta)&=-S^q_L\epsilon^{ij}_TD^i_kD^j_b=S^q_L(\hat b_T\times\hat k_T)_L,\\
S^q_L\,B^{(2,2)}_{UL}(\hat k_T,\hat b_T;\hat P,\eta)&=-\eta\,S^q_L\epsilon^{ij}_TQ^{il}_kQ^{jl}_b=\eta\,S^q_L(\hat b_T\times\hat k_T)_L\,(\hat k_T\cdot\hat b_T).
\end{align}
None of these survive integration over $\uvec k_T$ or $\uvec b_T$. Both therefore represent completely new information which is  not accessible \emph{via} GPDs or TMDs at leading twist. The $\uvec k_T$-dipole in $\rho^{(1,1)}_{UL}$ signals the presence of a net flow in the polar direction $(\hat b_T\times\hat k_T)_L$, \emph{i.e.} a net longitudinal component of quark OAM, which can be seen as the projection of a 3-dimensional azimuthal flow $(\vec b\times\vec k)\cdot\hat P$ onto the transverse position space
\begin{equation}
\int\ud b_L\,[(\vec b\times\vec k)\cdot \hat P]\sim (\hat b_T\times\hat k_T)_L.
\end{equation}
By reversing the quark longitudinal polarization $S^q_L$, one reverses also the orbital flow. The coefficient function $C^{(1,1)}_{UL}$ then represents in some sense the strength of the correlation between the longitudinal components of quark polarization and OAM $\langle S^q_L\ell^q_L\rangle$~\cite{Lorce:2011kd,Lorce:2014mxa}. 
\newline

On the contrary, the contribution $\rho^{(2,2)}_{UL}$ does not modify the net quark flow. The effect of the $\uvec k_T$-quadrupoles is to globally tilt the $\uvec k_T$-distributions with respect to $\uvec b_T$, so that the preferred flow is now a spiral correlated with the quark longitudinal polarization, which can be seen as the projection of a 3-dimensional spiral flow onto the transverse position space
\begin{equation}
\int\ud b_L\,[(\vec b\times\vec k)\cdot \hat P]\,(\vec k\cdot\vec b)\sim (\hat b_T\times\hat k_T)_L\,(\hat k_T\cdot\hat b_T).
\end{equation}
In other words, the contribution $\rho^{(2,2)}_{UL}$ gives the difference of radial flows between quarks with opposite $\langle S^q_L\ell^q_L\rangle$ correlations. The coefficient function $C^{(2,2)}_{UL}$ then represents in some sense the strength of the $\langle S^q_L\ell^q_L\rangle$-dependent part of the force felt by the quark due to initial- and final-state interactions.

\subsubsection{Transversely polarized quark}
\label{subsect:UT}

\begin{figure}[t]
\centerline{\includegraphics[width=7cm]{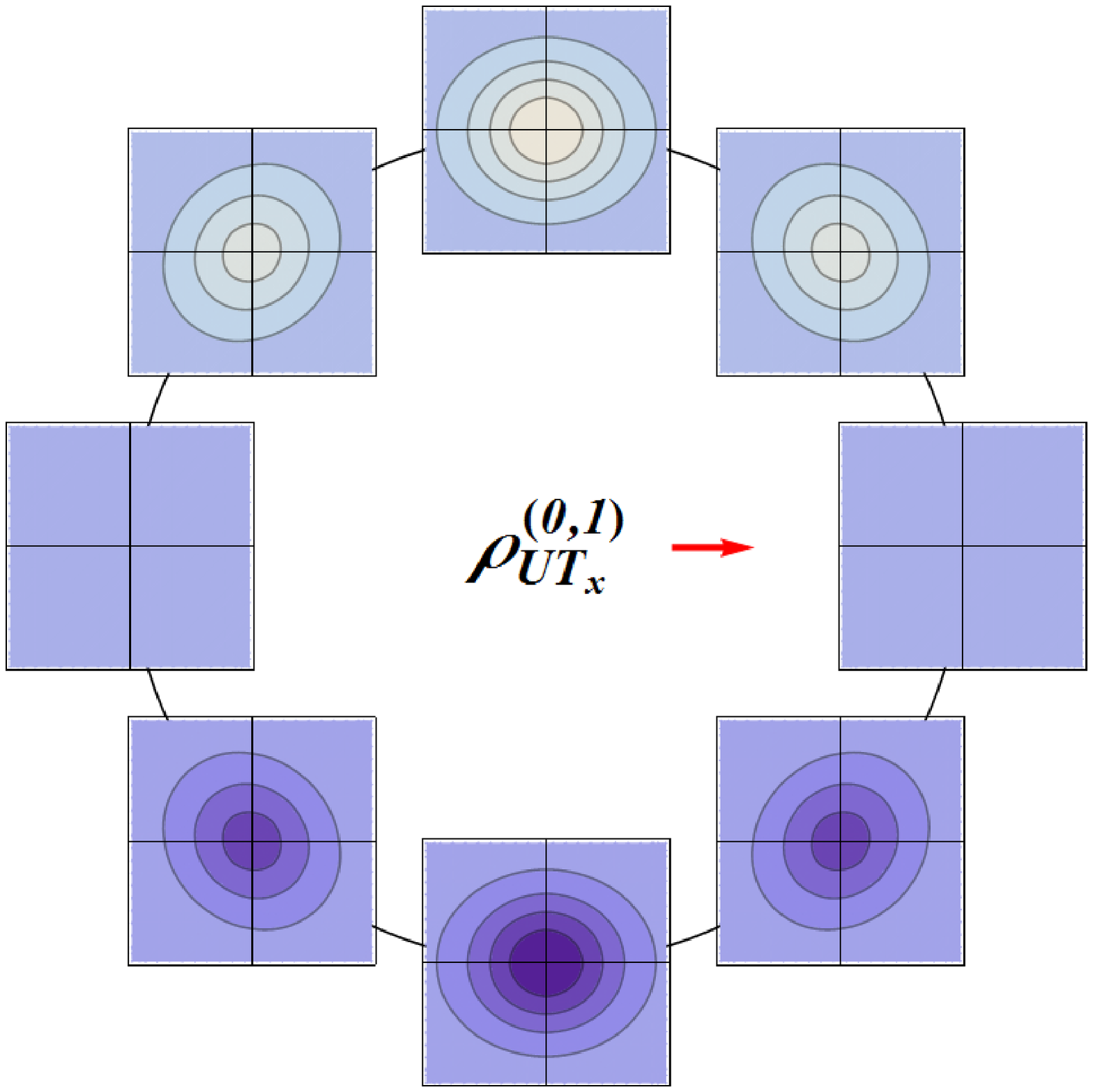}\hspace{1.5cm}\includegraphics[width=7cm]{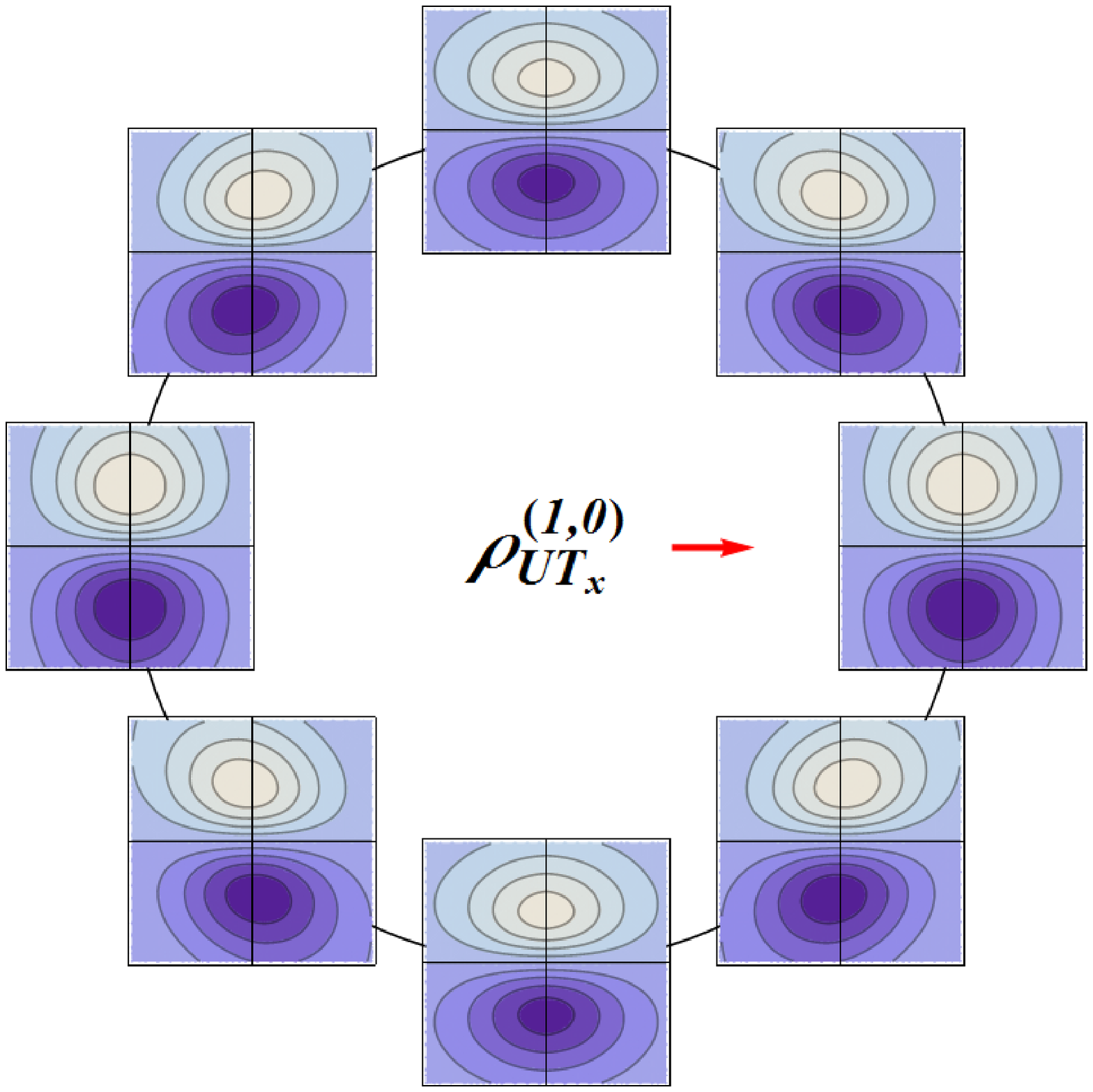}}
\vspace*{15pt}
\centerline{\includegraphics[width=7cm]{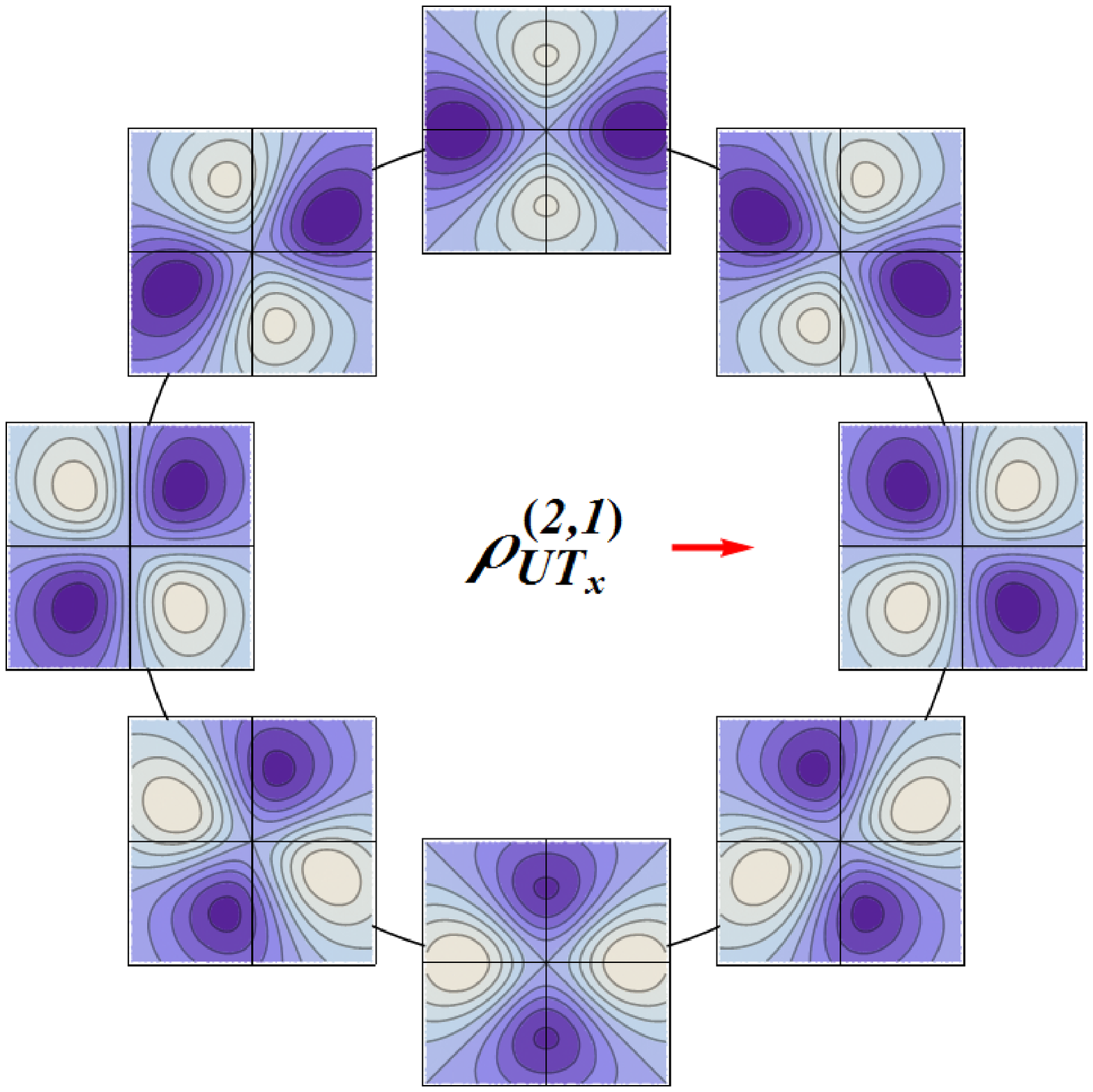}\hspace{1.5cm}\includegraphics[width=7cm]{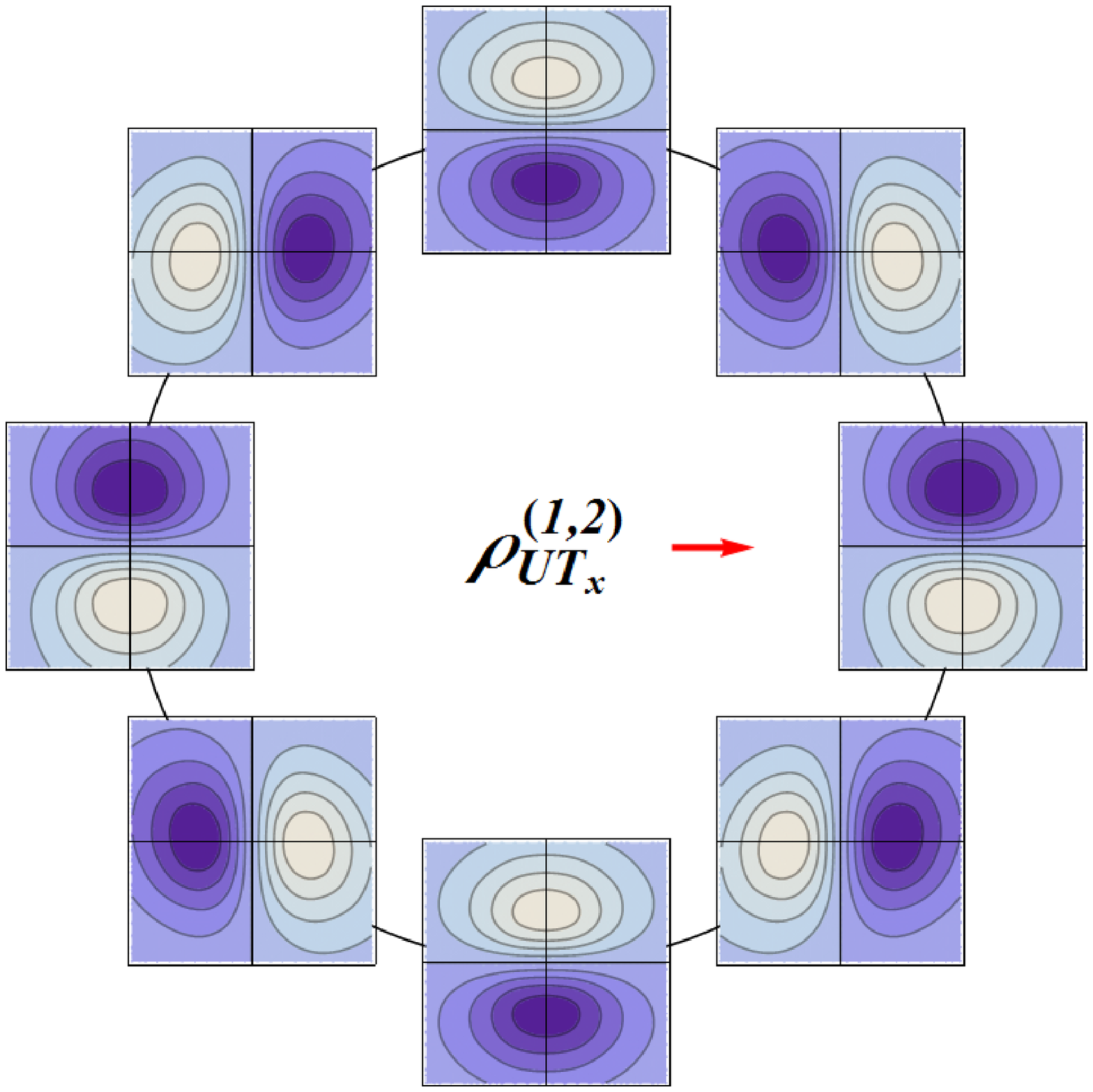}}
\vspace*{8pt}
\caption{Naive $\mathsf T$-even (left) and $\mathsf T$-odd (right) contributions to the transverse phase-space distribution $\rho_{UT}$ for the quark polarization $\vec S^q_T=\vec e_x$ (red arrow). See text for more details. \label{fig3}}
\end{figure}

The contribution $\rho_{UT^i}$ describes how the distribution of quarks inside an unpolarized target is affected by the quark transverse polarization. We find in total four phase-space distributions
\begin{equation}
\rho^e_{UT^i}=\rho^{(0,1)}_{UT^i}+\rho^{(2,1)}_{UT^i},\qquad\rho^o_{UT^i}=\rho^{(1,0)}_{UT^i}+\rho^{(1,2)}_{UT^i},
\end{equation}
which are represented in Fig.~\ref{fig3} for the quark polarization $\vec S^q_T=\vec e_x$.
The corresponding basic multipoles are
\begin{align}
S^{qi}_T\,B^{(0,1)}_{UT^i}(\hat k_T,\hat b_T;\hat P,\eta)&=S^{qi}_T\epsilon^{ij}_TM_kD^j_b=(\uvec S^q_T\times\hat b_T)_L,\label{UT1}\\
S^{qi}_T\,B^{(2,1)}_{UT^i}(\hat k_T,\hat b_T;\hat P,\eta)&=S^{qi}_T\epsilon^{ij}_TQ^{jl}_kD^l_b=(\uvec S^q_T\times\hat k_T)_L\,(\hat k_T\cdot\hat b_T)-\tfrac{1}{2}\,(\uvec S^q_T\times\hat b_T)_L,\label{UT2}\\
S^{qi}_T\,B^{(1,0)}_{UT^i}(\hat k_T,\hat b_T;\hat P,\eta)&=\eta\,S^{qi}_T\epsilon^{ij}_TD^j_kM_b=\eta\,(\uvec S^q_T\times\hat k_T)_L,\label{UT3}\\
S^{qi}_T\,B^{(1,2)}_{UT^i}(\hat k_T,\hat b_T;\hat P,\eta)&=\eta\,S^{qi}_T\epsilon^{jl}_TD^l_kQ^{ij}_b=\eta\left[(\uvec S^q_T\cdot\hat b_T)\,(\hat b_T\times\hat k_T)_L-\tfrac{1}{2}\,(\uvec S^q_T\times\hat k_T)_L\right].\label{UT4}
\end{align}
The contribution $\rho^{(0,1)}_{UT^i}$ is the only one surviving the integration over $\uvec k_T$ and is then naturally related to the GPD combination $2\tilde H_T+E_T$~\cite{Diehl:2005jf,Lorce:2011dv,Pasquini:2007xz}. The dipole in $\uvec b_T$-space indicates the presence of a spatial separation between quarks with opposite transverse polarizations. This transverse shift is actually an effect related to the light-front imaging due to the fact that the light-front densities are defined in terms of the $j^+=\tfrac{1}{\sqrt{2}}(j^0+j^3)$ component of the current instead of the $j^0$ component, and finds its physical origin in the correlation between the transverse components of quark polarization and OAM $\langle \uvec S^q_T\cdot\uvec\ell^q_T\rangle$~\cite{Burkardt:2005hp}. Indeed, because of transverse OAM, quarks situated at opposite sides tend to have opposite longitudinal momenta $k_L\,\hat P$, \emph{i.e.} opposite $j^3$ components, and are then associated with different light-front densities $j^+$. From a slightly different perspective, the transverse shift can also be understood from the fact that the position of the relativistic center-of-mass of a rotating body is frame-dependent~\cite{Moller:1949,Moller:1972}.

There are actually two independent transverse correlations, say $\langle S^q_x\ell^q_x\rangle$ and $\langle S^q_y\ell^q_y\rangle$. The contribution $\rho^{(0,1)}_{UT^i}$ gives us information about only one particular combination. The other combination is given by  the other naive $\mathsf T$-even contribution $\rho^{(2,1)}_{UT^i}$ which is  not accessible \emph{via} GPDs or TMDs at leading twist. Indeed, let us consider the projection of a 3-dimensional $\langle (\vec S^q_T\cdot\vec n_T)(\vec\ell^q_T\cdot\vec n_T)\rangle$ correlation onto the transverse position space, where $\vec n_T$ is some transverse vector. For $\vec n_T=\vec k_T$ and $\vec n_T=(\vec k_T\times\hat P)$, we respectively find
\begin{align}
\label{eq:36}
\int\ud b_L\,(\vec S^q_T\cdot\vec k_T)\,[(\vec b\times\vec k)_T\cdot\vec k_T]&\sim (\uvec S^q_T\cdot\hat k_T)\,(\hat k_T\times\hat b_T)_L,\\
\label{eq:37}
\int\ud b_L\,[\vec S^q_T\cdot(\vec k_T\times\hat P)]\,[(\vec b\times\vec k)_T\cdot(\vec k_T\times\hat P)]&\sim(\uvec S^q_T\times\hat k_T)_L\,(\hat k_T\cdot\hat b_T).
\end{align}
Noting that 
\begin{equation}
\label{eq:38}
(\uvec S^q_T\cdot\hat k_T)\,(\hat k_T\times\hat b_T)_L+(\uvec S^q_T\times\hat k_T)_L\,(\hat k_T\cdot\hat b_T)=(\uvec S^q_T\times\hat b_T)_L
\end{equation}
and comparing with the basic multipoles~\eqref{UT1} and~\eqref{UT2}, we can see that the two coefficient functions $C^{(0,1)}_{UT^i}$ and $C^{(2,1)}_{UT^i}$ are related to the strength of  two different combinations of the transverse correlations $\langle  S^q_x\ell^q_x\rangle$ and $\langle  S^q_y\ell^q_y\rangle$.
\newline

Similarly, the contribution $\rho^{(1,0)}_{UT^i}$ is the only one surviving the integration over $\uvec b_T$ and is then naturally related to the Boer-Mulders TMD $h^\perp_1$. The dipole in $\uvec k_T$-space  indicates the presence of a net transverse flow orthogonal to the quark transverse polarization. Interestingly, this phenomenon is reminiscent of the spin Hall effect in spintronics and the Magnus effect in fluid mechanics~\cite{Dyakonov:1971a,Dyakonov:1971b}. Such a net transverse flow can only arise from initial- and/or final-state interactions, in accordance with the naive $\mathsf T$-odd nature of $\rho^{(1,0)}_{UT^i}$. 

The contribution $\rho^{(1,2)}_{UT^i}$ corresponds to a completely new information which is  not accessible \emph{via} GPDs or TMDs at leading twist. Combined with $\rho^{(1,0)}_{UT^i}$, it tells us how the initial- and final-state interactions depend  on the two transverse correlations, say $\langle S^q_x\ell^q_x\rangle$ and $\langle S^q_y\ell^q_y\rangle$. Indeed, let us consider the projection of a 3-dimensional transverse spiral flow $(\vec S^q_T\cdot\vec n_T)\,[(\vec b\times\vec k)_T\cdot\vec n_T]\,(\vec k\cdot\vec b)$ onto the transverse position space. For $\vec n_T=\vec b_T$ and $\vec n_T=(\vec b_T\times\hat P)$, we respectively find
\begin{align}
\label{eq:39}
\int\ud b_L\,(\vec S^q_T\cdot\vec b_T)\,[(\vec b\times\vec k)_T\cdot\vec b_T]\,(\vec k\cdot\vec b)&\sim (\uvec S^q_T\cdot\hat b_T)\,(\hat b_T\times\hat k_T)_L,\\
\label{eq:40}
\int\ud b_L\,[\vec S^q_T\cdot(\vec b_T\times\hat P)]\,[(\vec b\times\vec k)_T\cdot(\vec b_T\times\hat P)]\,(\vec k\cdot\vec b)&\sim(\uvec S^q_T\times\hat b_T)_L\,(\hat b_T\cdot\hat k_T).
\end{align}
Noting that 
\begin{equation}\label{eq:41}
(\uvec S^q_T\cdot\hat b_T)\,(\hat b_T\times\hat k_T)_L+(\uvec S^q_T\times\hat b_T)_L\,(\hat b_T\cdot\hat k_T)=(\uvec S^q_T\times\hat k_T)_L
\end{equation}
and comparing with the basic multipoles~\eqref{UT3} and~\eqref{UT4}, we can see that the two coefficient functions $C^{(1,0)}_{UT^i}$ and $C^{(1,2)}_{UT^i}$ are related to the strength of the $\langle S^q_x\ell^q_x\rangle$- and $\langle S^q_y\ell^q_y\rangle$-dependent parts of the force felt by the quark due to initial- and final-state interactions. In other words, the contributions $\rho^{(1,0)}_{UT^i}$ and $\rho^{(1,2)}_{UT^i}$ describe the difference of radial flows between quarks with opposite $\langle S^q_x\ell^q_x\rangle$ or $\langle S^q_y\ell^q_y\rangle$ correlations.

As a final remark, it has been suggested by Burkardt~\cite{Burkardt:2005hp} that $\int\ud^2k_T\,\rho^e_{UT^i}$ and $\int\ud^2b_T\,\rho^o_{UT^i}$ could be related by some lensing effect. We cannot unfortunately confirm this suggestion, because such a relation relies on a dynamical mechanism which goes beyond the general constraints considered in the present paper.

\subsection{Longitudinally polarized target}

\subsubsection{Unpolarized quark}

\begin{figure}[t]
\centerline{\includegraphics[width=7cm]{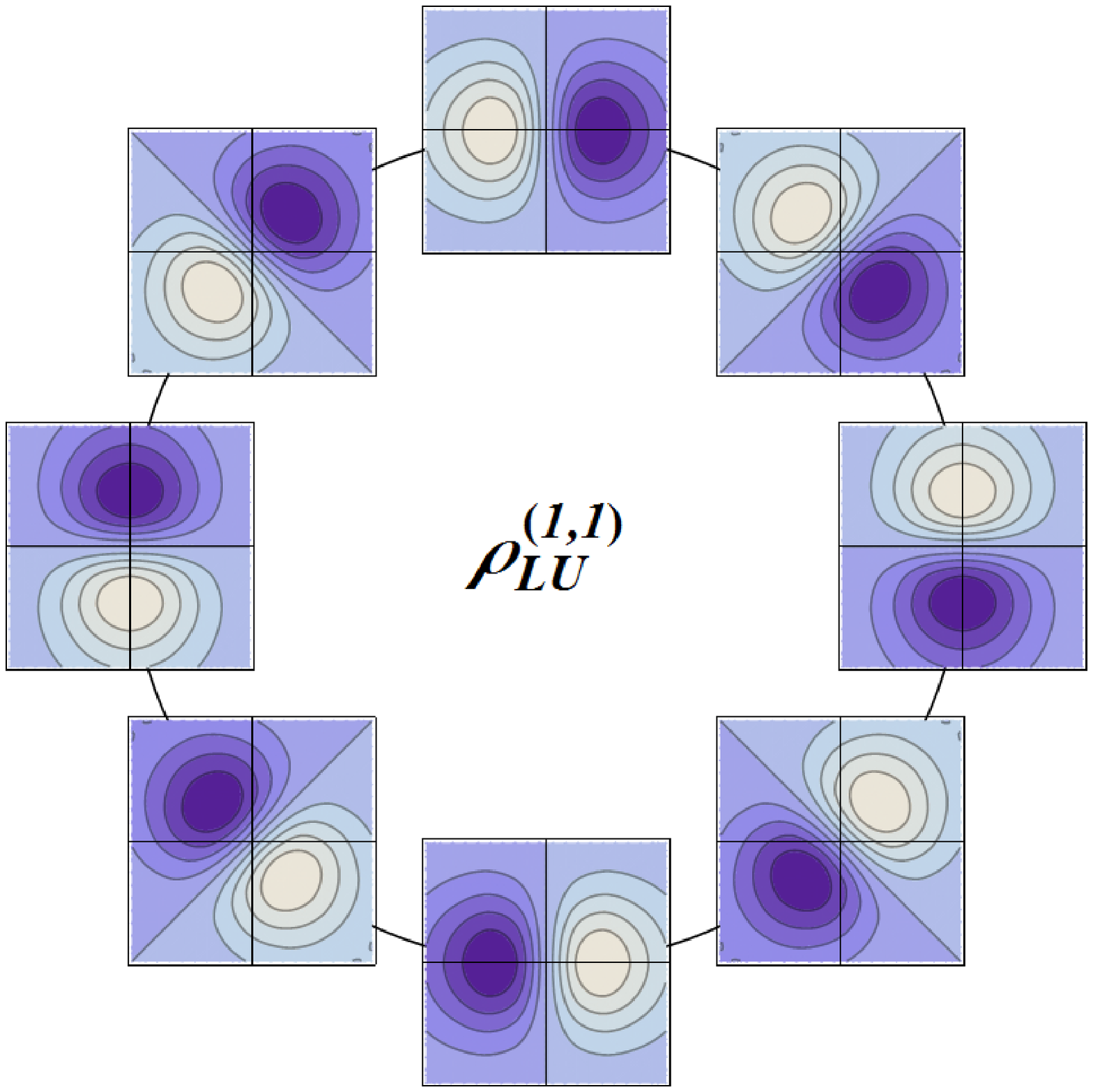}\hspace{1.5cm}\includegraphics[width=7cm]{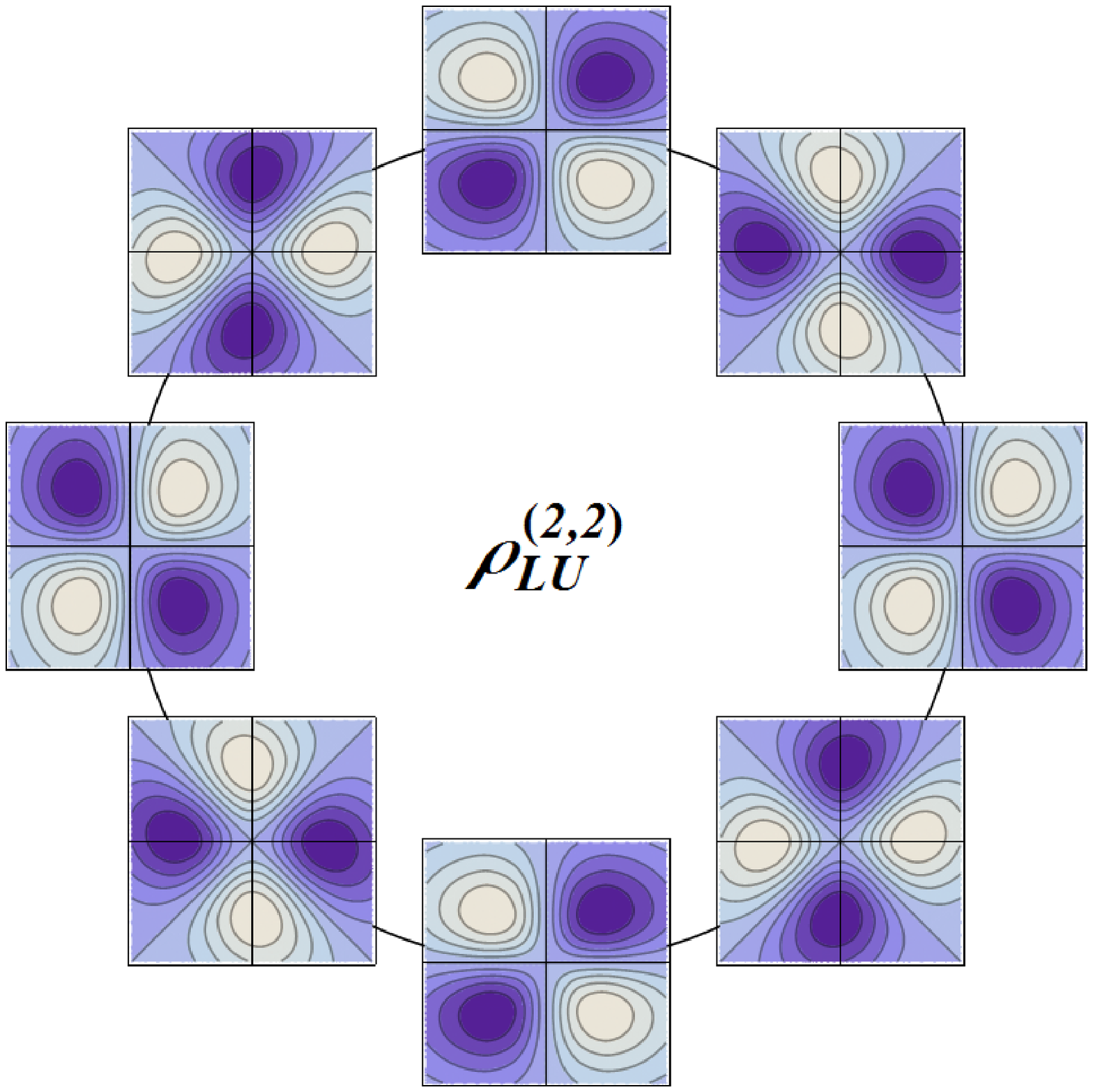}}
\vspace*{8pt}
\caption{Naive $\mathsf T$-even (left) and $\mathsf T$-odd (right) contributions to the transverse phase-space distribution $\rho_{LU}$. See text for more details. \label{fig4}}
\end{figure}

The contribution $\rho_{LU}$ describes how the distribution of unpolarized quarks is affected by the target longitudinal polarization. Its structure is very similar to $\rho_{UL}$ because one just exchanges the roles of quark and target polarizations. We then find only two phase-space distributions
\begin{equation}
\rho^e_{LU}=\rho^{(1,1)}_{LU},\qquad\rho^o_{LU}=\rho^{(2,2)}_{LU},
\end{equation}
which are represented in Fig.~\ref{fig4}. 
None of these survive integration over $\uvec k_T$ or $\uvec b_T$. Both therefore represent completely new information which is  not accessible \emph{via} GPDs or TMDs at leading twist. 

Following the same arguments as in Sec.~\ref{subsect:UL}, with $S_{L}^{q}$ replaced by $S_L$, we can relate the $\rho^{(1,1)}_{LU}$ contribution to  the presence of a net longitudinal component of quark OAM correlated with the target longitudinal polarization $S_L$, with the  coefficient function $C^{(1,1)}_{LU}$ giving  the amount of longitudinal quark OAM in a longitudinally polarized target $\langle S_L\ell^q_L\rangle$~\cite{Lorce:2011kd}.
Similarly, the contribution $\rho^{(2,2)}_{LU}$ gives the difference of radial flows between quarks with opposite OAM $\langle S_L\ell^q_L\rangle$, with  the coefficient function $C^{(2,2)}_{LU}$ representing in some sense the strength of the $\langle S_L\ell^q_L\rangle$-dependent part of the force felt by the quark due to initial- and final-state interactions.

The corresponding basic multipoles are
\begin{align}
S_L\,B^{(1,1)}_{LU}(\hat k_T,\hat b_T;\hat P,\eta)&=-S_L\epsilon^{ij}_TD^i_kD^j_b=S_L(\hat b_T\times\hat k_T)_L,\\
S_L\,B^{(2,2)}_{LU}(\hat k_T,\hat b_T;\hat P,\eta)&=-\eta\,S_L\epsilon^{ij}_TQ^{il}_kQ^{jl}_b=\eta\,S_L(\hat b_T\times\hat k_T)_L\,(\hat k_T\cdot\hat b_T).
\end{align}

\subsubsection{Longitudinally polarized quark}

\begin{figure}[t]
\centerline{\includegraphics[width=7cm]{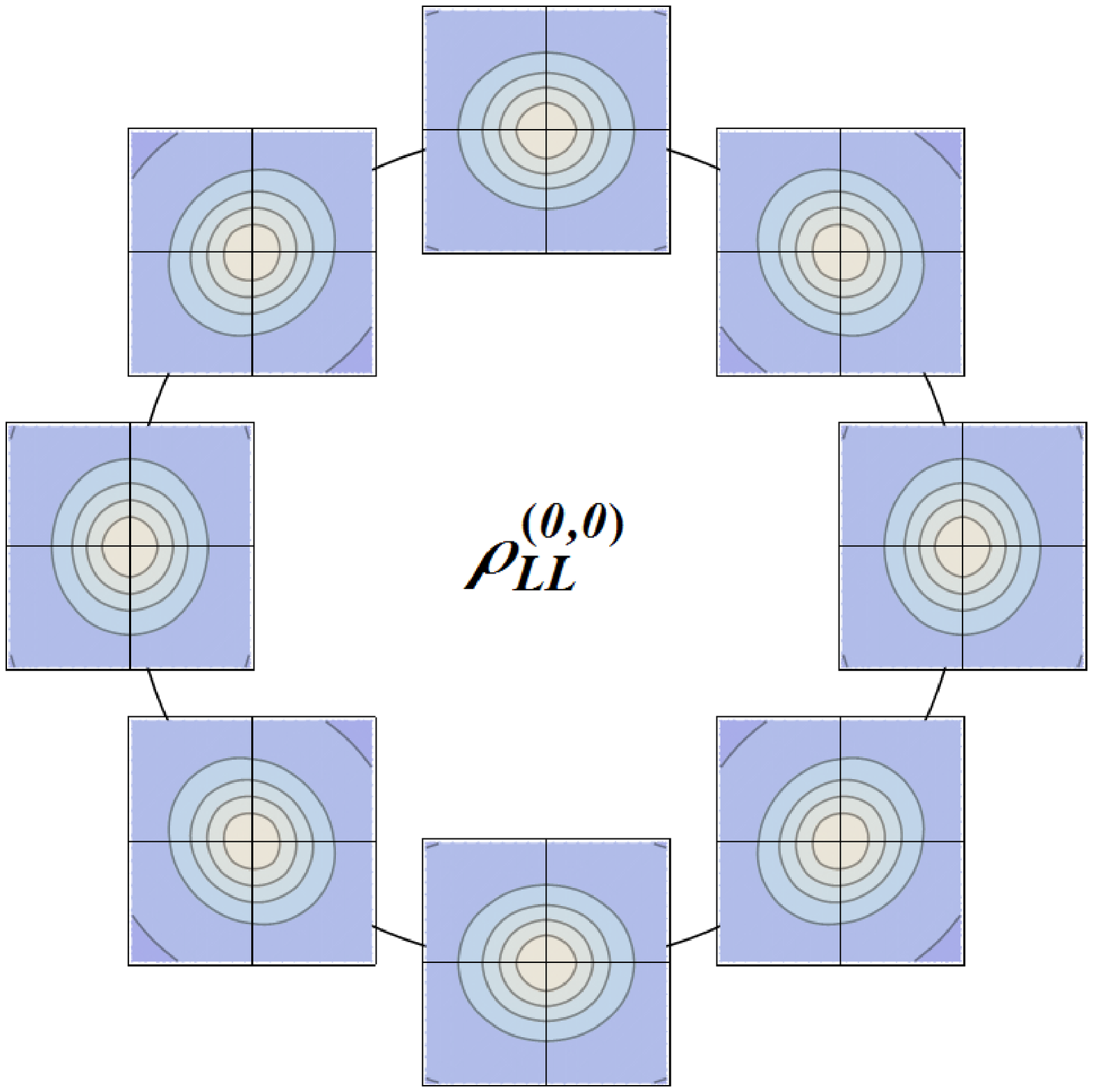}\hspace{1.5cm}\includegraphics[width=7cm]{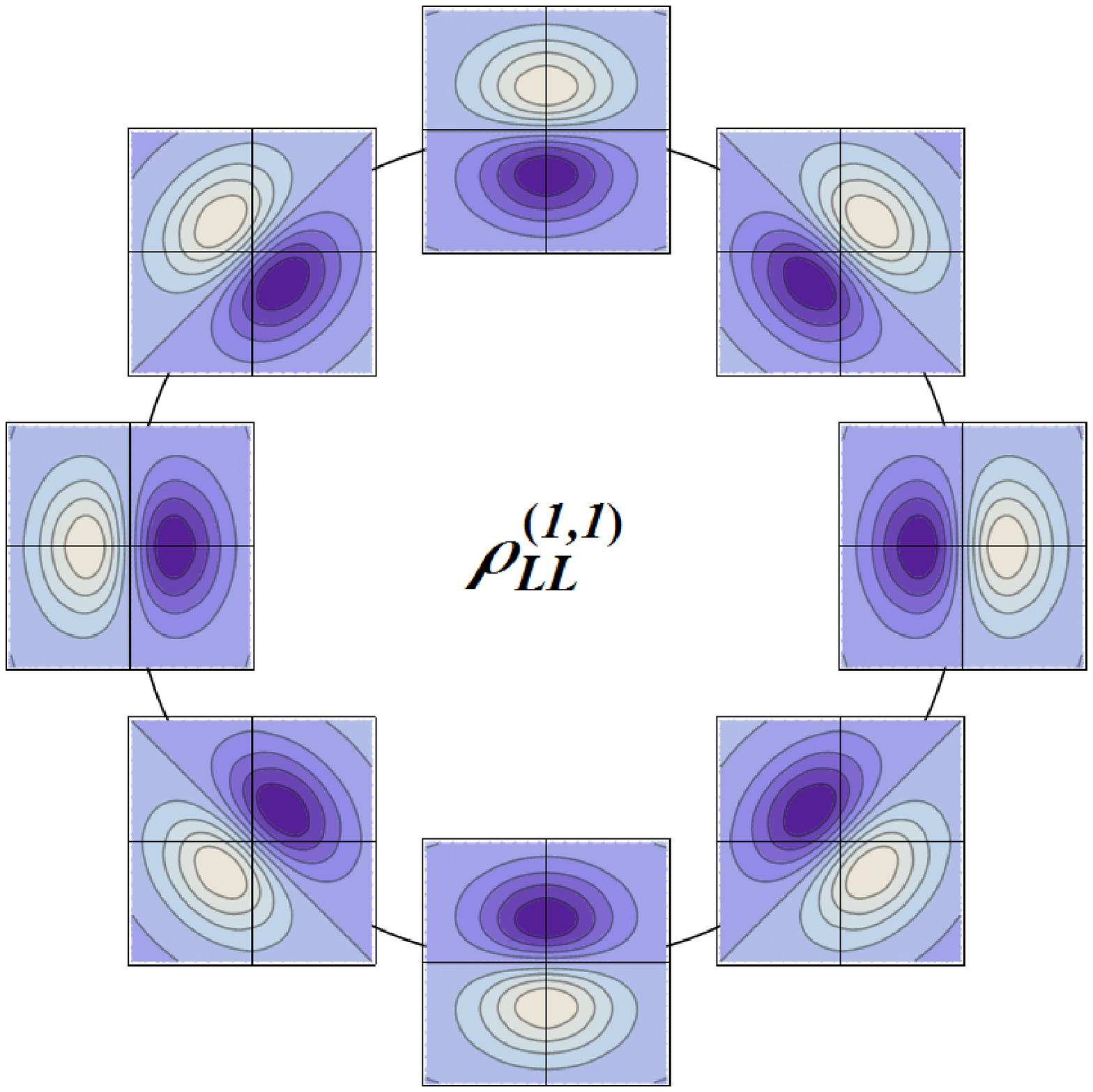}}
\vspace*{8pt}
\caption{Naive $\mathsf T$-even (left) and $\mathsf T$-odd (right) contributions to the transverse phase-space distribution $\rho_{LL}$. See text for more details. \label{fig5}}
\end{figure}

The contribution $\rho_{LL}$ describes how the quark distribution is affected by the correlation between the quark and target longitudinal polarizations. Since the product $S_LS^q_L$ is invariant under parity and time-reversal, the contribution $\rho_{LL}$ turns out to be very similar to $\rho_{UU}$. We then find only two phase-space distributions
\begin{equation}
\rho^e_{LL}=\rho^{(0,0)}_{LL},\qquad\rho^o_{LL}=\rho^{(1,1)}_{LL},
\end{equation}
which are represented in Fig.~\ref{fig5}. 
The corresponding basic multipoles are
\begin{align}
S_LS^q_L\,B^{(0,0)}_{LL}(\hat k_T,\hat b_T;\hat P,\eta)&=S_LS^q_LM_kM_b=S_LS^q_L,\\
S_LS^q_L\,B^{(1,1)}_{LL}(\hat k_T,\hat b_T;\hat P,\eta)&=\eta\,S_LS^q_LD^i_kD^i_b=\eta\,S_LS^q_L(\hat k_T\cdot\hat b_T).
\end{align}
Only $\rho^{(0,0)}_{LL}$ survives integration over $\uvec k_T$ or $\uvec b_T$ and is then naturally related to both the helicity GPD $\tilde H$ and the helicity TMD $g_{1L}$~\cite{Diehl:2005jf,Lorce:2011dv}. Contrary to its $\uvec k_T$- and $\uvec b_T$-integrated versions, $\rho^{(0,0)}_{LL}$ is not circularly symmetric. The reason is that $\rho^{(0,0)}_{LL}$ also contains information about the \emph{correlation} between $\uvec k_T$ and $\uvec b_T$, which is lost under integration over one of the transverse variables~\cite{Lorce:2011kd}. 
\newline
Following the same arguments as in Sec.~\ref{subsect:UU}, with now all expressions multiplied by $S_LS_{L}^{q}$, we can relate the coefficient function $C^{(0,0)}_{LL}$ to the strength of the correlation between the longitudinal component of quark and target polarizations $\langle S_LS^q_L\rangle$. Similarly, the contribution $\rho^{(1,1)}_{LL}$ gives the difference of radial flows between quarks with opposite $\langle S_LS^q_L\rangle$ correlations, with  the coefficient function $C^{(1,1)}_{LL}$ representing in some sense the strength of the $\langle S_LS^q_L\rangle$-dependent part of the force felt by the quark due to initial- and final-state interactions.

\subsubsection{Transversely polarized quark}
\label{subsect:LT}

\begin{figure}[t]
\centerline{\includegraphics[width=7cm]{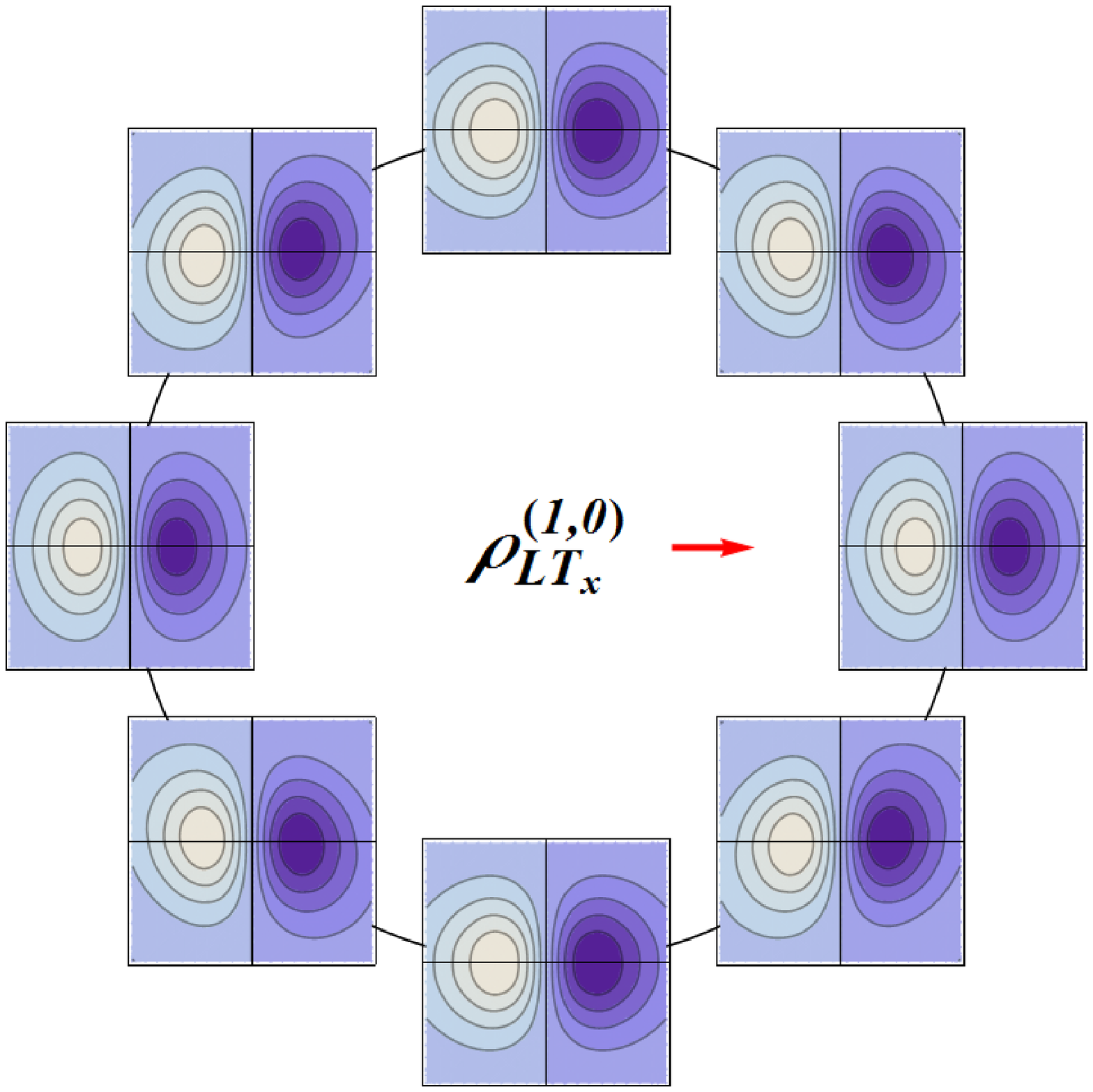}\hspace{1.5cm}\includegraphics[width=7cm]{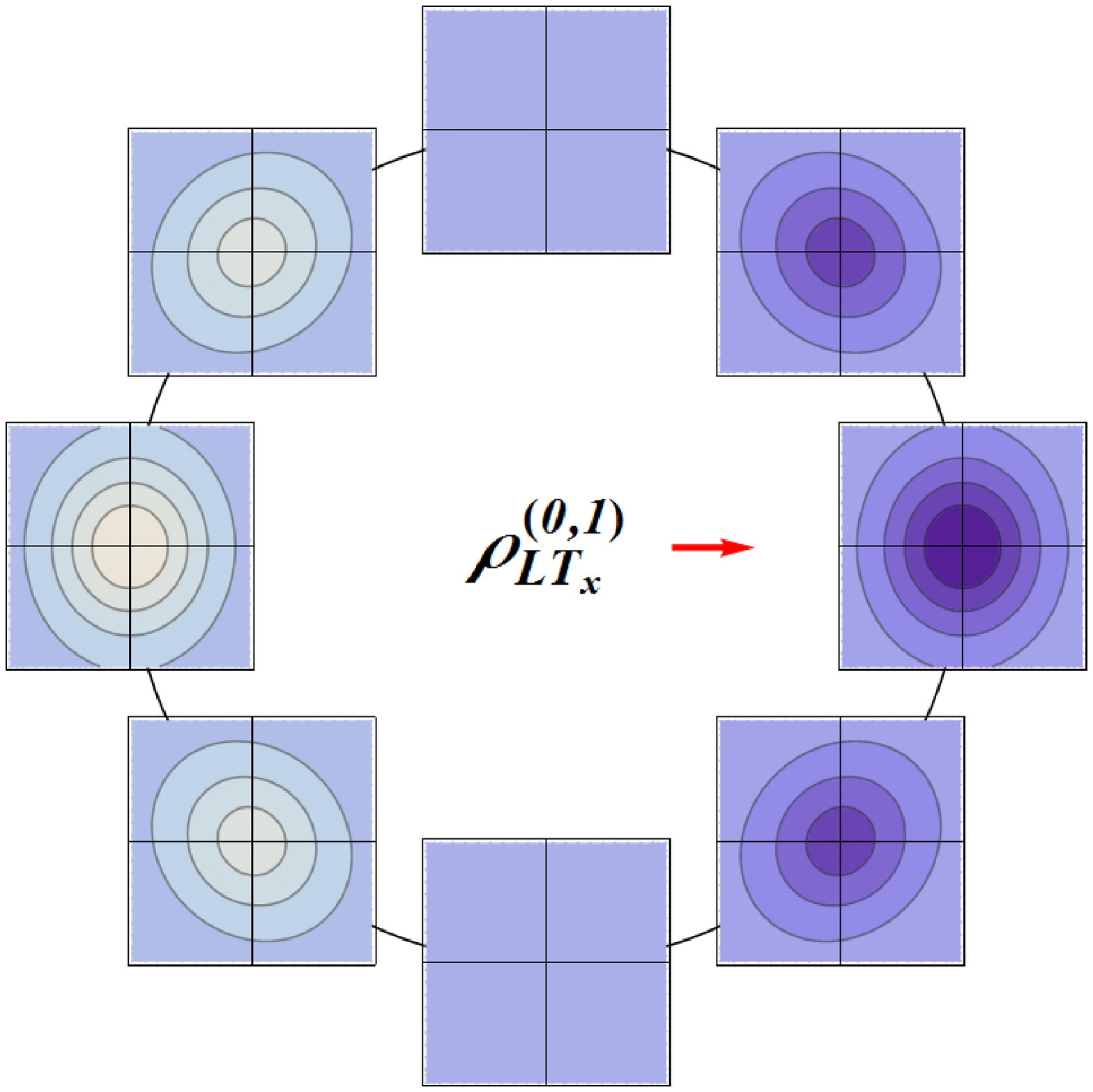}}
\vspace*{15pt}
\centerline{\includegraphics[width=7cm]{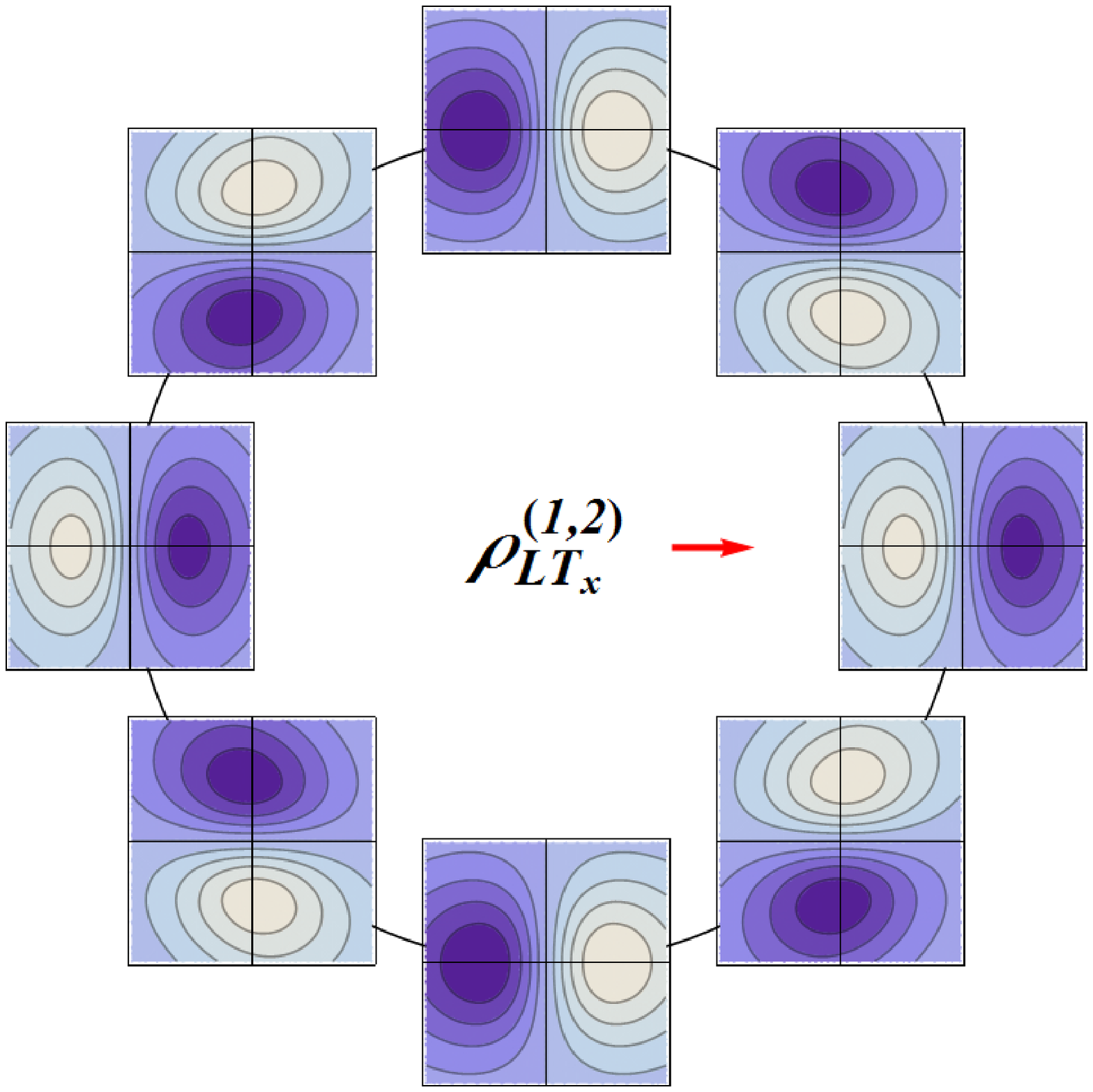}\hspace{1.5cm}\includegraphics[width=7cm]{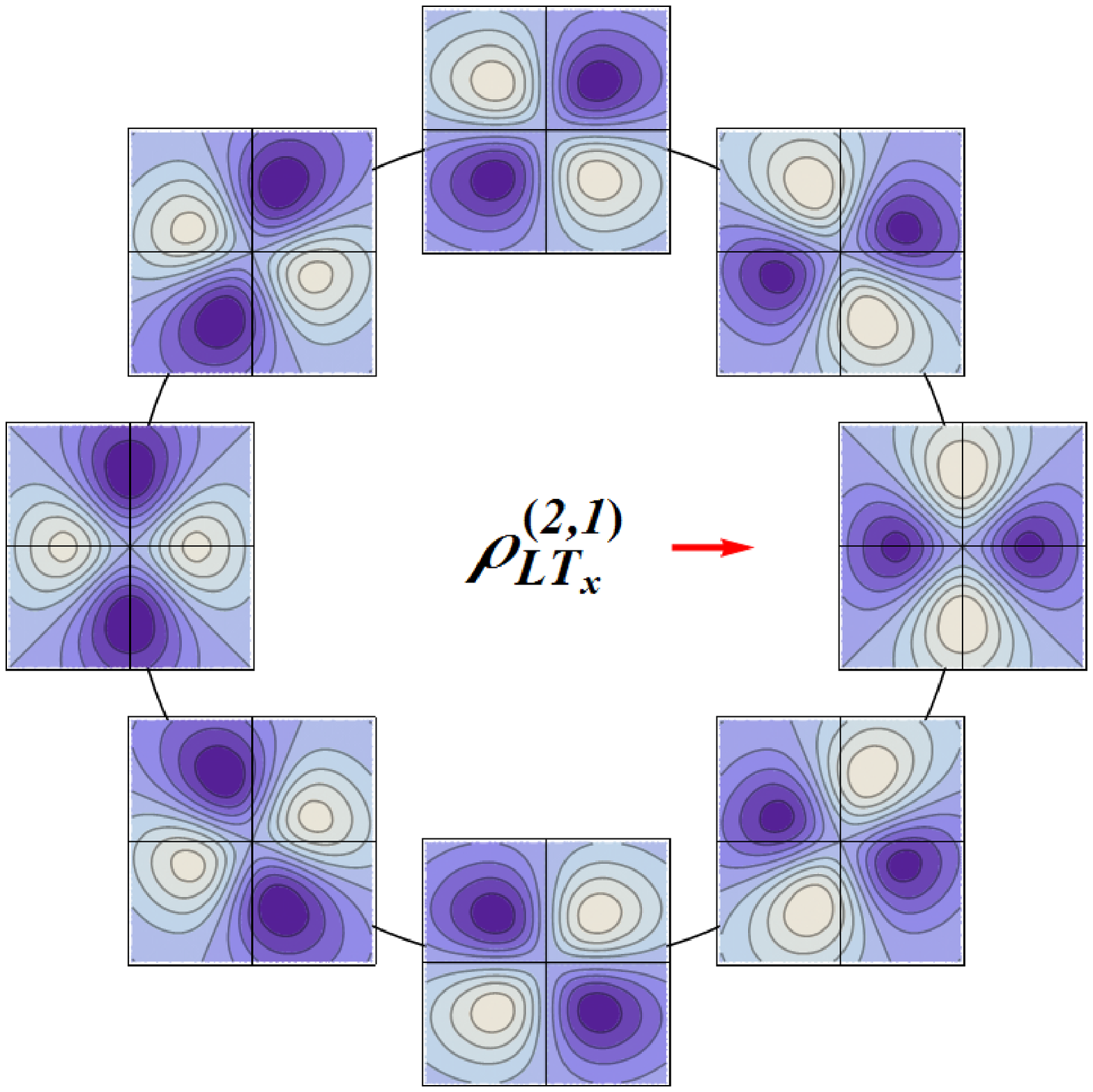}}
\vspace*{8pt}
\caption{Naive $\mathsf T$-even (left) and $\mathsf T$-odd (right) contributions to the transverse phase-space distribution $\rho_{LT}$ for the quark polarization $\vec S^q_T=\vec e_x$ (red arrow). See text for more details. \label{fig6}}
\end{figure}

The contribution $\rho_{LT^i}$ describes how the distribution of quarks is affected by the combination of quark transverse polarization and target longitudinal polarization. We find in total four phase-space distributions
\begin{equation}
\rho^e_{LT^i}=\rho^{(1,0)}_{LT^i}+\rho^{(1,2)}_{LT^i},\qquad\rho^o_{LT^i}=\rho^{(0,1)}_{LT^i}+\rho^{(2,1)}_{LT^i},
\end{equation}
which are represented in Fig.~\ref{fig6} for the quark polarization $\vec S^q_T=\vec e_x$. 
The corresponding basic multipoles are
\begin{align}
S_LS^{qi}_T\,B^{(1,0)}_{LT^i}(\hat k_T,\hat b_T;\hat P,\eta)&=S_LS^{qi}_TD^i_kM_b=S_L(\uvec S^q_T\cdot\hat k_T),\label{LT1}\\
S_LS^{qi}_T\,B^{(1,2)}_{LT^i}(\hat k_T,\hat b_T;\hat P,\eta)&=S_LS^{qi}_TD^j_kQ^{ij}_b=S_L\!\left[(\uvec S^q_T\cdot\hat b_T)\,(\hat b_T\cdot\hat k_T)-\tfrac{1}{2}\,(\uvec S^q_T\cdot\hat k_T)\right],\label{LT2}\\
S_LS^{qi}_T\,B^{(0,1)}_{LT^i}(\hat k_T,\hat b_T;\hat P,\eta)&=\eta\,S_LS^{qi}_TM_kD^i_b=\eta\,S_L(\uvec S^q_T\cdot\hat b_T),\label{LT3}\\
S_LS^{qi}_T\,B^{(2,1)}_{LT^i}(\hat k_T,\hat b_T;\hat P,\eta)&=\eta\,S_LS^{qi}_TQ^{ij}_kD^j_b=\eta\,S_L\!\left[(\uvec S^q_T\cdot\hat k_T)\,(\hat k_T\cdot\hat b_T)-\tfrac{1}{2}\,(\uvec S^q_T\cdot\hat b_T)\right].\label{LT4}
\end{align}
The contribution $\rho^{(1,0)}_{LT^i}$ is the only one surviving the integration over $\uvec b_T$ and is then naturally related to  the worm-gear TMD $h^\perp_{1L}$~\cite{Diehl:2005jf,Lorce:2011dv}. The dipole in $\uvec k_T$-space indicates the presence of a net transverse flow parallel to the quark transverse polarization. This transverse flow is actually an effect due to the light-front imaging, once again associated with the fact that the light-front densities are defined in terms of the $j^+=\tfrac{1}{\sqrt{2}}(j^0+j^3)$ component of the current instead of the $j^0$ component. As we will soon see, it turns out that the transverse flow finds its physical origin in the correlation between the longitudinal component of quark OAM and the transverse spin-orbit coupling $\langle S_L\ell^q_L(\uvec S^q_T\cdot\uvec\ell^q_T)\rangle$.

The contribution $\rho^{(1,2)}_{LT^i}$ corresponds to a completely new information which is  not accessible \emph{via} GPDs or TMDs at leading twist. Combined with $\rho^{(1,0)}_{LT^i}$, it tells us how the quark distribution is affected by the two longitudinal-transverse worm-gear correlations, say $\langle S_L\ell^q_LS^q_x\ell^q_x\rangle$ and $\langle S_L\ell^q_LS^q_y\ell^q_y\rangle$. Indeed, let us consider the projection of a 3-dimensional \mbox{$\langle S_L\ell^q_L (\vec S^q_T\cdot\vec n_T)(\vec\ell^q_T\cdot\vec n_T)\rangle$} correlation onto the transverse position space. For $\vec n_T=\vec b_T$ and $\vec n_T=(\vec b_T\times\hat P)$, we respectively find
\begin{align}
\label{eq:57}
\int\ud b_L\,[(\vec b\times\vec k)\cdot\hat P]\,(\vec S^q_T\cdot\vec b_T)\,[(\vec b\times\vec k)_T\cdot\vec b_T]&\sim (\uvec S^q_T\cdot\hat b_T)\,(\hat b_T\cdot\hat k_T),\\
\label{eq:58}
\int\ud b_L\,[(\vec b\times\vec k)\cdot\hat P]\,[\vec S^q_T\cdot(\vec b_T\times\hat P)]\,[(\vec b\times\vec k)_T\cdot(\vec b_T\times\hat P)]&\sim(\uvec S^q_T\times\hat b_T)_L\,(\hat b_T\times\hat k_T)_L.
\end{align}
Noting that 
\begin{equation}
\label{eq:59}
(\uvec S^q_T\cdot\hat b_T)\,(\hat b_T\cdot\hat k_T)-(\uvec S^q_T\times\hat b_T)_L\,(\hat b_T\times\hat k_T)_L=(\uvec S^q_T\cdot\hat k_T)
\end{equation}
and comparing with the basic multipoles~\eqref{LT1} and~\eqref{LT2}, we can see that the two coefficient functions $C^{(1,0)}_{LT^i}$ and $C^{(1,2)}_{LT^i}$ are related to the strength of two different combinations of the two longitudinal-transverse worm-gear correlations $\langle S_L\ell^q_LS^q_x\ell^q_x\rangle$ and $\langle S_L\ell^q_LS^q_y\ell^q_y\rangle$.
\newline

Similarly, the contribution $\rho^{(0,1)}_{LT^i}$ is the only one surviving the integration over $\uvec k_T$. It cannot however be related to the GPD $\tilde E_T$~\cite{Diehl:2005jf,Lorce:2011dv,Pasquini:2007xz} since the latter is $\eta$-independent\footnote{Moreover, the GPD $\tilde E_T$ is $\xi$-odd and cannot therefore appear in our multipole decomposition based on $\xi=0$.}. It then corresponds to a completely new information. Once again, the dipole in $\uvec b_T$-space indicates the presence of a spatial separation between quarks with opposite correlations. This is again an effect related to  the light-front imaging. 

The contribution $\rho^{(2,1)}_{LT^i}$ corresponds to another completely new information which is  not accessible \emph{via} GPDs or TMDs at leading twist. Combined with $\rho^{(0,1)}_{LT^i}$, it tells us how the initial- and final state-interactions depend on the two longitudinal-transverse worm-gear correlations, say $\langle S_L\ell^q_LS^q_x\ell^q_x\rangle$ and $\langle S_L\ell^q_LS^q_y\ell^q_y\rangle$. Indeed, let us consider the projection of a 3-dimensional spiral worm-gear flow \mbox{$[(\vec b\times\vec k)\cdot\hat P]\,(\vec S^q_T\cdot\vec n_T)\,[(\vec b\times\vec k)_T\cdot\vec n_T]\,(\vec k\cdot\vec b)$} onto the transverse position space. For $\vec n_T=\vec k_T$ and $\vec n_T=(\vec k_T\times\hat P)$, we respectively find
\begin{align}
\label{eq:60}
\int\ud b_L\,[(\vec b\times\vec k)\cdot\hat P]\,(\vec S^q_T\cdot\vec k_T)\,[(\vec b\times\vec k)_T\cdot\vec k_T]\,(\vec k\cdot\vec b)&\sim (\uvec S^q_T\cdot\hat k_T)\,(\hat k_T\cdot\hat b_T),\\
\label{eq:61}
\int\ud b_L\,[(\vec b\times\vec k)\cdot\hat P]\,[\vec S^q_T\cdot(\vec k_T\times\hat P)]\,[(\vec b\times\vec k)_T\cdot(\vec k_T\times\hat P)]\,(\vec k\cdot\vec b)&\sim(\uvec S^q_T\times\hat k_T)_L\,(\hat k_T\times\hat b_T)_L.
\end{align}
Noting that 
\begin{equation}
\label{eq:62}
(\uvec S^q_T\cdot\hat k_T)\,(\hat k_T\cdot\hat b_T)-(\uvec S^q_T\times\hat k_T)_L\,(\hat k_T\times\hat b_T)_L=(\uvec S^q_T\cdot\hat b_T)
\end{equation}
and comparing with the basic multipoles~\eqref{LT3} and~\eqref{LT4}, we can see that the two coefficient functions $C^{(0,1)}_{LT^i}$ and $C^{(2,1)}_{LT^i}$ are related to the strength of the $\langle S_L\ell^q_LS^q_x\ell^q_x\rangle$- and $\langle S_L\ell^q_LS^q_y\ell^q_y\rangle$-dependent parts of the force felt by the quark due to initial and final-state interactions. In other words, the contributions $\rho^{(0,1)}_{LT^i}$ and $\rho^{(2,1)}_{LT^i}$  describe the difference of radial flows between quarks with opposite $\langle S_L\ell^q_LS^q_x\ell^q_x\rangle$ or $\langle S_L\ell^q_LS^q_y\ell^q_y\rangle$ correlations.

\subsection{Transversely polarized target}

\subsubsection{Unpolarized quark}

\begin{figure}[t]
\centerline{\includegraphics[width=7cm]{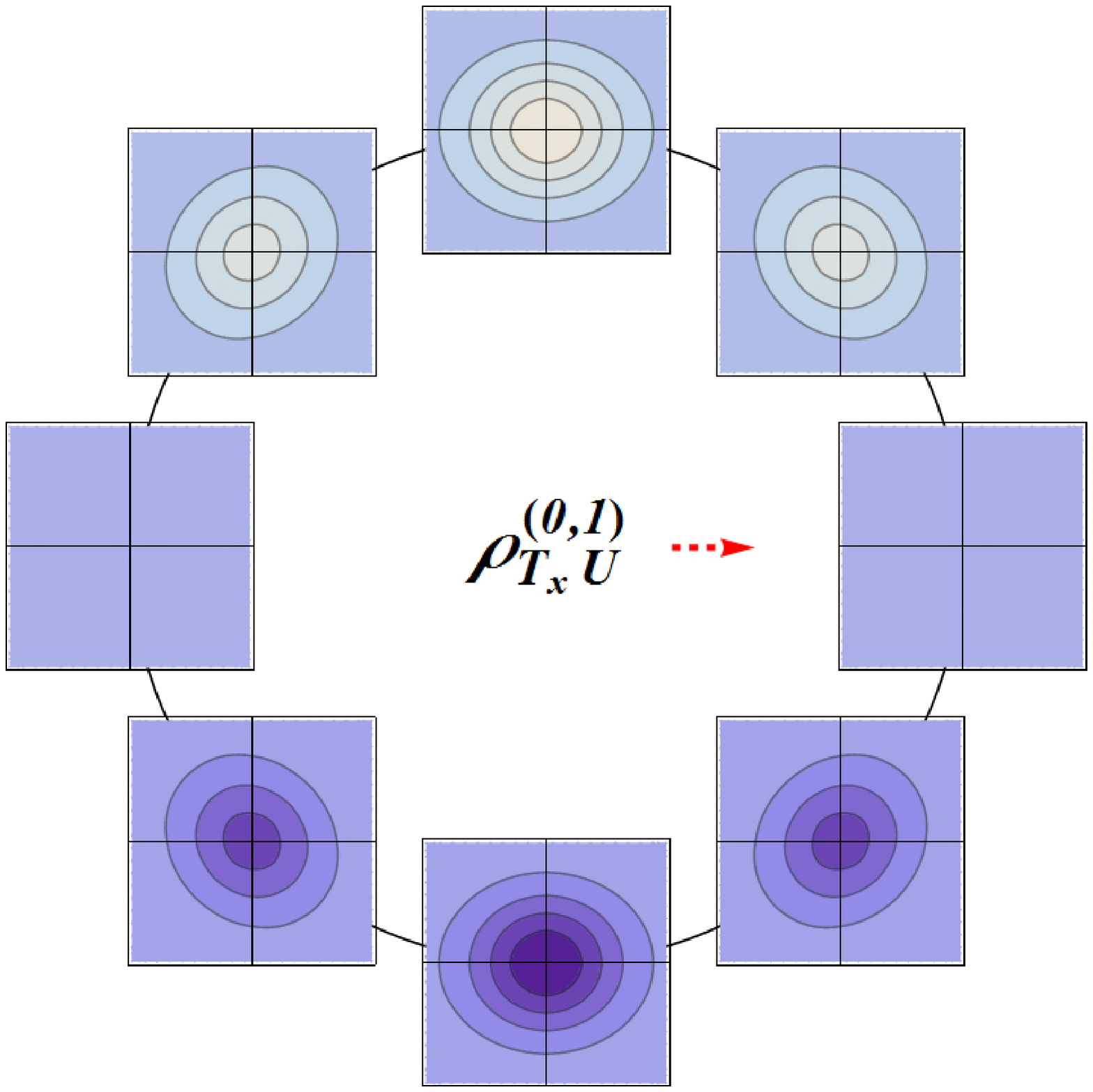}\hspace{1.5cm}\includegraphics[width=7cm]{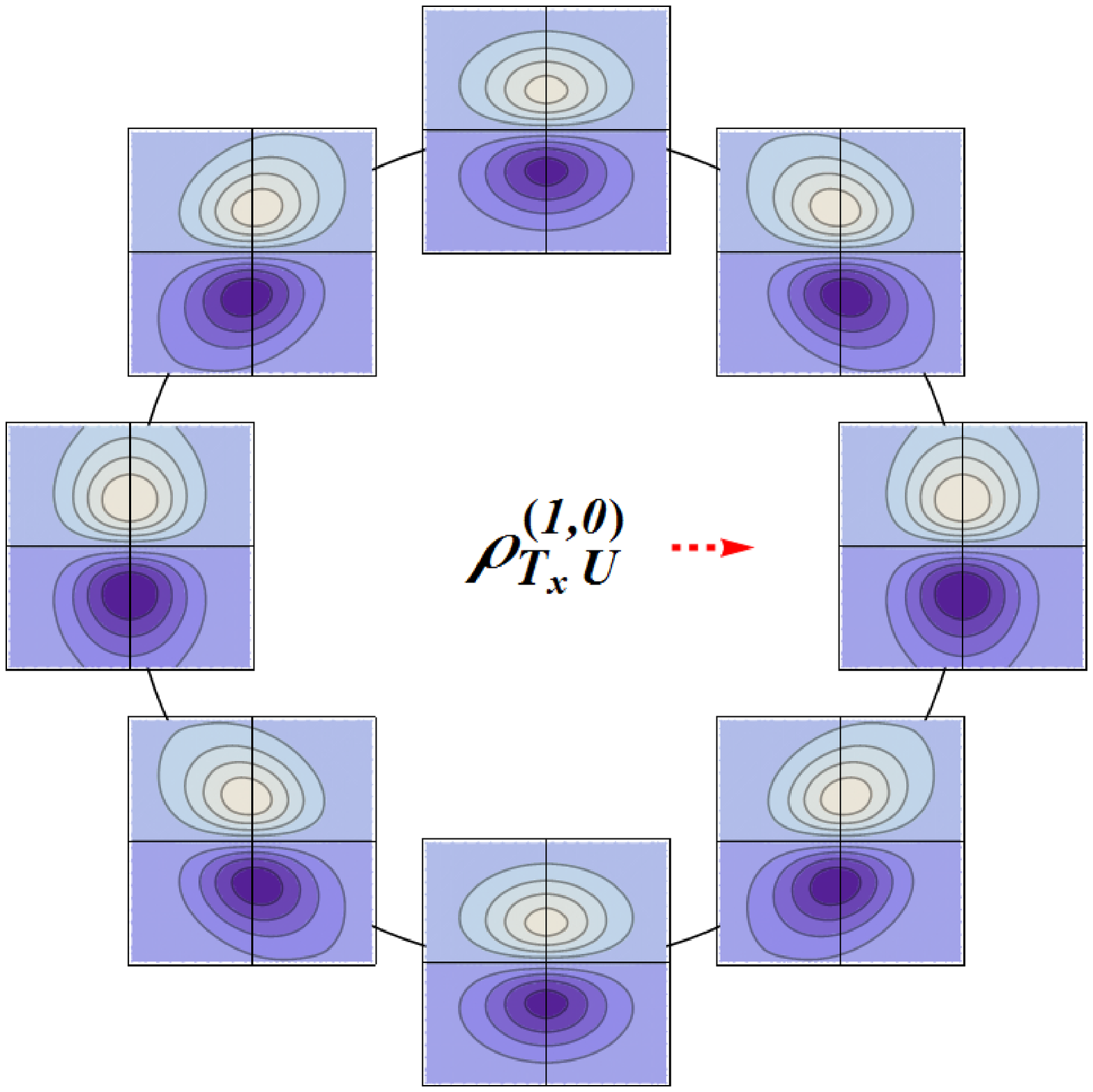}}
\vspace*{15pt}
\centerline{\includegraphics[width=7cm]{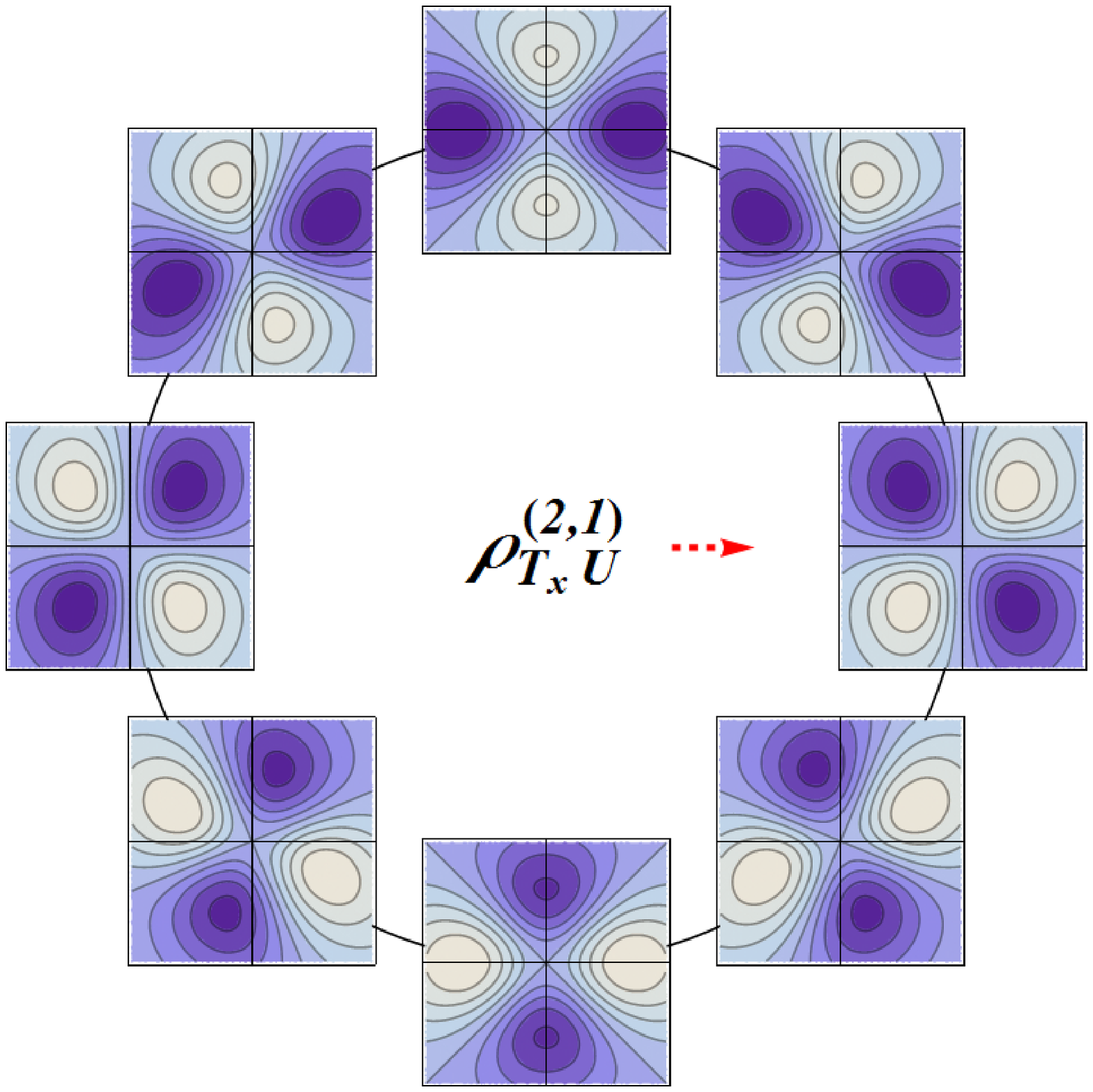}\hspace{1.5cm}\includegraphics[width=7cm]{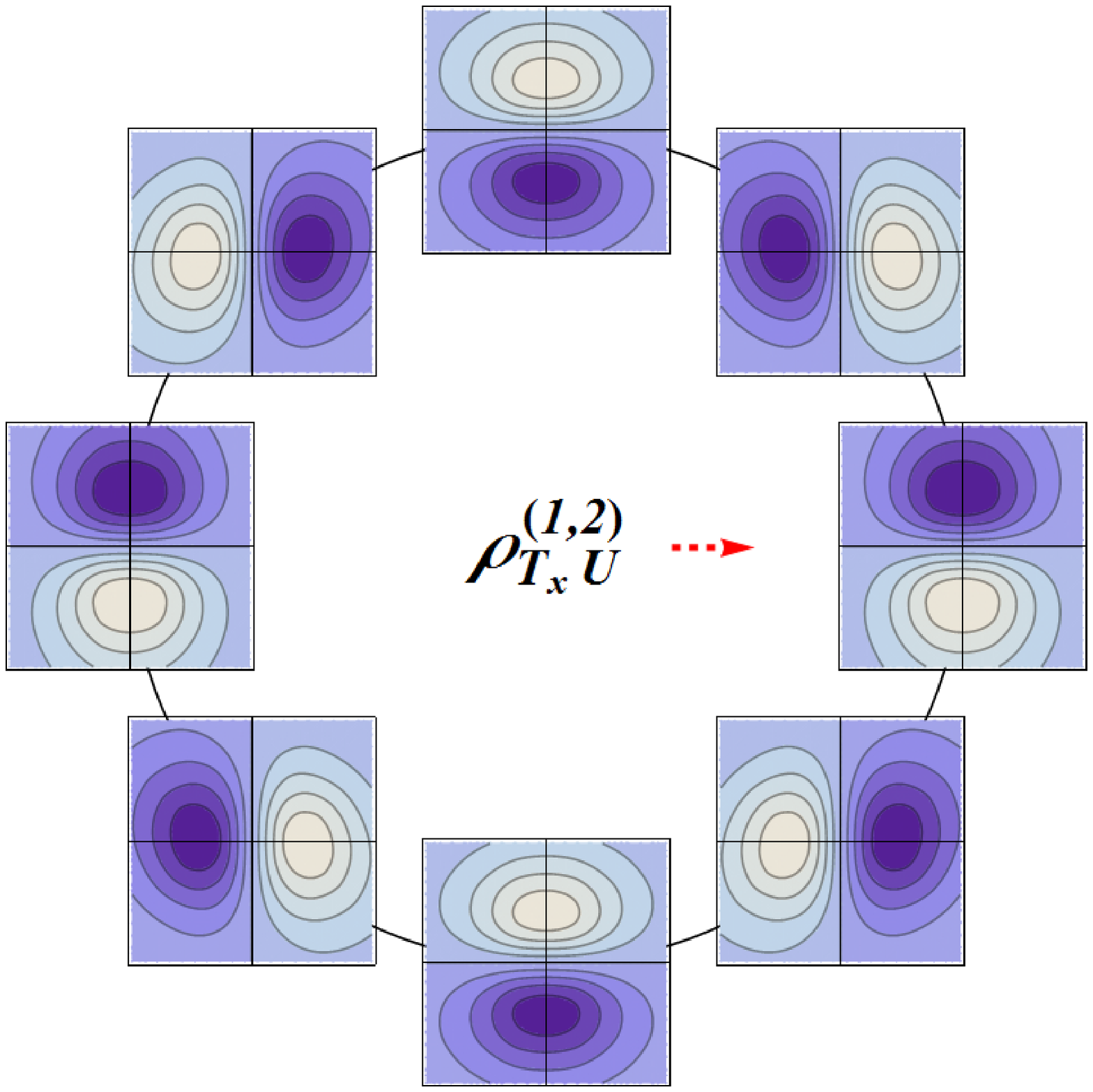}}
\vspace*{8pt}
\caption{Naive $\mathsf T$-even (left) and $\mathsf T$-odd (right) contributions to the transverse phase-space distribution $\rho_{TU}$ for the target polarization $\vec S_T=\vec e_x$ (red dashed arrow). See text for more details. \label{fig7}}
\end{figure}

The contribution $\rho_{T^iU}$ describes how the distribution of unpolarized quarks is affected by the target transverse polarization. Its structure is very similar to $\rho_{UT^i}$ because one just exchanges the roles of quark and target polarizations. We then find in total four phase-space distributions
\begin{equation}
\rho^e_{T^iU}=\rho^{(0,1)}_{T^iU}+\rho^{(2,1)}_{T^iU},\qquad\rho^o_{T^iU}=\rho^{(1,0)}_{T^iU}+\rho^{(1,2)}_{T^iU},
\end{equation}
which are represented in Fig.~\ref{fig7} for the target polarization $\vec S_T=\vec e_x$.
The corresponding basic multipoles are
\begin{align}
S^i_T\,B^{(0,1)}_{T^iU}(\hat k_T,\hat b_T;\hat P,\eta)&=S^i_T\epsilon^{ij}_TM_kD^j_b=(\uvec S_T\times\hat b_T)_L,\label{TU1}\\
S^i_T\,B^{(2,1)}_{T^iU}(\hat k_T,\hat b_T;\hat P,\eta)&=S^i_T\epsilon^{ij}_TQ^{jl}_kD^l_b=(\uvec S_T\times\hat k_T)_L\,(\hat k_T\cdot\hat b_T)-\tfrac{1}{2}\,(\uvec S_T\times\hat b_T)_L,\label{TU2}\\
S^i_T\,B^{(1,0)}_{T^iU}(\hat k_T,\hat b_T;\hat P,\eta)&=\eta\,S^i_T\epsilon^{ij}_TD^j_kM_b=\eta\,(\uvec S_T\times\hat k_T)_L,\label{TU3}\\
S^i_T\,B^{(1,2)}_{T^iU}(\hat k_T,\hat b_T;\hat P,\eta)&=\eta\,S^i_T\epsilon^{jl}_TD^l_kQ^{ij}_b=\eta\left[(\uvec S_T\cdot\hat b_T)\,(\hat b_T\times\hat k_T)_L-\tfrac{1}{2}\,(\uvec S_T\times\hat k_T)_L\right].\label{TU4}
\end{align}
The contribution $\rho^{(0,1)}_{T^iU}$ is the only one surviving the integration over $\uvec k_T$ and is then naturally related to the GPD $E$~\cite{Diehl:2005jf,Lorce:2011dv,Pasquini:2007xz}. The dipole in $\uvec b_T$-space indicates a spatial shift in the distribution of quarks due to the target transverse polarization. This is again a result of the light-front imaging associated with the fact that the light-front densities are defined in terms of the $j^+=\tfrac{1}{\sqrt{2}}(j^0+j^3)$ component of the current instead of the $j^0$ component. The spatial shift finds its physical origin in the transverse quark OAM $\langle \uvec S_T\cdot\uvec\ell^q_T\rangle$~\cite{Burkardt:2002hr}, and can also be understood from the fact that the position of the relativistic center-of-mass of a rotating body is frame-dependent~\cite{Moller:1949,Moller:1972}.

The contribution $\rho^{(2,1)}_{T^iU}$ corresponds to a completely new information which is  not accessible \emph{via} GPDs or TMDs at leading twist. Combined with $\rho^{(0,1)}_{T^iU}$, it tells us how the quark distribution is affected  by the two transverse components of quark OAM, say $\langle S_x\ell^q_x\rangle$ and $\langle S_y\ell^q_y\rangle$. Following the same arguments as in Sec.~\ref{subsect:UT}, with $\uvec S_{T}^q$ replaced by $\uvec S_T$, we can relate the two coefficient functions $C^{(0,1)}_{T^iU}$ and $C^{(2,1)}_{T^iU}$ to the amount of transverse quark OAM in a transversely polarized target $\langle  S_x\ell^q_x\rangle$ and $\langle  S_y\ell^q_y\rangle$.
\newline

Similarly, the contribution $\rho^{(1,0)}_{T^iU}$ is the only one surviving the integration over $\uvec b_T$ and is then naturally related to the Sivers TMD $f^\perp_{1T}$~\cite{Diehl:2005jf,Lorce:2011dv}. The dipole in $\uvec k_T$-space indicates the presence of a net transverse flow orthogonal to the quark transverse polarization. Such a net transverse flow can only arise from initial- and/or final-state interactions, in accordance with the naive $\mathsf T$-odd nature of $\rho^{(1,0)}_{T^iU}$. 

The contribution $\rho^{(1,2)}_{T^iU}$ corresponds to a completely new information which is  not accessible \emph{via} GPDs or TMDs at leading twist. Combined with $\rho^{(1,0)}_{T^iU}$, it tells us how the initial- and final-state interactions depend on the two transverse components of quark OAM, say $\langle S_x\ell^q_x\rangle$ and $\langle S_y\ell^q_y\rangle$. Following once again the same arguments as in Sec.~\ref{subsect:UT}, with $\uvec S_{T}^q$ replaced by $\uvec S_T$, we can relate the two coefficient functions $C^{(1,0)}_{T^iU}$ and $C^{(1,2)}_{T^iU}$  to the strength of the $\langle S_x\ell^q_x\rangle$- and $\langle S_y\ell^q_y\rangle$-dependent parts of the force felt by the quark due to initial- and final-state interactions. 
In other words, the contributions $\rho^{(1,0)}_{T^iU}$ and $\rho^{(1,2)}_{T^iU}$ describe the difference of radial flows between quarks with opposite transverse components of OAM $\langle S_x\ell^q_x\rangle$ or $\langle S_y\ell^q_y\rangle$.

Note that it has been suggested that $\int\ud^2k_T\,\rho^e_{T^iU}$ and $\int\ud^2b_T\,\rho^o_{T^iU}$ could be related by some lensing effect~\cite{Burkardt:2002hr,Bacchetta:2011gx}. We cannot unfortunately confirm this suggestion, because such a relation relies on a dynamical mechanism which goes beyond the general  constraints considered in the present paper.

\subsubsection{Longitudinally polarized quark}

\begin{figure}[t]
\centerline{\includegraphics[width=7cm]{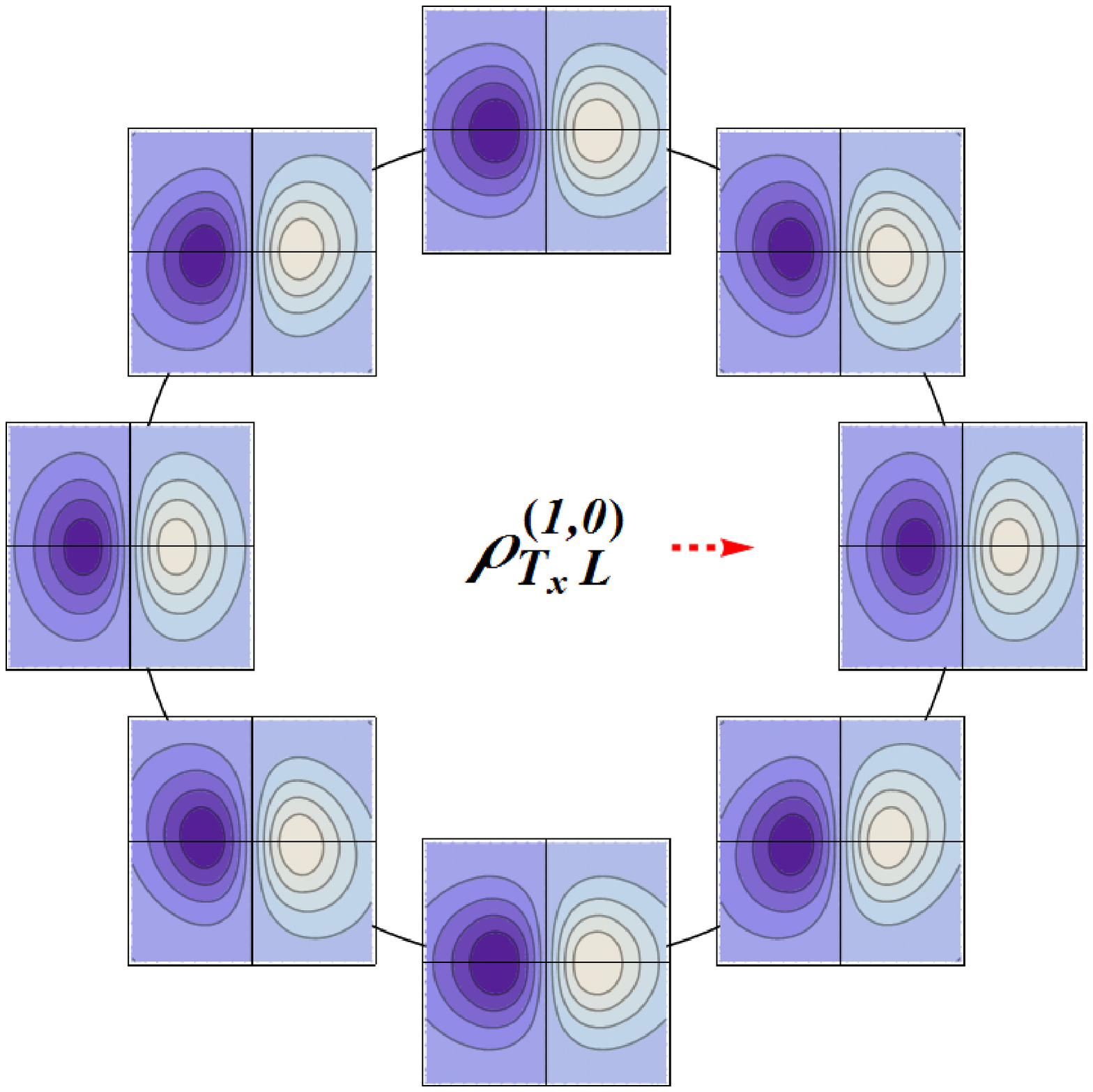}\hspace{1.5cm}\includegraphics[width=7cm]{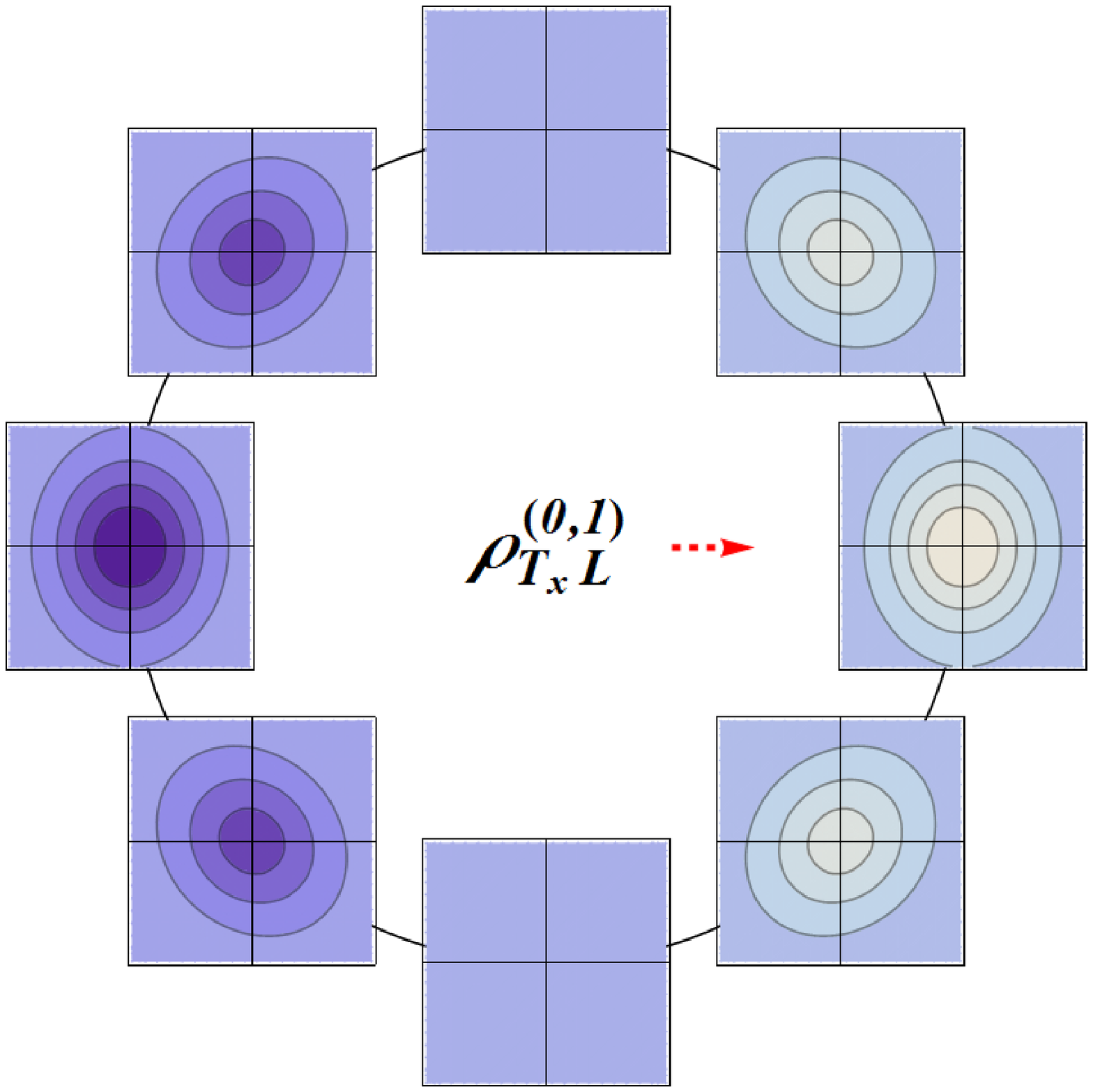}}
\vspace*{15pt}
\centerline{\includegraphics[width=7cm]{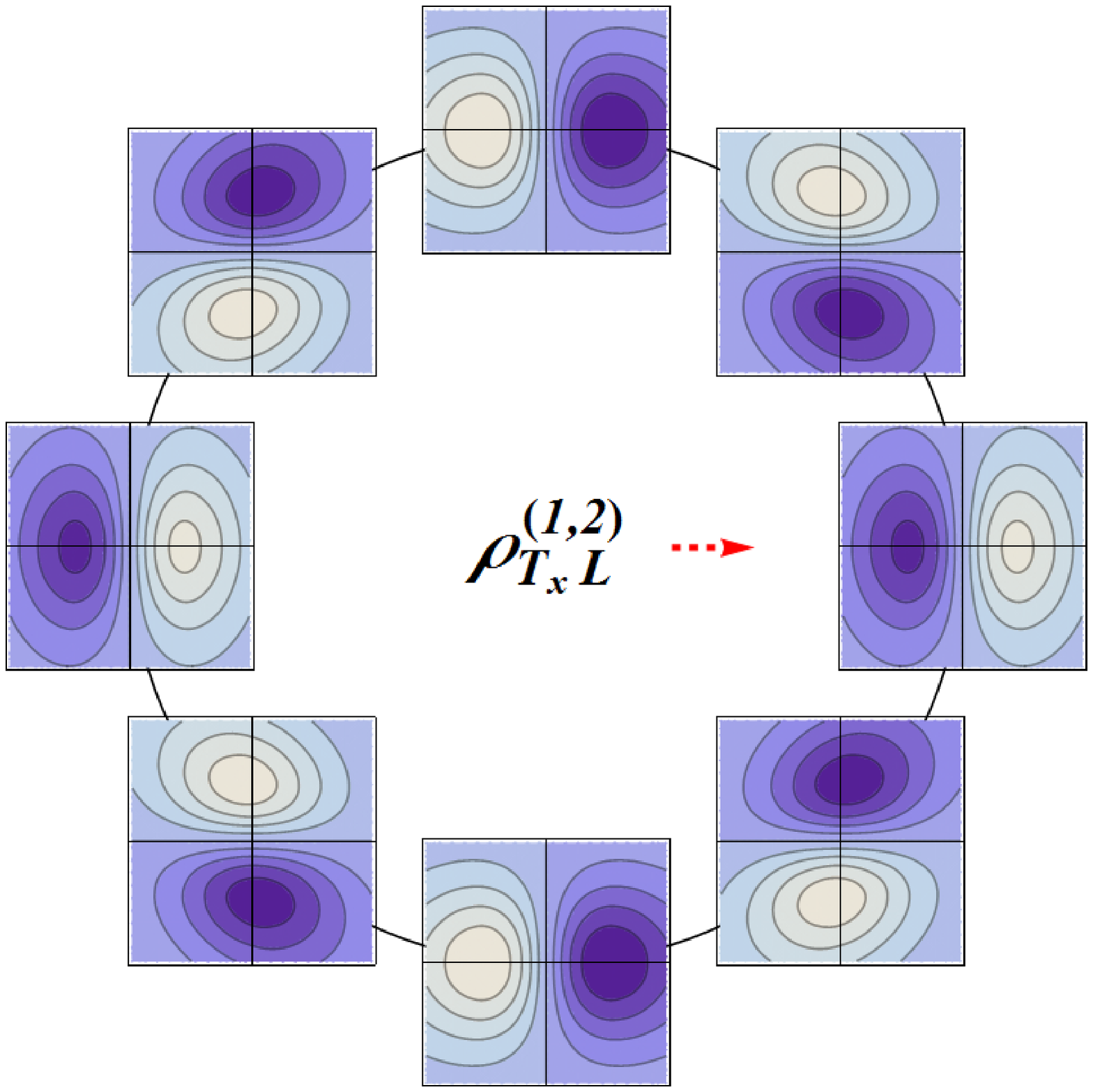}\hspace{1.5cm}\includegraphics[width=7cm]{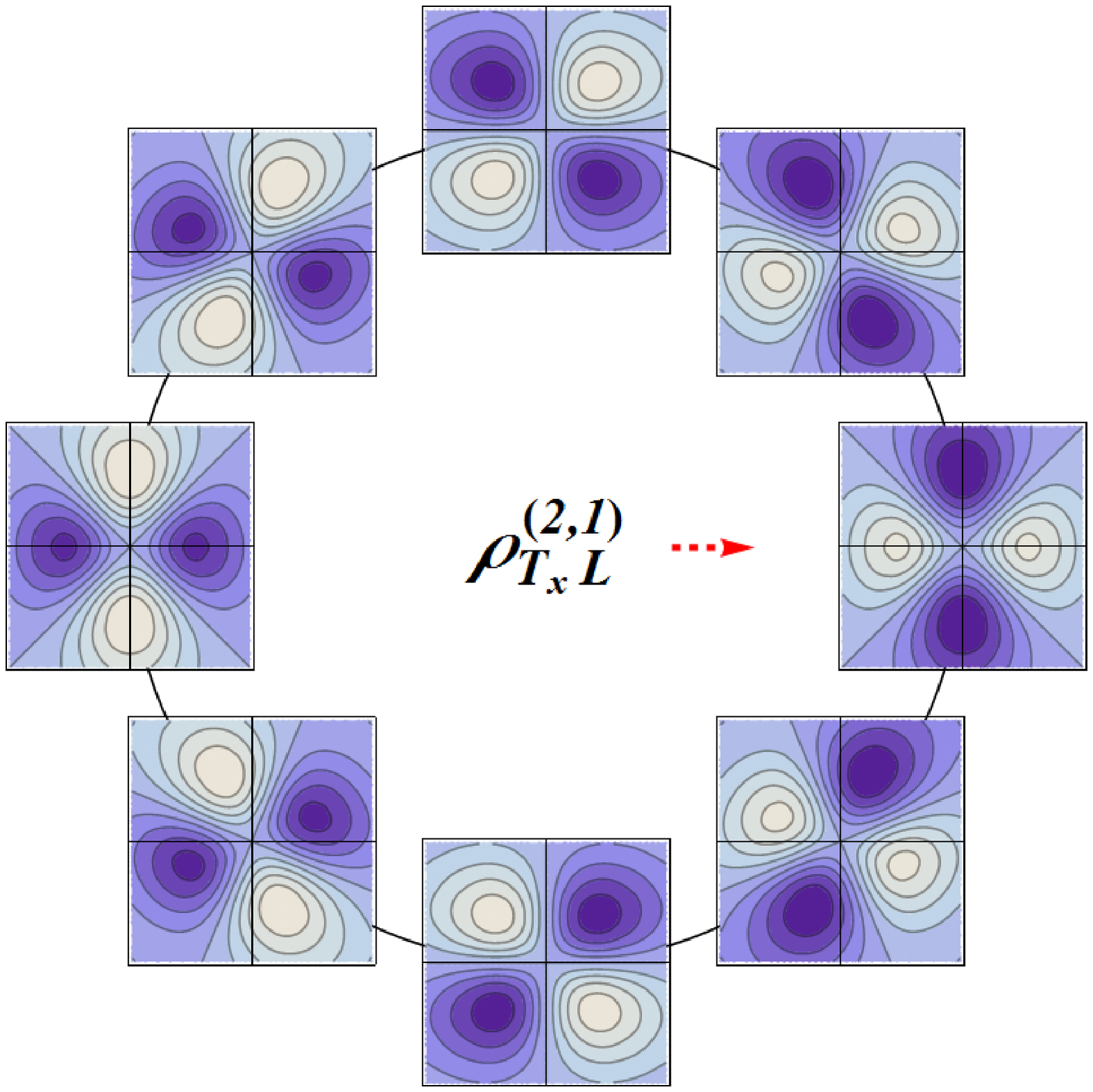}}
\vspace*{8pt}
\caption{Naive $\mathsf T$-even (left) and $\mathsf T$-odd (right) contributions to the transverse phase-space distribution $\rho_{TL}$ for the target polarization $\vec S_T=\vec e_x$ (red dashed arrow). See text for more details. \label{fig8}}
\end{figure}

The contribution $\rho_{T^iL}$ describes how the distribution of quarks is affected by the combination of quark longitudinal polarization and target transverse polarization. Its structure is very similar to $\rho_{LT^i}$ because one just exchanges the roles of quark and target polarizations. We then find in total four phase-space distributions
\begin{equation}
\rho^e_{T^iL}=\rho^{(1,0)}_{T^iL}+\rho^{(1,2)}_{T^iL},\qquad\rho^o_{T^iL}=\rho^{(0,1)}_{T^iL}+\rho^{(2,1)}_{T^iL},
\end{equation}
which are represented in Fig.~\ref{fig8} for the target polarization $\vec S_T=\vec e_x$. 
The corresponding basic multipoles are
\begin{align}
S^i_TS^q_L\,B^{(1,0)}_{T^iL}(\hat k_T,\hat b_T;\hat P,\eta)&=S^i_TS^q_LD^i_kM_b=S^q_L(\uvec S_T\cdot\hat k_T),\label{TL1}\\
S^i_TS^q_L\,B^{(1,2)}_{T^iL}(\hat k_T,\hat b_T;\hat P,\eta)&=S^i_TS^q_LD^j_kQ^{ij}_b=S^q_L\!\left[(\uvec S_T\cdot\hat b_T)\,(\hat b_T\cdot\hat k_T)-\tfrac{1}{2}\,(\uvec S_T\cdot\hat k_T)\right],\label{TL2}\\
S^i_TS^q_L\,B^{(0,1)}_{T^iL}(\hat k_T,\hat b_T;\hat P,\eta)&=\eta\,S^i_TS^q_LM_kD^i_b=\eta\,S^q_L(\uvec S_T\cdot\hat b_T),\label{TL3}\\
S^i_TS^q_L\,B^{(2,1)}_{T^iL}(\hat k_T,\hat b_T;\hat P,\eta)&=\eta\,S^i_TS^q_LQ^{ij}_kD^j_b=\eta\,S^q_L\!\left[(\uvec S_T\cdot\hat k_T)\,(\hat k_T\cdot\hat b_T)-\tfrac{1}{2}\,(\uvec S_T\cdot\hat b_T)\right].\label{TL4}
\end{align}
The contribution $\rho^{(1,0)}_{T^iL}$ is the only one surviving the integration over $\uvec b_T$ and is then naturally related to  the worm-gear TMD $g_{1T}$~\cite{Diehl:2005jf,Lorce:2011dv}. The dipole in $\uvec k_T$-space indicates the presence of a net transverse flow parallel to the quark transverse polarization. This transverse flow is once again due to the light-front imaging, associated with the fact that the light-front densities are defined in terms of the $j^+=\tfrac{1}{\sqrt{2}}(j^0+j^3)$ component of the current instead of the $j^0$ component. As we will soon see, it turns out that the transverse flow finds its physical origin in the correlation between the transverse component of quark OAM and the longitudinal spin-orbit coupling $\langle (\uvec S_T\cdot\uvec\ell^q_T)S^q_L\ell^q_L\rangle$.

The contribution $\rho^{(1,2)}_{T^iL}$ corresponds to a completely new information which is  not accessible \emph{via} GPDs or TMDs at leading twist. Combined with $\rho^{(1,0)}_{T^iL}$, it tells us how the quark distribution is affected  by the two transverse-longitudinal worm-gear correlations, say $\langle S_x\ell^q_xS^q_L\ell^q_L\rangle$ and $\langle S_y\ell^q_yS^q_L\ell^q_L\rangle$. Following the same arguments as in Sec.~\ref{subsect:LT}, with $S_L\uvec S_{T}^q$ replaced by $\uvec S_TS^q_L$,
we can relate the two coefficient functions $C^{(1,0)}_{T^iL}$ and $C^{(1,2)}_{T^iL}$  to the strength of the two transverse-longitudinal worm-gear correlations $\langle S_x\ell^q_xS^q_L\ell^q_L\rangle$ and $\langle S_y\ell^q_yS^q_L\ell^q_L\rangle$.
\newline

Similarly, the contribution $\rho^{(0,1)}_{T^iL}$ is the only one surviving the integration over $\uvec k_T$. It cannot however be related to the GPD $\tilde E$~\cite{Diehl:2005jf,Lorce:2011dv,Pasquini:2007xz} since the latter is $\eta$-independent\footnote{Moreover, while the GPD $\tilde E$ is $\xi$-even, it enters the amplitude with an explicit $\xi$ factor and cannot therefore appear in our multipole decomposition based on $\xi=0$. It then corresponds to a completely new information.}.  Once again, the dipole in $\uvec b_T$-space indicates the presence of a spatial separation between quarks with opposite correlations. This is likely  another effect due to the light-front imaging.

The contribution $\rho^{(2,1)}_{T^iL}$ corresponds to a completely new information which is  not accessible \emph{via} GPDs or TMDs at leading twist. Combined with $\rho^{(0,1)}_{T^iL}$, it tells us how the initial- and final-state interactions depend on the two transverse-longitudinal worm-gear correlations, say $\langle S_x\ell^q_xS^q_L\ell^q_L\rangle$ and $\langle S_y\ell^q_yS^q_L\ell^q_L\rangle$.

Following once again the same arguments as in Sec.~\ref{subsect:LT}, with $S_L\uvec S_{T}^q$ replaced by $\uvec S_TS^q_L$,
 we can relate the two coefficient functions $C^{(0,1)}_{T^iL}$ and $C^{(2,1)}_{T^iL}$  to the strength of the $\langle S_x\ell^q_xS^q_L\ell^q_L\rangle$- and $\langle S_y\ell^q_yS^q_L\ell^q_L\rangle$-dependent parts of the force felt by the quark due to initial- and final-state interactions. In other words, the contributions $\rho^{(0,1)}_{T^iL}$ and $\rho^{(2,1)}_{T^iL}$ describe the difference of radial flows between quarks with opposite $\langle S_x\ell^q_xS^q_L\ell^q_L\rangle$ or $\langle S_y\ell^q_yS^q_L\ell^q_L\rangle$ correlations.

\subsubsection{Transversely polarized quark}

\begin{figure}[t]
\centerline{\includegraphics[width=7cm]{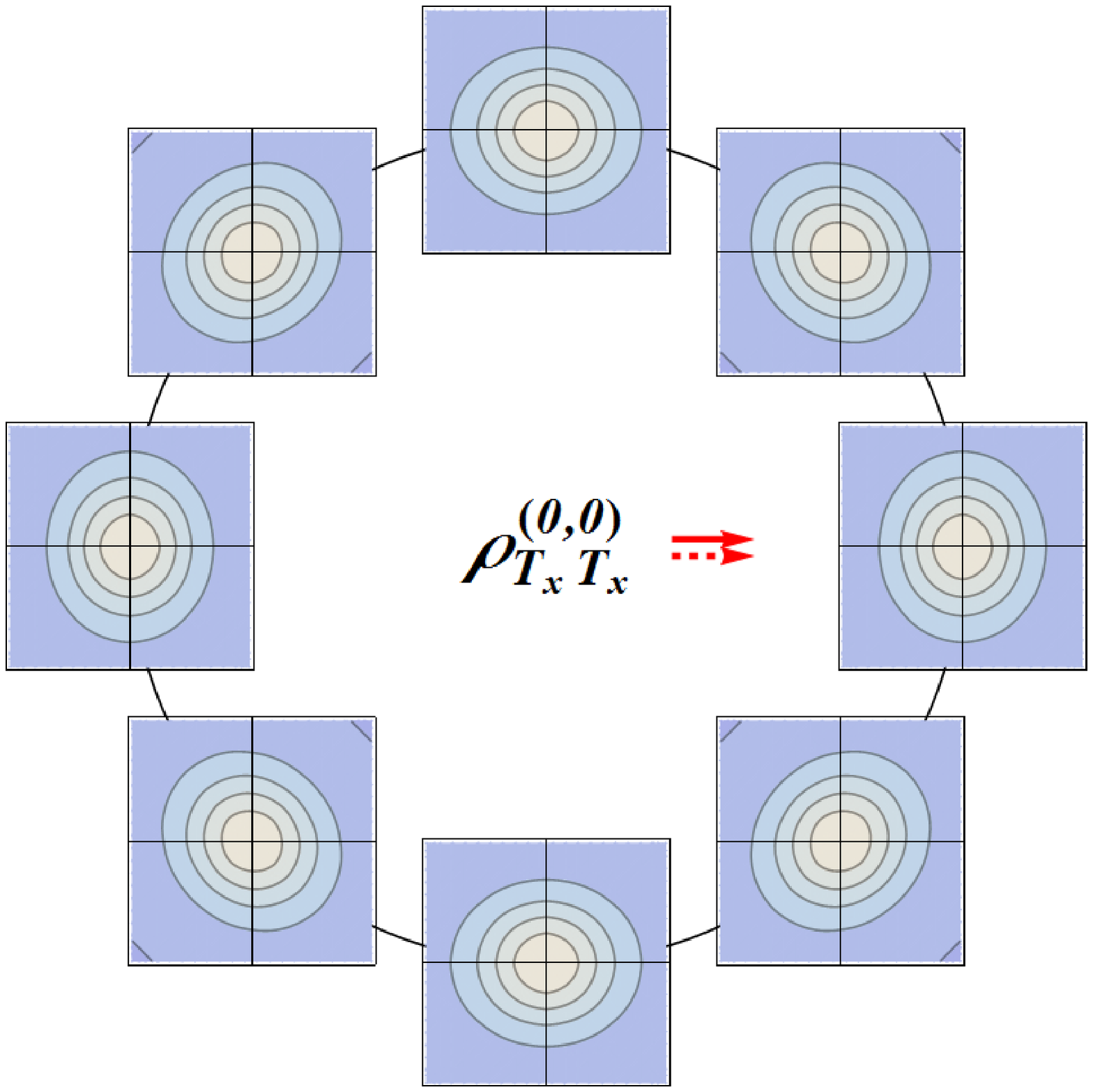}\hspace{1.5cm}\includegraphics[width=7cm]{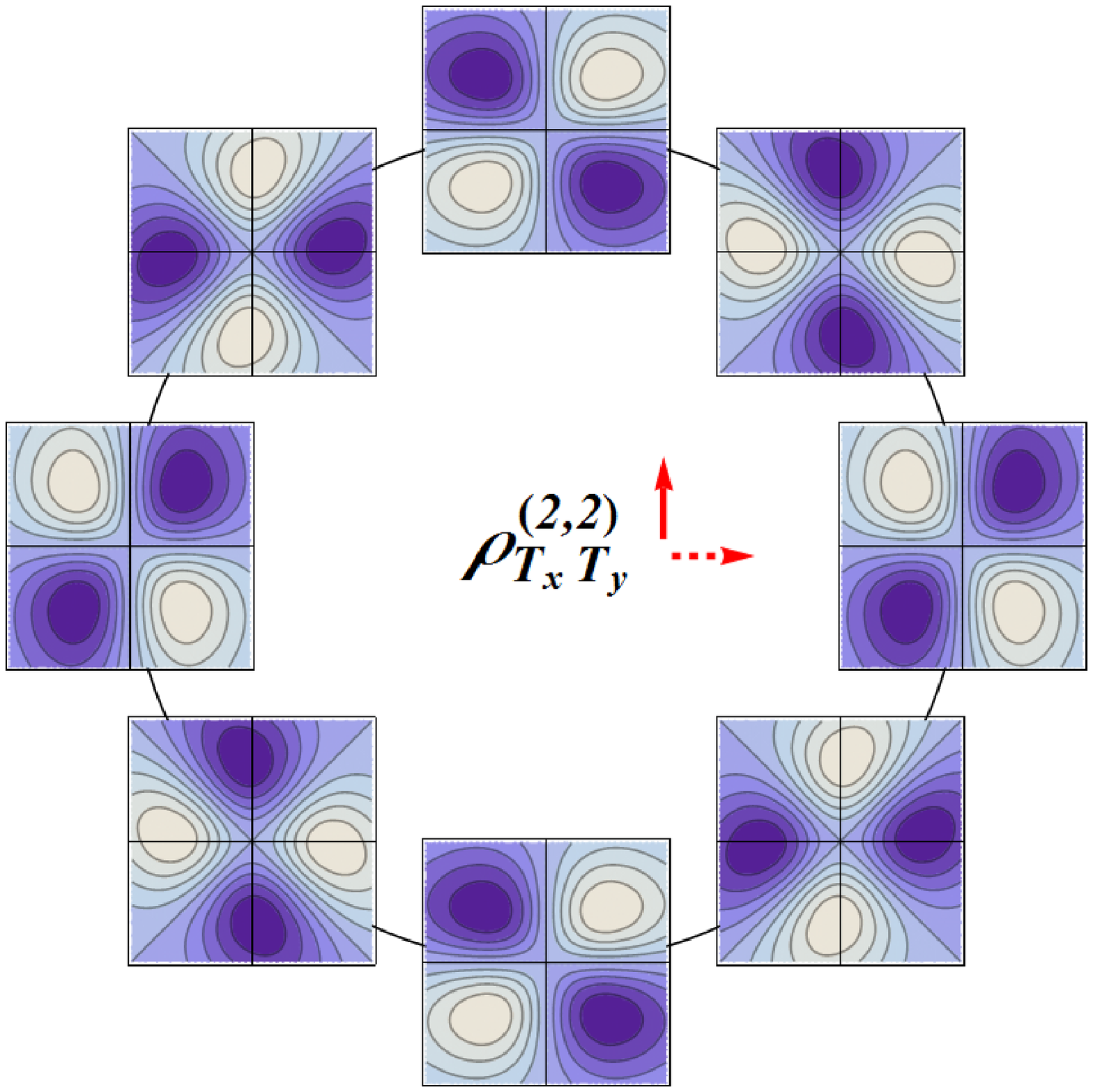}}
\vspace*{15pt}
\centerline{\includegraphics[width=7cm]{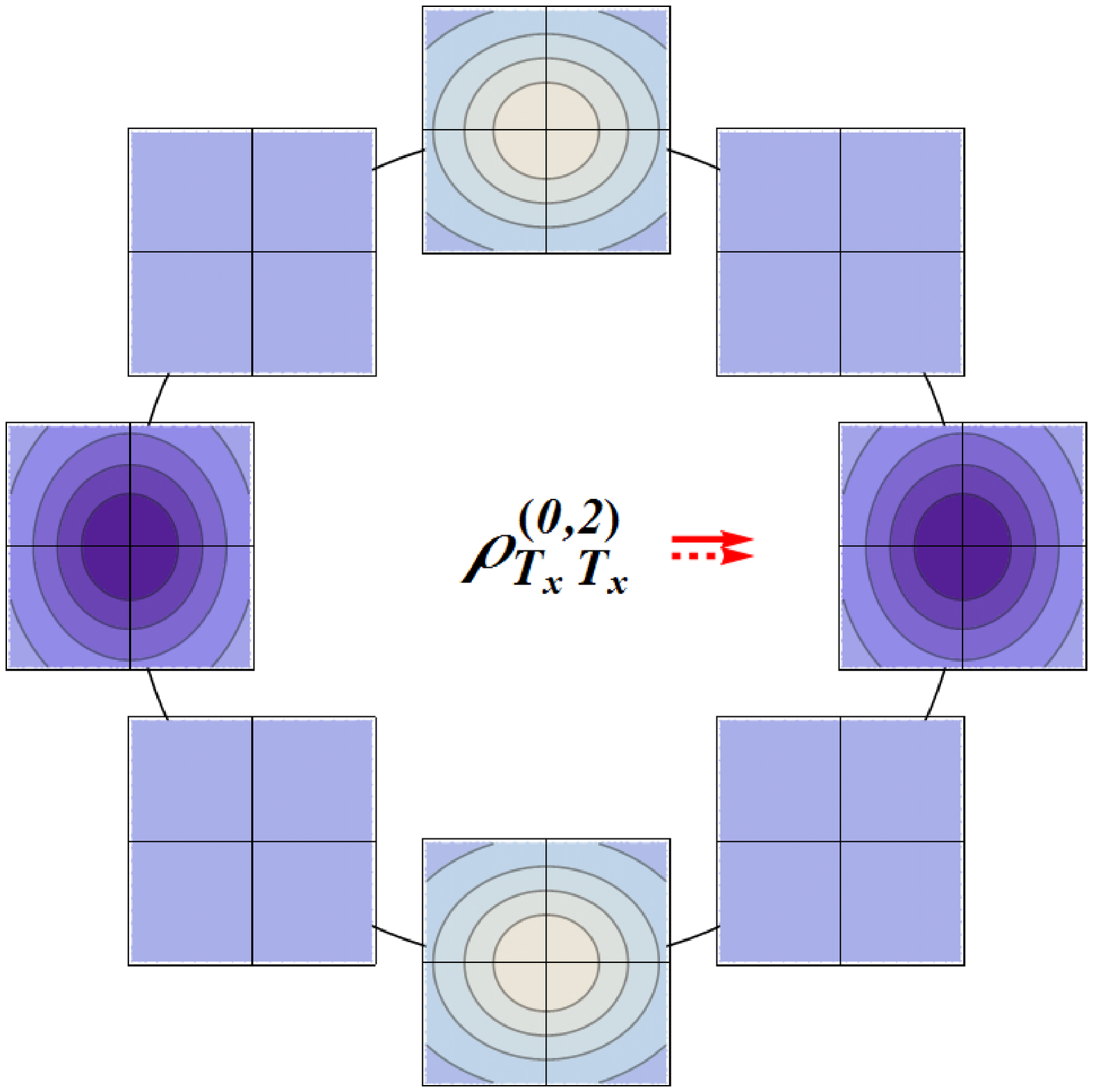}\hspace{1.5cm}\includegraphics[width=7cm]{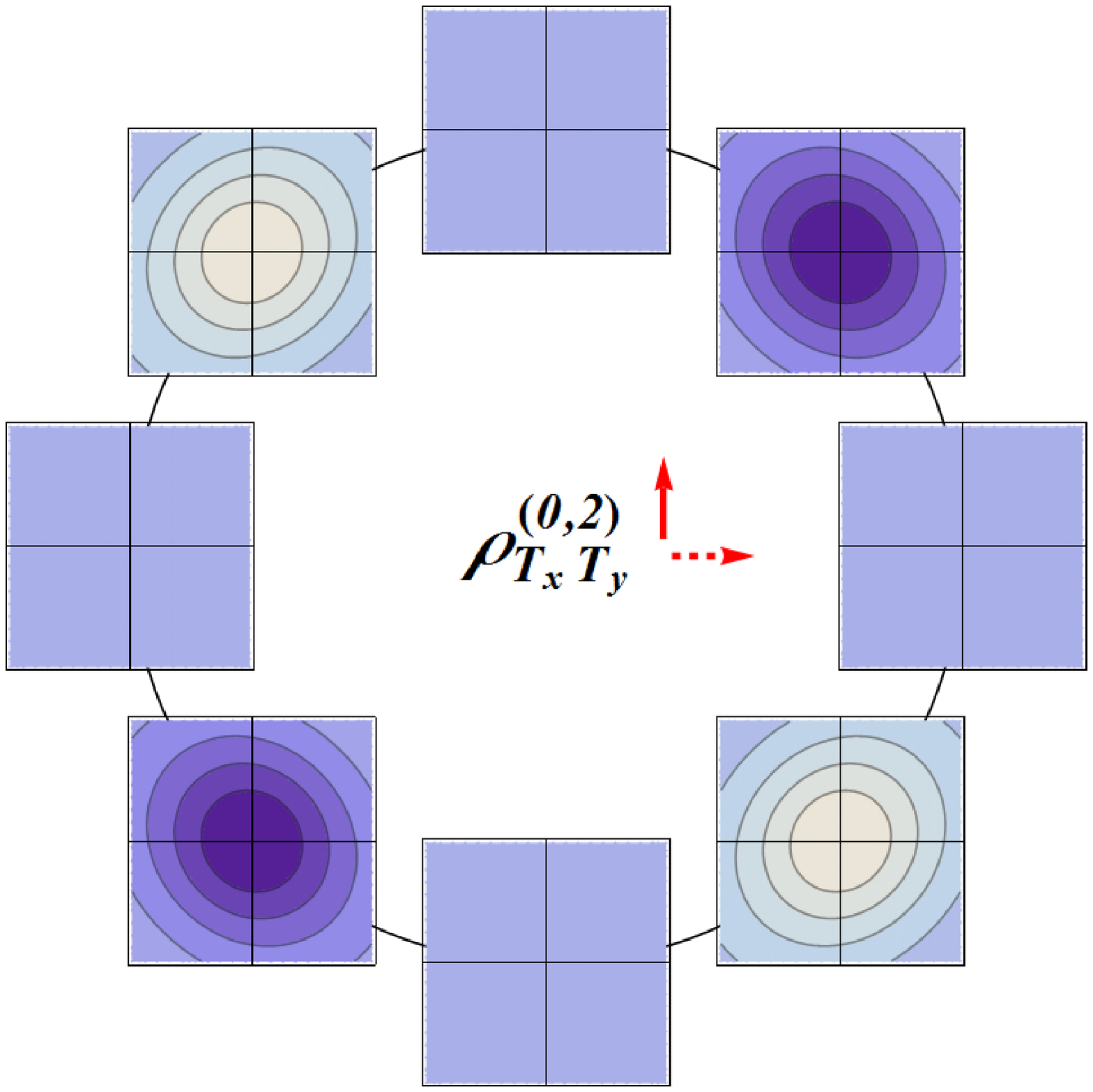}}
\vspace*{15pt}
\centerline{\includegraphics[width=7cm]{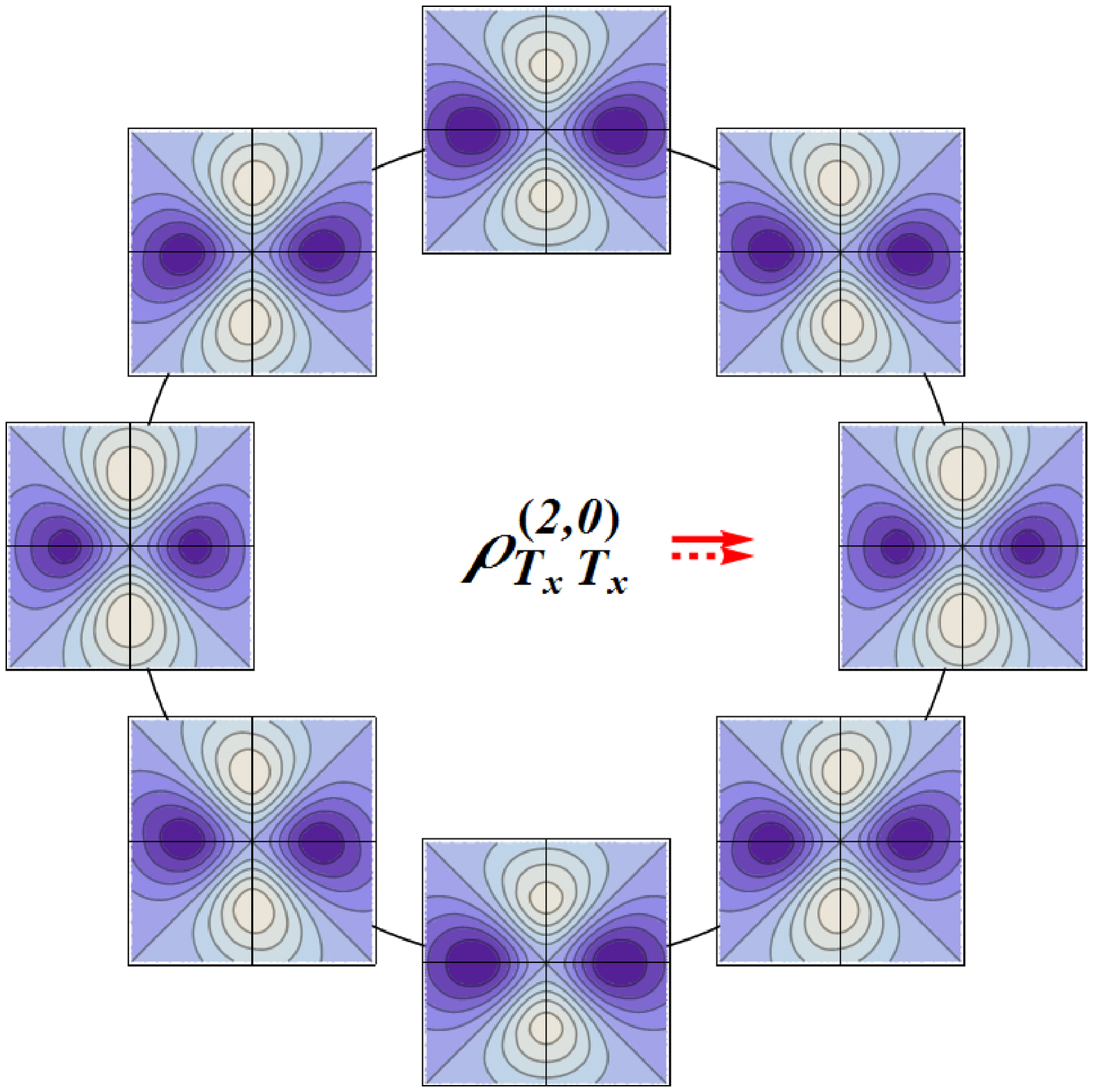}\hspace{1.5cm}\includegraphics[width=7cm]{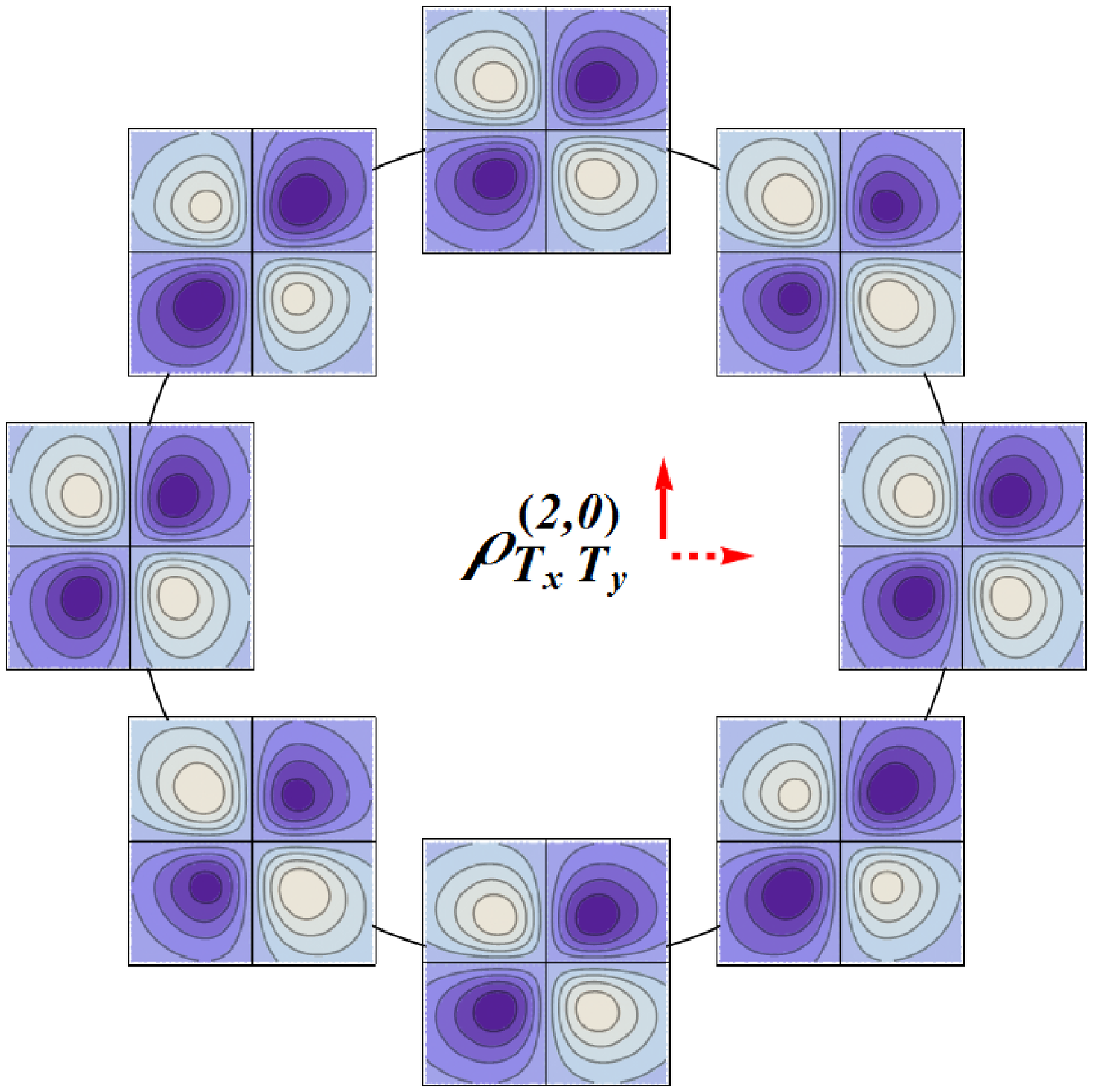}}
\vspace*{8pt}
\caption{Naive $\mathsf T$-even contributions to the transverse phase-space distribution $\rho_{TT}$ for the target polarization $\vec S_T=\vec e_x$ (red dashed arrow) and for the two quark polarizations (red solid arrow) $\vec S^q_T=\vec e_x$ (left) and $\vec S^q_T=\vec e_y$ (right). See text for more details. \label{fig9}}
\end{figure}

The contribution $\rho_{T^iT^j}$ describes how the quark distribution is affected by the correlation between the quark and target transverse polarizations. Focusing on the naive $\mathsf T$-even sector, we find four phase-space distributions
\begin{equation}
\rho^e_{T^iT^j}=\rho^{(0,0)}_{T^iT^j}+\rho^{(0,2)}_{T^iT^j}+\rho^{(2,0)}_{T^iT^j}+\rho^{(2,2)}_{T^iT^j},
\end{equation}
which are represented in Fig.~\ref{fig9} for the target polarization $\vec S_T=\vec e_x$ and for the two quark polarizations $\vec S^q_T=\vec e_{x,y}$.
The corresponding basic multipoles are
\begin{align}
S^i_TS^{qj}_T\,B^{(0,0)}_{T^iT^j}(\hat k_T,\hat b_T;\hat P,\eta)&=S^i_TS^{qi}_LM_kM_b=(\uvec S_T\cdot\uvec S^q_T),\label{TT1}\\
S^i_TS^{qj}_T\,B^{(0,2)}_{T^iT^j}(\hat k_T,\hat b_T;\hat P,\eta)&=S^i_TS^{qj}_TM_kQ^{ij}_b=(\uvec S_T\cdot\hat b_T)\,(\hat b_T\cdot\uvec S^q_T)-\tfrac{1}{2}\,(\uvec S_T\cdot\uvec S^q_T),\label{TT2}\\
S^i_TS^{qj}_T\,B^{(2,0)}_{T^iT^j}(\hat k_T,\hat b_T;\hat P,\eta)&=S^i_TS^{qj}_TQ^{ij}_kM_b=(\uvec S_T\cdot\hat k_T)\,(\hat k_T\cdot\uvec S^q_T)-\tfrac{1}{2}\,(\uvec S_T\cdot\uvec S^q_T),\label{TT3}\\
S^i_TS^{qj}_T\,B^{(2,2)}_{T^iT^j}(\hat k_T,\hat b_T;\hat P,\eta)&=-S^i_TS^{qj}_T\epsilon^{ij}_T\epsilon^{mn}_TQ^{lm}_kQ^{ln}_b=(\uvec S_T\times\uvec S^q_T)_L\,(\hat b_T\times\hat k_T)_L\,(\hat k_T\cdot\hat b_T).\label{TT4}
\end{align}
The contribution $\rho^{(0,0)}_{T^iT^j}$ is the only one surviving both integrations over $\uvec b_T$ and $\uvec k_T$, and is then naturally related to both the transversity GPD combination $H_T+\tfrac{\uvec\Delta^2_T}{4M^2}\,\tilde H_T$ and the transversity TMD $h_1$~\cite{Diehl:2005jf,Lorce:2011dv}. Contrary to its $\uvec k_T$- and $\uvec b_T$-integrated versions, $\rho^{(0,0)}_{T^iT^j}$ is not circularly symmetric. The reason is that $\rho^{(0,0)}_{T^iT^j}$ also contains information about the \emph{correlation} between $\uvec k_T$ and $\uvec b_T$, which is lost under integration over one of the transverse variables~\cite{Lorce:2011kd}.
\newline
Following the same arguments as in Sec.~\ref{subsect:UU} for $\rho_{UU}^{(0,0)}$, with now the corresponding expressions multiplied by $\uvec S_T\cdot\uvec S^q_T$, we can relate the coefficient function $C^{(0,0)}_{T^iT^j}$ to the strength of the correlation between the transverse component of quark and target polarizations $\langle \uvec S_T\cdot\uvec S^q_T\rangle$.

The contribution $\rho^{(0,2)}_{T^iT^j}$ is the only other contribution surviving integration over $\uvec k_T$ and is then naturally related to the the GPD $\tilde H_T$~\cite{Diehl:2005jf,Lorce:2011dv,Pasquini:2007xz}. Similarly, the contribution $\rho^{(2,0)}_{T^iT^j}$ is the only other contribution surviving integration over $\uvec b_T$ and is then naturally related to the the pretzelosity TMD $h^\perp_{1T}$~\cite{Diehl:2005jf,Miller:2007ae,She:2009jq,Avakian:2010br,Lorce:2011kn}. Combined with $\rho^{(0,0)}_{T^iT^j}$, these two contributions tell us how the quark distribution is affected  by the two transverse spin-spin correlations, say $\langle S_xS^q_x\rangle$ and $\langle S_yS^q_y\rangle$. Indeed, let us consider the projection of a 3-dimensional $\langle (\vec S_T\cdot\vec n_T)(\vec S^q_T\cdot\vec n_T)\rangle$ correlation onto the transverse position space. For $\vec n_T=\vec b_T$ and $\vec n_T=(\vec b_T\times\hat P)$, we respectively find
\begin{align}
\int\ud b_L\,(\vec S_T\cdot\vec b_T)\,(\vec S^q_T\cdot\vec b_T)&\sim (\uvec S_T\cdot\hat b_T)\,(\uvec S^q_T\cdot\hat b_T),\\
\int\ud b_L\,[\vec S_T\cdot(\vec b_T\times\hat P)]\,[\vec S^q_T\cdot(\vec b_T\times\hat P)]&\sim(\uvec S_T\times\hat b_T)_L\,(\uvec S^q_T\times\hat b_T)_L,
\end{align}
and similarly for $\vec n_T=\vec k_T$ and $\vec n_T=(\vec k_T\times\hat P)$. Now, noting that for any unit transverse vector $\hat n_T$
\begin{equation}
(\uvec S_T\cdot\hat n_T)\,(\uvec S^q_T\cdot\hat n_T)+(\uvec S_T\times\hat n_T)_L\,(\uvec S^q_T\times\hat n_T)_L=(\uvec S_T\cdot\uvec S^q_T)
\end{equation}
and comparing with the basic multipoles~\eqref{TT1},~\eqref{TT2} and~\eqref{TT3}, we can see that the three coefficient functions $C^{(0,0)}_{T^iT^j}$, $C^{(0,2)}_{T^iT^j}$ and $C^{(2,0)}_{T^iT^j}$ are related to the strength of the two transverse spin-spin correlations $\langle S_xS^q_x\rangle$ and $\langle S_yS^q_y\rangle$.

It may seem weird that we need three contributions to determine two transverse spin-spin correlations. The reason is that the two contributions $\rho^{(0,2)}_{T^iT^j}$ and $\rho^{(2,0)}_{T^iT^j}$ also contain information about another type of correlation. Combined with $\rho^{(2,2)}_{T^iT^j}$, which corresponds to a completely new information  not accessible \emph{via} GPDs or TMDs at leading twist, they also tell us how the quark distribution is affected  by the two transverse-transverse worm-gear correlations, say $\langle S_x\ell^q_xS^q_y\ell^q_y\rangle$ and $\langle S_y\ell^q_yS^q_x\ell^q_x\rangle$. Indeed, let us consider the projection of a 3-dimensional $\langle (\vec S_T\cdot\vec n_T)(\vec\ell^q_T\cdot\vec n_T)[\vec S^q_T\cdot(\vec n_T\times\hat P)][\vec \ell^q_T\cdot(\vec n_T\times\hat P)]\rangle$ correlation onto the transverse position space. For $\vec n_T=\vec b_T$ and $\vec n_T=(\vec b_T\times\hat P)$, we respectively find
\begin{align}
\int\ud b_L\,(\vec S_T\cdot\vec b_T)\,[(\vec b\times\vec k)_T\cdot\vec b_T]\,[\vec S^q_T\cdot(\vec b_T\times\hat P)]\,[(\vec b\times\vec k)_T\cdot(\vec b_T\times\hat P)]&\sim (\uvec S_T\cdot\hat b_T)\,(\uvec S^q_T\times\hat b_T)_L\,(\hat b_T\times\hat k_T)_L\,(\hat k_T\cdot\hat b_T),\\
\int\ud b_L\,[\vec S_T\cdot(\vec b_T\times\hat P)]\,[(\vec b\times\vec k)_T\cdot(\vec b_T\times\hat P)]\,(\vec S^q_T\cdot\vec b_T)\,[(\vec b\times\vec k)_T\cdot\vec b_T)]&\sim(\uvec S_T\times\hat b_T)_L\,(\uvec S^q_T\cdot\hat b_T)\,(\hat b_T\times\hat k_T)_L\,(\hat k_T\cdot\hat b_T),
\end{align}
and similarly for $\vec n_T=\vec k_T$ and $\vec n_T=(\vec k_T\times\hat P)$. Now, noting that for any unit transverse vector $\hat n_T$
\begin{align}
(\uvec S_T\times\hat n_T)_L\,(\uvec S^q_T\cdot\hat n_T)-&(\uvec S_T\cdot\hat n_T)\,(\uvec S^q_T\times\hat n_T)_L=(\uvec S_T\times\uvec S^q_T)_L,\\
[(\uvec S_T\times\hat n_T)_L\,(\uvec S^q_T\cdot\hat n_T)+&(\uvec S_T\cdot\hat n_T)\,(\uvec S^q_T\times\hat n_T)_L]\,(\hat b_T\times\hat k_T)_L\,(\hat k_T\cdot\hat b_T)\nonumber\\
&=[(\hat k_T\cdot\hat n_T)^2-(\hat k_T\times\hat n_T)^2_L]\,[(\uvec S_T\cdot\hat b_T)\,(\hat b_T\cdot\uvec S^q_T)-\tfrac{1}{2}\,(\uvec S_T\cdot\uvec S^q_T)]\nonumber\\
-&[(\hat b_T\cdot\hat n_T)^2-(\hat b_T\times\hat n_T)^2_L]\,[(\uvec S_T\cdot\hat k_T)\,(\hat k_T\cdot\uvec S^q_T)-\tfrac{1}{2}\,(\uvec S_T\cdot\uvec S^q_T)],
\end{align}
and comparing with the basic multipoles~\eqref{TT2},~\eqref{TT3} and~\eqref{TT4}, we can see that the three coefficient functions $C^{(0,2)}_{T^iT^j}$, $C^{(2,0)}_{T^iT^j}$ and $C^{(2,2)}_{T^iT^j}$ are related to the strength of the two transverse-transverse worm-gear correlations $\langle S_x\ell^q_xS^q_y\ell^q_y\rangle$ and $\langle S_y\ell^q_yS^q_x\ell^q_x\rangle$.
\newline 

\begin{figure}[t]
\centerline{\includegraphics[width=7cm]{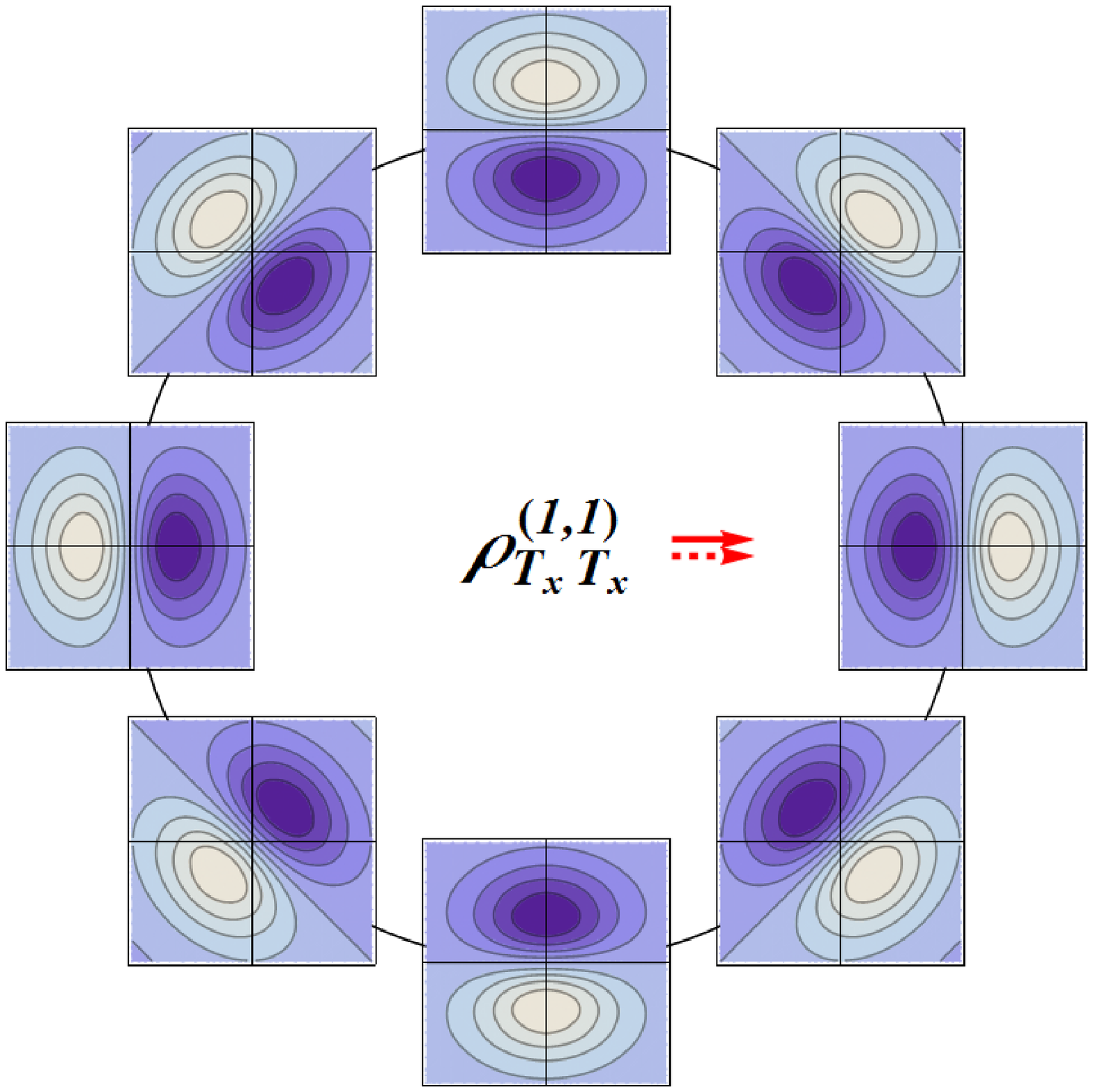}\hspace{1.5cm}\includegraphics[width=7cm]{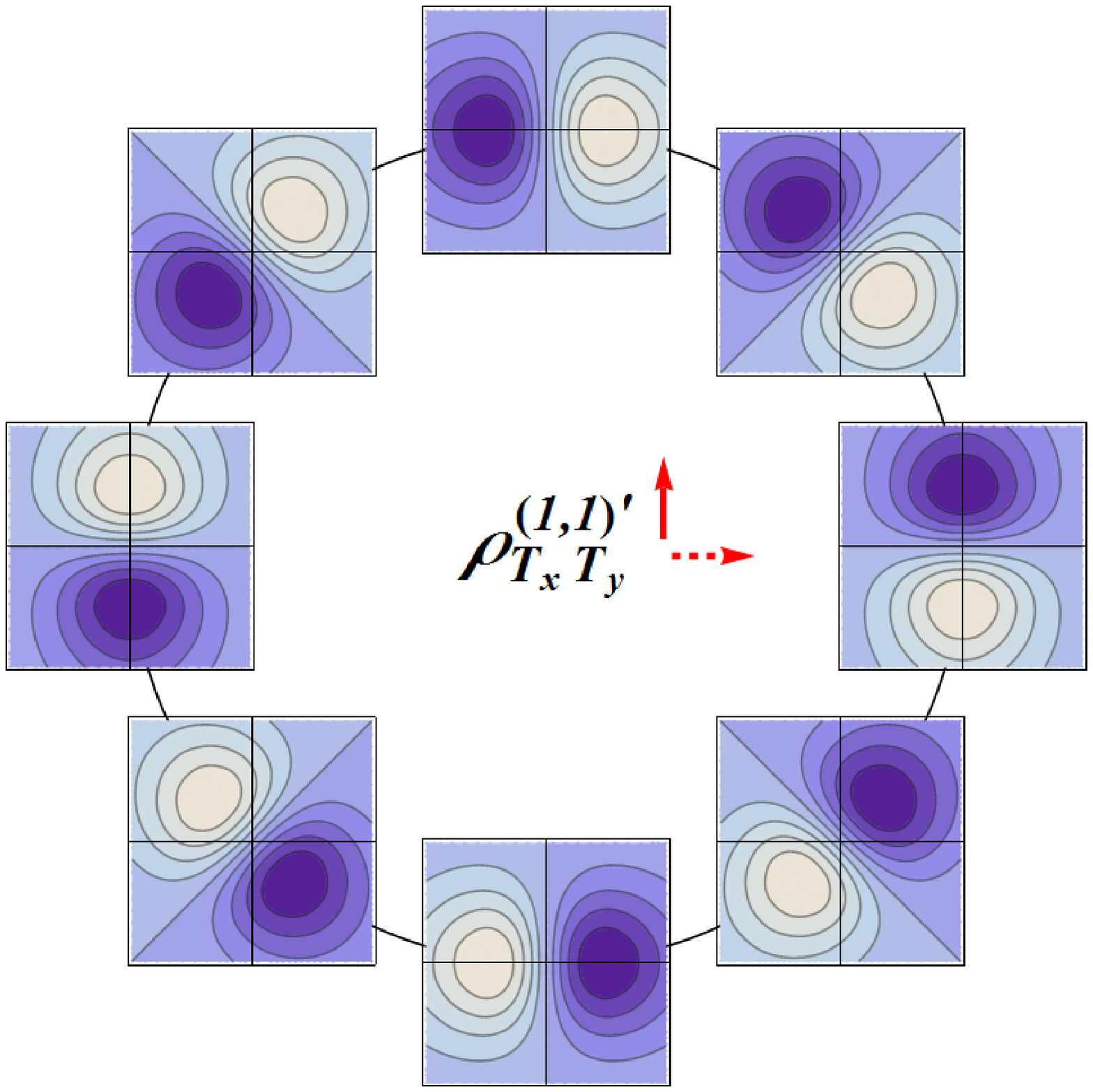}}
\vspace*{15pt}
\centerline{\includegraphics[width=7cm]{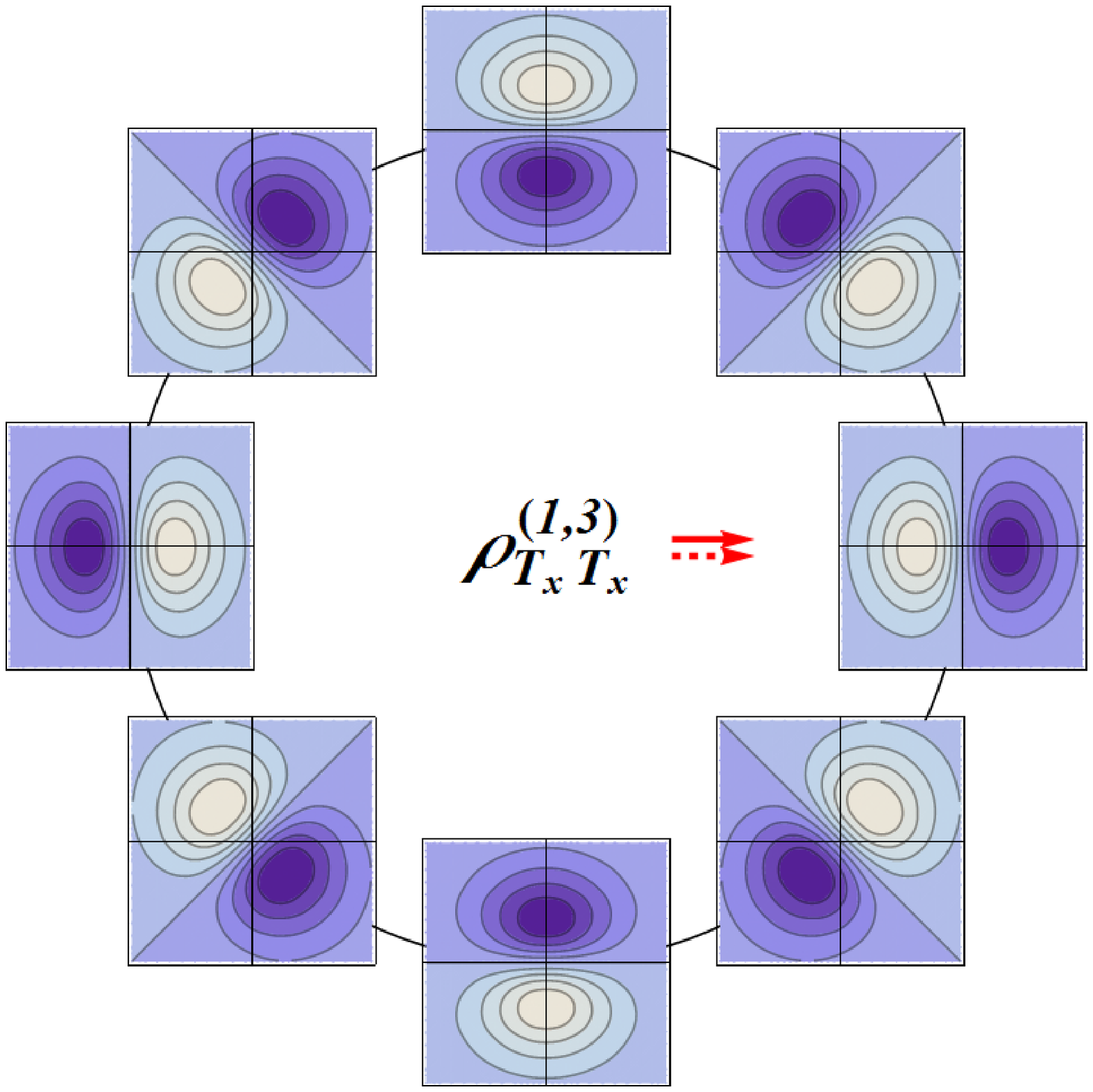}\hspace{1.5cm}\includegraphics[width=7cm]{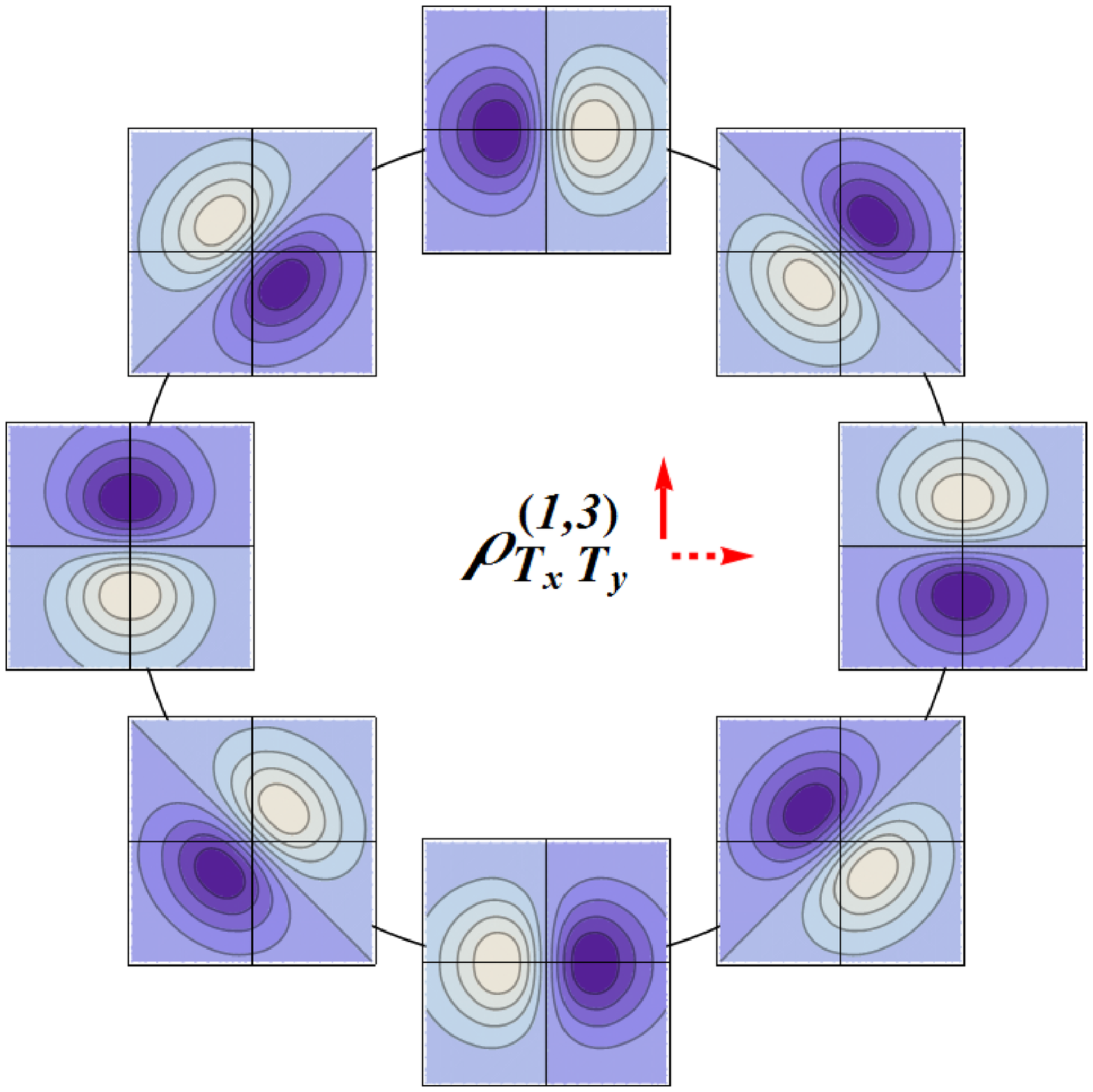}}
\vspace*{15pt}
\centerline{\includegraphics[width=7cm]{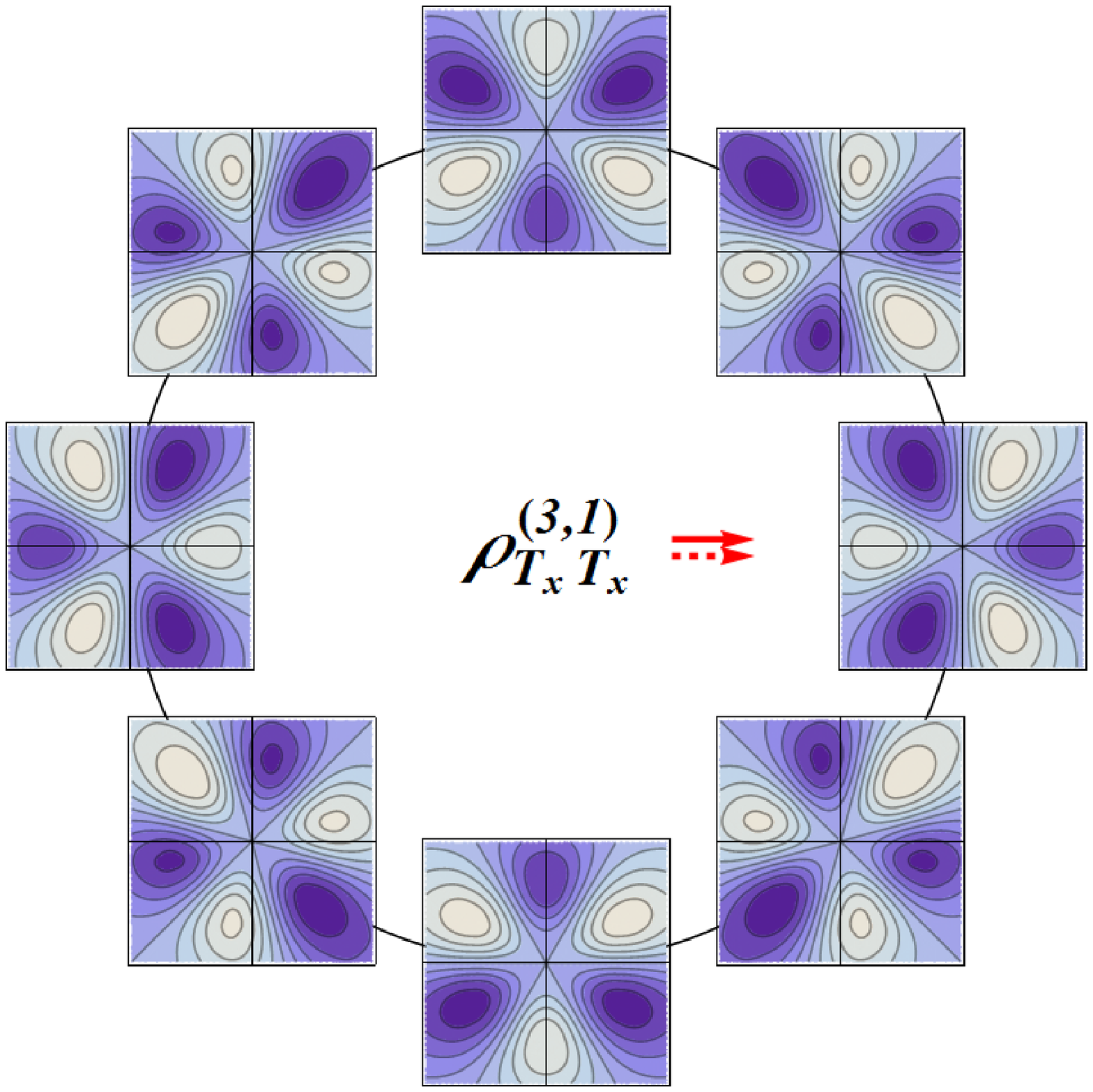}\hspace{1.5cm}\includegraphics[width=7cm]{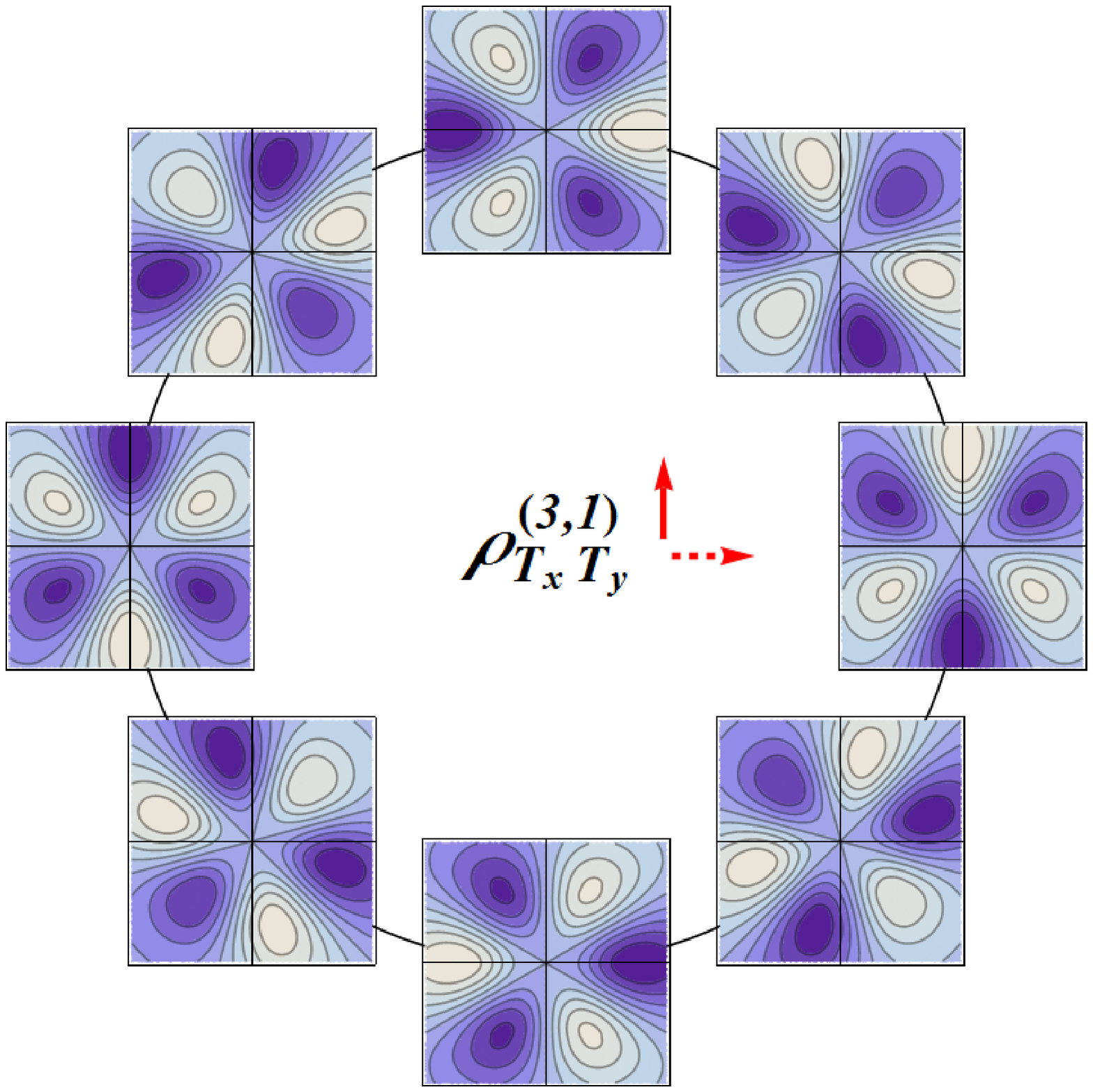}}
\vspace*{8pt}
\caption{Naive $\mathsf T$-odd contributions to the transverse phase-space distribution $\rho_{TT}$ for the target polarization $\vec S_T=\vec e_x$ (red dashed arrow) and for the two quark polarizations (red solid arrow) $\vec S^q_T=\vec e_x$ (left) and $\vec S^q_T=\vec e_y$ (right). See text for more details. \label{fig10}}
\end{figure}

Focusing now on the naive $\mathsf T$-odd sector, we also find four phase-space distributions
\begin{equation}
\rho^o_{T^iT^j}=\rho^{(1,1)}_{T^iT^j}+\rho^{(1,3)}_{T^iT^j}+\rho^{(3,1)}_{T^iT^j}+\rho^{(1,1)'}_{T^iT^j},
\end{equation}
which are represented in Fig.~\ref{fig10} for the target polarization $\vec S_T=\vec e_x$ and for the two quark polarizations $\vec S^q_T=\vec e_{x,y}$.
The corresponding basic multipoles are
\begin{align}
S^i_TS^{qj}_T\,B^{(1,1)}_{T^iT^j}(\hat k_T,\hat b_T;\hat P,\eta)&=S^i_TS^{qi}_LD^j_kD^j_b=(\uvec S_T\cdot\uvec S^q_T)\,(\hat k_T\cdot\hat b_T),\label{TT5}\\
S^i_TS^{qj}_T\,B^{(1,3)}_{T^iT^j}(\hat k_T,\hat b_T;\hat P,\eta)&=S^i_TS^{qj}_TD^l_kO^{ijl}_b\nonumber\\
=(\uvec S_T\cdot\hat b_T)\,(\uvec S^q_T\cdot\hat b_T)&\,(\hat k_T\cdot\hat b_T)-\tfrac{1}{4}\,[(\uvec S_T\cdot\uvec S^q_T)\,(\hat k_T\cdot\hat b_T)+(\uvec S_T\cdot\hat b_T)\,(\uvec S^q_T\cdot\hat k_T)+(\uvec S_T\cdot\hat k_T)\,(\uvec S^q_T\cdot\hat b_T)],\label{TT6}\\
S^i_TS^{qj}_T\,B^{(3,1)}_{T^iT^j}(\hat k_T,\hat b_T;\hat P,\eta)&=S^i_TS^{qj}_TO^{ijl}_kD^l_b\nonumber\\
=(\uvec S_T\cdot\hat k_T)\,(\uvec S^q_T\cdot\hat k_T)&\,(\hat k_T\cdot\hat b_T)-\tfrac{1}{4}\,[(\uvec S_T\cdot\uvec S^q_T)\,(\hat k_T\cdot\hat b_T)+(\uvec S_T\cdot\hat b_T)\,(\uvec S^q_T\cdot\hat k_T)+(\uvec S_T\cdot\hat k_T)\,(\uvec S^q_T\cdot\hat b_T)],\label{TT7}\\
S^i_TS^{qj}_T\,B^{(1,1)'}_{T^iT^j}(\hat k_T,\hat b_T;\hat P,\eta)&=-S^i_TS^{qj}_T\epsilon^{ij}_T\epsilon^{lm}_TD^l_kD^m_b=(\uvec S_T\times\uvec S^q_T)_L\,(\hat b_T\times\hat k_T)_L.\label{TT8}
\end{align}
None of these survive integration over $\uvec k_T$ or $\uvec b_T$. They therefore represent completely new information which is  not accessible \emph{via} GPDs or TMDs at leading twist. 

Following the same arguments as in Sec.~\ref{subsect:UU} for $\rho_{UU}^{(1,1)}$, with now the corresponding expressions multiplied by $\uvec S_T\cdot\uvec S^q_T$, we can relate the coefficient function $C^{(1,1)}_{T^iT^j}$ to the strength of the correlation between the transverse component of quark and target polarizations $\langle \uvec S_T\cdot\uvec S^q_T\rangle$. Combining $\rho^{(1,1)}_{T^iT^j}$ with $\rho^{(1,3)}_{T^iT^j}$ and $\rho^{(3,1)}_{T^iT^j}$ tells us how the initial- and final-state interactions depend separately on the two transverse spin-spin correlations, say $\langle S_xS^q_x\rangle$ and $\langle S_yS^q_y\rangle$. Indeed, let us consider the projection of a 3-dimensional radial flow \mbox{$(\vec S_T\cdot\vec n_T)\,(\vec S^q_T\cdot\vec n_T)\,(\vec k\cdot\vec b)$} onto the transverse position space. For $\vec n_T=\vec b_T$ and $\vec n_T=(\vec b_T\times\hat P)$, we respectively find
\begin{align}
\int\ud b_L\,(\vec S_T\cdot\vec b_T)\,(\vec S^q_T\cdot\vec b_T)\,(\vec k\cdot\vec b)&\sim (\uvec S_T\cdot\hat b_T)\,(\uvec S^q_T\cdot\hat b_T)\,(\hat k_T\cdot\hat b_T),\\
\int\ud b_L\,[\vec S_T\cdot(\vec b_T\times\hat P)]\,[\vec S^q_T\cdot(\vec b_T\times\hat P)]\,(\vec k\cdot\vec b)&\sim(\uvec S_T\times\hat b_T)_L\,(\uvec S^q_T\times\hat b_T)_L\,(\hat k_T\cdot\hat b_T),
\end{align}
and similarly for $\vec n_T=\vec k_T$ and $\vec n_T=(\vec k_T\times\hat P)$. Now, noting that for any unit transverse vectors $\hat n_T$ and $\hat n'_T$
\begin{align}
(\uvec S_T\cdot\hat n_T)\,(\uvec S^q_T\cdot\hat n_T)+(\uvec S_T\times\hat n_T)_L\,(\uvec S^q_T\times\hat n_T)_L&=(\uvec S_T\cdot\uvec S^q_T),\\
(\uvec S_T\cdot\hat n_T)\,(\uvec S^q_T\cdot\hat n_T)(\hat n_T\cdot\hat n'_T)+(\uvec S_T\cdot\hat n'_T)\,(\uvec S^q_T\cdot\hat n'_T)(\hat n'_T\cdot\hat n_T)& \nonumber\\
=[(\uvec S_T\cdot\hat n_T)\,(\uvec S^q_T\cdot\hat n'_T)+(\uvec S_T\cdot\hat n'_T)\,(\uvec S^q_T\cdot\hat n_T)]\,(\hat n_T\cdot\hat n'_T)^2&+[(\uvec S_T\cdot\uvec S^q_T)(\hat n_T\cdot\hat n'_T)]\,(\hat n_T\times\hat n'_T)^2_L,
\end{align}
and comparing with the basic multipoles~\eqref{TT5},~\eqref{TT6} and~\eqref{TT7}, we can see that the three coefficient functions $C^{(1,1)}_{T^iT^j}$, $C^{(1,3)}_{T^iT^j}$ and $C^{(3,1)}_{T^iT^j}$ are related to the strength of the $\langle S_xS^q_x\rangle$- and $\langle S_yS^q_y\rangle$-dependent parts of the force felt by the quark due to initial- and final-state interactions. In other words, the contributions $\rho^{(1,1)}_{T^iT^j}$, $\rho^{(1,3)}_{T^iT^j}$ and $\rho^{(3,1)}_{T^iT^j}$ describe the difference of radial flows between quarks with opposite $\langle S_xS^q_x\rangle$ or $\langle S_yS^q_y\rangle$ correlations.

Like in the naive $\mathsf T$-even sector, it may seem weird that we need three contributions to determine the dependence of initial- and final-state interactions on two transverse spin-spin correlations. The reason is that the two contributions $\rho^{(1,3)}_{T^iT^j}$ and $\rho^{(3,1)}_{T^iT^j}$ also contain information about another type of dependence. Combined with $\rho^{(1,1)'}_{T^iT^j}$, they also tell us how the initial- and final-state interactions depend separately on the two transverse-transverse worm-gear correlations, say $\langle S_x\ell^q_xS^q_y\ell^q_y\rangle$ and $\langle S_y\ell^q_yS^q_x\ell^q_x\rangle$. Indeed, let us consider the projection of a 3-dimensional spiral worm-gear flow \mbox{$(\vec S_T\cdot\vec n_T)(\vec\ell^q_T\cdot\vec n_T)[\vec S^q_T\cdot(\vec n_T\times\hat P)][\vec \ell^q_T\cdot(\vec n_T\times\hat P)](\vec k\cdot\vec b)$} onto the transverse position space. For $\vec n_T=\vec b_T$ and $\vec n_T=(\vec b_T\times\hat P)$, we respectively find
\begin{align}
\int\ud b_L\,(\vec S_T\cdot\vec b_T)\,[(\vec b\times\vec k)_T\cdot\vec b_T]\,[\vec S^q_T\cdot(\vec b_T\times\hat P)]\,[(\vec b\times\vec k)_T\cdot(\vec b_T\times\hat P)]\,(\vec k\cdot\vec b)&\sim (\uvec S_T\cdot\hat b_T)\,(\uvec S^q_T\times\hat b_T)_L\,(\hat b_T\times\hat k_T)_L,\\
\int\ud b_L\,[\vec S_T\cdot(\vec b_T\times\hat P)]\,[(\vec b\times\vec k)_T\cdot(\vec b_T\times\hat P)]\,(\vec S^q_T\cdot\vec b_T)\,[(\vec b\times\vec k)_T\cdot\vec b_T]\,(\vec k\cdot\vec b)&\sim(\uvec S_T\times\hat b_T)_L\,(\uvec S^q_T\cdot\hat b_T)\,(\hat b_T\times\hat k_T)_L,
\end{align}
and similarly for $\vec n_T=\vec k_T$ and $\vec n_T=(\vec k_T\times\hat P)$. Noting that for any unit transverse vectors $\hat n_T$ and $\hat n'_T$
\begin{align}
(\uvec S_T\times\hat n_T)_L\,(\uvec S^q_T\cdot\hat n_T)-&(\uvec S_T\cdot\hat n_T)\,(\uvec S^q_T\times\hat n_T)_L=(\uvec S_T\times\uvec S^q_T)_L,\\
4(\uvec S_T\cdot\hat n_T)\,(\uvec S^q_T\cdot\hat n_T)\,(\hat n_T\cdot\hat n'_T)-&[(\uvec S_T\cdot\uvec S^q_T)\,(\hat n_T\cdot\hat n'_T)+(\uvec S_T\cdot\hat n_T)\,(\uvec S^q_T\cdot\hat n'_T)+(\uvec S_T\cdot\hat n'_T)\,(\uvec S^q_T\cdot\hat n_T)]\nonumber\\
&=[(\uvec S_T\cdot\hat n_T)\,(\uvec S^q_T\cdot\hat n_T)-(\uvec S_T\times\hat n_T)_L\,(\uvec S^q_T\times\hat n_T)_L]\,(\hat n_T\cdot\hat n'_T)\nonumber\\
+&[(\uvec S_T\cdot\hat n_T)\,(\uvec S^q_T\times\hat n_T)_L+(\uvec S_T\times\hat n_T)_L\,(\uvec S^q_T\cdot\hat n_T)]\,(\hat n_T\times\hat n'_T)_L,
\end{align}
and comparing with the basic multipoles~\eqref{TT6},~\eqref{TT7} and~\eqref{TT8}, we can see that the three coefficient functions $C^{(1,3)}_{T^iT^j}$, $C^{(3,1)}_{T^iT^j}$ and $C^{(1,1)'}_{T^iT^j}$ are related to the strength of the $\langle S_x\ell^q_xS^q_y\ell^q_y\rangle$- and $\langle S_y\ell^q_yS^q_x\ell^q_x\rangle$-dependent parts of the force felt by the quark due to initial- and final-state interactions. In other words, the contributions $\rho^{(1,3)}_{T^iT^j}$, $\rho^{(3,1)}_{T^iT^j}$ and $\rho^{(1,1)'}_{T^iT^j}$ describe the difference  of radial flows between quarks with opposite $\langle S_x\ell^q_xS^q_y\ell^q_y\rangle$ or $\langle S_y\ell^q_yS^q_x\ell^q_x\rangle$ correlations.

\section{Conclusions}
\label{section:5}
We presented for the first time a systematic study of the complete set of the leading-twist quark Wigner distributions in the nucleon, introducing a  multipole analysis 
in the transverse phase space. 
In this approach each distribution is represented as combination of basic multipoles structures multiplied by coefficient functions giving the corresponding strengths. 
The multipole structures are obtained for each configuration of the nucleon and target polarizations, taking into account the constraints from hermiticity,  parity and time-reversal transformations, while the coefficient functions depend on $\mathsf P$- and $\mathsf T$-invariant hermitian variables only.
There are several advantages in using this representation.
First, it provides a clear interpretation of all the amplitudes in terms of the possible correlations between target and quark angular momenta in the transverse phase space.
Second, it provides a convenient basis to make a direct connection with GPDs in impact-parameter space and TMD in transverse-momentum space after integration over the transverse-momentum and the transverse-position space, respectively. In order to emphasize these multipole structures, we also proposed a new graphical representation of the transverse phase-space distributions.

We presented results for both the naive $\mathsf T$-even and naive $\mathsf T$-odd contributions.
The first ones describe the contributions to the \emph{intrinsic} distribution of quarks inside the target, whereas the naive $\mathsf T$-odd contributions describe how initial- and final-state interactions modify this distribution. 
We have explicitly calculated the naive $\mathsf T-$even contributions adopting a light-front quark model, whereas 
the naive $\mathsf T$-odd contributions have been obtained by extracting the coefficient functions from the naive $\mathsf T$-even part and multiplying them by the appropriate basic multipoles.
In this way, the global sign of the naive $\mathsf T$-odd contributions has been  chosen arbitrarily. Only a proper calculation taking into account the dynamics of the initial- and/or final-state interactions can determine the global signs.
However,  these global signs are not important for the purpose of the present paper since we wanted to emphasize the general features related to the multipole structure of the distribution, and to identify the physical (angular) correlation encoded in each distribution.

\section*{Acknowledgements}

For a part of this work, C.L. was supported by the Belgian Fund F.R.S.-FNRS \emph{via} the contract of Charg\'e de recherches.


\begin{thebibliography}{99}

\bibitem{Wigner:1932eb} 
  E.~P.~Wigner,
  Phys.\ Rev.\  {\bf 40}, 749 (1932).

\bibitem{Balazs:1983hk} 
  N.~L.~Balazs and B.~K.~Jennings,
  Phys.\ Rept.\  {\bf 104}, 347 (1984).

\bibitem{Hillery:1983ms} 
  M.~Hillery, R.~F.~O'Connell, M.~O.~Scully and E.~P.~Wigner,
  Phys.\ Rept.\  {\bf 106}, 121 (1984).

\bibitem{Lee:1995}
  H.-W.~Lee, 
  Phys.\ Rept.\ {\bf 259}, 147 (1995).

\bibitem{Carruthers:1982fa} 
  P.~Carruthers and F.~Zachariasen,
  Rev.\ Mod.\ Phys.\  {\bf 55}, 245 (1983).
\bibitem{Hakim:1976bn} 
  R.~Hakim,
  Riv.\ Nuovo Cim.\  {\bf 1N6}, 1 (1978).
  doi:10.1007/BF02724474

\bibitem{DeGroot:1980dk} 
  S.~R.~De Groot, W.~A.~Van Leeuwen and C.~G.~Van Weert,
  Amsterdam, Netherlands: North-holland ( 1980) 417p
  
\bibitem{Elze:1986qd} 
  H.~T.~Elze, M.~Gyulassy and D.~Vasak,
  Nucl.\ Phys.\ B {\bf 276}, 706 (1986).


\bibitem{Ochs:1998qj} 
  S.~Ochs and U.~W.~Heinz,
  Annals Phys.\  {\bf 266}, 351 (1998).

\bibitem{Heinz:1983nx} 
  U.~W.~Heinz,
  Phys.\ Rev.\ Lett.\  {\bf 51}, 351 (1983).
  doi:10.1103/PhysRevLett.51.351

\bibitem{Heinz:1984yq} 
  U.~W.~Heinz,
  Annals Phys.\  {\bf 161}, 48 (1985).
  doi:10.1016/0003-4916(85)90336-7

\bibitem{Elze:1986hq} 
  H.~T.~Elze, M.~Gyulassy and D.~Vasak,
  Phys.\ Lett.\ B {\bf 177}, 402 (1986).
  doi:10.1016/0370-2693(86)90778-1

\bibitem{Ji:2003ak} 
  X.~d.~Ji,
  Phys.\ Rev.\ Lett.\  {\bf 91}, 062001 (2003).

\bibitem{Belitsky:2003nz} 
  A.~V.~Belitsky, X.~d.~Ji and F.~Yuan,
  Phys.\ Rev.\ D {\bf 69}, 074014 (2004).

\bibitem{Soper:1976jc} 
  D.~E.~Soper,
  Phys.\ Rev.\ D {\bf 15}, 1141 (1977).

\bibitem{Burkardt:2000za} 
  M.~Burkardt,
  Phys.\ Rev.\ D {\bf 62}, 071503 (2000)
  [Erratum-ibid.\ D {\bf 66}, 119903 (2002)].

\bibitem{Burkardt:2002hr} 
  M.~Burkardt,
  Int.\ J.\ Mod.\ Phys.\ A {\bf 18}, 173 (2003).

\bibitem{Burkardt:2005hp} 
  M.~Burkardt,
  Phys.\ Rev.\ D {\bf 72}, 094020 (2005).

\bibitem{Lorce:2011kd} 
  C.~Lorc\'e and B.~Pasquini,
  Phys.\ Rev.\ D {\bf 84}, 014015 (2011).

\bibitem{Meissner:2009ww} 
  S.~Meissner, A.~Metz and M.~Schlegel,
  JHEP {\bf 0908}, 056 (2009).

\bibitem{Lorce:2011dv} 
  C.~Lorc\'e, B.~Pasquini and M.~Vanderhaeghen,
  JHEP {\bf 1105}, 041 (2011).

\bibitem{Lorce:2013pza} 
  C.~Lorc\'e and B.~Pasquini,
  JHEP {\bf 1309}, 138 (2013).

\bibitem{Hatta:2011ku} 
  Y.~Hatta,
  Phys.\ Lett.\ B {\bf 708}, 186 (2012).


\bibitem{Lorce:2011ni} 
  C.~Lorc\'e, B.~Pasquini, X.~Xiong and F.~Yuan,
  Phys.\ Rev.\ D {\bf 85}, 114006 (2012).



\bibitem{Liu:2015xha} 
  K.~F.~Liu and C.~Lorc\'e,
  arXiv:1508.00911 [hep-ph].
  
\bibitem{Martin:1999wb} 
  A.~D.~Martin, M.~G.~Ryskin and T.~Teubner,
  Phys.\ Rev.\ D {\bf 62}, 014022 (2000).

\bibitem{Khoze:2000cy} 
  V.~A.~Khoze, A.~D.~Martin and M.~G.~Ryskin,
  Eur.\ Phys.\ J.\ C {\bf 14}, 525 (2000).

\bibitem{Martin:2001ms} 
  A.~D.~Martin and M.~G.~Ryskin,
  Phys.\ Rev.\ D {\bf 64}, 094017 (2001).

\bibitem{Albrow:2008pn} 
  M.~G.~Albrow {\it et al.}  [FP420 R and D Collaboration],
  JINST {\bf 4}, T10001 (2009).

\bibitem{Martin:2009ku} 
  A.~D.~Martin, M.~G.~Ryskin and V.~A.~Khoze,
  Acta Phys.\ Polon.\ B {\bf 40}, 1841 (2009).

\bibitem{Kanazawa:2014nha} 
  K.~Kanazawa, C.~Lorc\'e, A.~Metz, B.~Pasquini and M.~Schlegel,
  Phys.\ Rev.\ D {\bf 90}, 014028 (2014).

\bibitem{Mukherjee:2014nya} 
  A.~Mukherjee, S.~Nair and V.~K.~Ojha,
  Phys.\ Rev.\ D {\bf 90}, 014024 (2014).

\bibitem{Liu:2014vwa} 
  T.~Liu,
  arXiv:1406.7709 [hep-ph].

\bibitem{Liu:2015eqa} 
  T.~Liu and B.~Q.~Ma,
  Phys.\ Rev.\ D {\bf 91}, 034019 (2015).

\bibitem{Miller:2014vla} 
  G.~A.~Miller,
  Phys.\ Rev.\ D {\bf 90},  113001 (2014).
  
\bibitem{Mukherjee:2015aja} 
  A.~Mukherjee, S.~Nair and V.~K.~Ojha,
  Phys.\ Rev.\ D {\bf 91},  054018 (2015).
\bibitem{Burkardt:2015qoa} 
  M.~Burkardt and B.~Pasquini,
  arXiv:1510.02567 [hep-ph].
  
\bibitem{Ji:2013dva} 
  X.~Ji,
  Phys.\ Rev.\ Lett.\  {\bf 110}, 262002 (2013).

\bibitem{Lorce:2011zta} 
  C.~Lorc\'e and B.~Pasquini,
  Phys.\ Rev.\ D {\bf 84}, 034039 (2011).


\bibitem{Soper:1972xc} 
  D.~E.~Soper,
  Phys.\ Rev.\ D {\bf 5}, 1956 (1972).

\bibitem{Carlson:2003je} 
  C.~E.~Carlson and C.~R.~Ji,
  Phys.\ Rev.\ D {\bf 67}, 116002 (2003).


\bibitem{Brodsky:2006ez} 
  S.~J.~Brodsky, S.~Gardner and D.~S.~Hwang,
  Phys.\ Rev.\ D {\bf 73}, 036007 (2006).




\bibitem{Diehl:2005jf} 
  M.~Diehl and P.~Hagler,
  Eur.\ Phys.\ J.\ C {\bf 44}, 87 (2005).


\bibitem{Lorce:2015lna} 
  C.~Lorc\'e,
  JHEP {\bf 1508}, 045 (2015).

\bibitem{Lorce:2012ce} 
  C.~Lorc\'e,
  Phys.\ Lett.\ B {\bf 719}, 185 (2013).

\bibitem{paper} 
  M.G.~Echevarria \emph{et al.}, in preparation.
 
\bibitem{Lorce:2014mxa} 
  C.~Lorc\'e,
  Phys.\ Lett.\ B {\bf 735}, 344 (2014).

 
  \bibitem{Pasquini:2007xz} 
  B.~Pasquini and S.~Boffi,
  Phys.\ Lett.\ B {\bf 653}, 23 (2007).


\bibitem{Moller:1949}
  C.~M\o ller,
  Commun.\ Dublin\ Inst.\ Adv.\ Stud.\ A {\bf 5}, 1 (1949). 

\bibitem{Moller:1972}
  C.~M\o ller,
  \emph{The Theory of Relativity}, 2nd ed., Oxford Univ. Press, Oxford, 1972, p. 176.

\bibitem{Dyakonov:1971a}
  M.~I.~Dyakonov and V.~I.~Perel,
  Sov.\ Phys.\ JETP Lett.\ {\bf 13}, 467 (1971).

\bibitem{Dyakonov:1971b}
  M.~I.~Dyakonov and V.~I.~Perel,
  Phys.\ Lett.\ A {\bf 35}, 459 (1971).

\bibitem{Ji:2012ba} 
  X.~Ji, X.~Xiong and F.~Yuan,
  Phys.\ Rev.\ D {\bf 88}, no. 1, 014041 (2013).


\bibitem{Hatta:2012cs} 
  Y.~Hatta and S.~Yoshida,
  JHEP {\bf 1210}, 080 (2012).


\bibitem{Bacchetta:2011gx} 
  A.~Bacchetta and M.~Radici,
  Phys.\ Rev.\ Lett.\  {\bf 107}, 212001 (2011).


       \bibitem{Miller:2007ae} 
  G.~A.~Miller,
  Phys.\ Rev.\ C {\bf 76}, 065209 (2007).

\bibitem{She:2009jq} 
  J.~She, J.~Zhu and B.~Q.~Ma,
  Phys.\ Rev.\ D {\bf 79}, 054008 (2009).
\bibitem{Avakian:2010br} 
  H.~Avakian, A.~V.~Efremov, P.~Schweitzer and F.~Yuan,
  Phys.\ Rev.\ D {\bf 81}, 074035 (2010).
  
\bibitem{Lorce:2011kn} 
  C.~Lorc\'e and B.~Pasquini,
  Phys.\ Lett.\ B {\bf 710}, 486 (2012).
  















\end{thebibliography}
\end{document}